\documentclass[%
reprint,
superscriptaddress,
 amsmath,amssymb,
 aps,
]{revtex4-1}

\usepackage{mathtools}
\usepackage{graphicx}
\usepackage{dcolumn}
\usepackage{bm}

\usepackage{amsmath}
\usepackage{amssymb}
\usepackage{amsthm}
\usepackage{graphicx}
\usepackage{algorithmic}
\usepackage{mathrsfs}
\usepackage{braket}
\usepackage{xcolor}
\usepackage{tikz}
\usepackage[roman]{complexity}
\usepackage[caption=false]{subfig}

\newcommand{\dbar}{d\hspace*{-0.08em}\bar{}\hspace*{0.1em}}
\newtheorem{theorem}{Theorem}

\newtheorem{definition}{Definition}

\usepackage{pdfpages}
\usepackage{pgffor}
\makeatletter
\AtBeginDocument{\let\LS@rot\@undefined}
\makeatother

\begin{document}

\title{Representation Learning via Quantum Neural Tangent Kernels}

\author{Junyu Liu}
\email{junyuliu@uchicago.edu}
\affiliation{Pritzker School of Molecular Engineering, The University of Chicago, Chicago, IL 60637, USA}
\affiliation{Chicago Quantum Exchange, Chicago, IL 60637, USA}
\affiliation{Kadanoff Center for Theoretical Physics, The University of Chicago, Chicago, IL 60637, USA}

\author{Francesco Tacchino}
\email{fta@zurich.ibm.com}
\affiliation{IBM Quantum, IBM Research -- Zurich, 8803 R{\"u}schlikon, Switzerland}

\author{Jennifer R. Glick}
\email{jennifer.r.glick@ibm.com}
\affiliation{IBM Quantum, IBM T. J. Watson Research Center, Yorktown Heights, NY 10598, USA}

\author{Liang Jiang}
\email{liang.jiang@uchicago.edu}
\affiliation{Pritzker School of Molecular Engineering, The University of Chicago, Chicago, IL 60637, USA}
\affiliation{Chicago Quantum Exchange, Chicago, IL 60637, USA}

\author{Antonio Mezzacapo}
\email{mezzacapo@ibm.com}
\affiliation{IBM Quantum, IBM T. J. Watson Research Center, Yorktown Heights, NY 10598, USA}

\date{\today}

\begin{abstract}
Variational quantum circuits are used in quantum machine learning and variational quantum simulation tasks. Designing good variational circuits or predicting how well they perform for given learning or optimization tasks is still unclear. Here we discuss these problems, analyzing variational quantum circuits using the theory of neural tangent kernels. We define quantum neural tangent kernels, and derive dynamical equations for their associated loss function in optimization and learning tasks. We analytically solve the dynamics in the frozen limit, or lazy training regime, where variational angles change slowly and a linear perturbation is good enough. We extend the analysis to a dynamical setting, including quadratic corrections in the variational angles. We then consider hybrid quantum-classical architecture and define a large-width limit for hybrid kernels, showing that a hybrid quantum-classical neural network can be approximately Gaussian. The results presented here show limits for which analytical understandings of the training dynamics for variational quantum circuits, used for quantum machine learning and optimization problems, are possible. These analytical results are supported by numerical simulations of quantum machine learning experiments. 

\end{abstract}

\maketitle


\section{Introduction}
The idea of using quantum computers for machine learning has recently received attention both in academia and industry~\cite{harrow2009quantum,wiebe2012quantum,lloyd2014quantum,wittek2014quantum,wiebe2014quantum,rebentrost2014quantum,biamonte2017quantum,mcclean2018barren,schuld2019quantum,tang2019quantum,havlivcek2019supervised,huang2021power,liu2021rigorous}. While proof of principle study have shown that some problems of mathematical interest quantum computers are useful~\cite{liu2021rigorous}, quantum advantage in machine learning algorithms for practical applications is still unclear \cite{huang2021information}. On classical architectures, a first-principle theory of machine learning, especially the so-called deep learning that uses a large number of layers, is still in development. Early developments of the statistical learning theory provide rigorous guarantees on the learning capability in generic learning algorithms, but theoretical bounds obtained from information theory are sometimes weak in practical settings. 

The theory of neural tangent kernel (NTK) has been deemed an important tool to understand deep neural networks \cite{lee2017deep,jacot2018neural,lee2019wide,arora2019exact,sohl2020infinite,yang2020feature,yaida2020non}. In the large-width limit, a generic neural network becomes nearly Gaussian when averaging over the initial weights and biases, and the learning capabilities become predictable. The NTK theory allows to derive analytical understanding of the neural networks dynamics, improving on statistical learning theory and shedding light on the underlying principle of deep learning \cite{dyer2019asymptotics,halverson2021neural,roberts2021ai,roberts2021principles,Liu:2021ohs}.
In the quantum machine learning community, a similar first principle theory would help in understanding the training dynamics and selecting appropriate variational quantum circuits to target specific problems. A step in this direction has been onsidered recently for quantum classical neural networks~\cite{nakaji2021quantumenhanced}. However in the framework considered there no variational parameters were considered in the quantum circuits, leaving the problem of understanding and designing the quantum dynamical training not addressed.

\begin{figure}[h]
    \centering
    \includegraphics[width=0.5\textwidth]{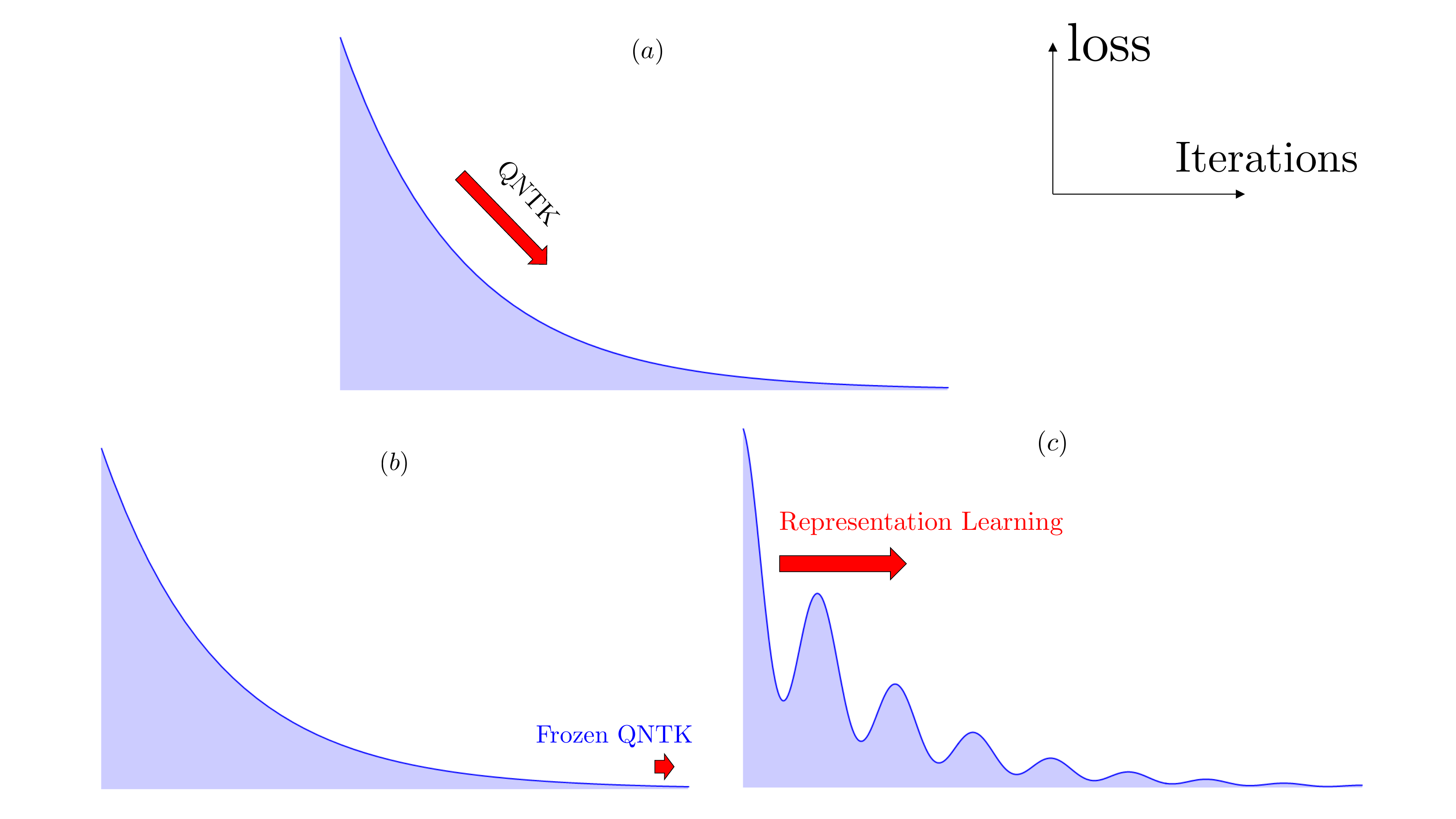}
    \caption{A cartoon illustration of the QNTK theory. $(a)$: the QNTK characterizes the gradient descent dynamics in the variational quantum circuit. The quantum state modifies according to the QNTK prediction. $(b)$: Around the end of the training, the QNTK is \emph{frozen} and almost a constant. $(c)$: The gradient descent dynamics could be highly non-linear, and the QNTK is running during gradient descent, which is a property of representation learning.}
    \label{fig:cartoon}
\end{figure}

In this paper, we address this problem, focusing on the limit where the learning rate is sufficiently small, inspired by the classical theory of NTK. Following the framework and results from \cite{roberts2021ai,roberts2021principles,summer}, we first define a quantum analogue of a classical NTK. In the limit where the variational angles do not change much, the so-called \emph{lazy training}~\cite{chizat2018lazy}, the \emph{frozen} QNTK leads to an exponential decaying of the loss function used on the training set. We furthermore compute the leading order perturbation above the static limit, where we define a quantum version of the classical \emph{meta-kernel}. We derive closed-form formulas for the dynamics of the training in terms of parameters of variational quantum circuits, see Fig. \ref{fig:cartoon}).

We then move to a hybrid quantum-classical neural network framework, and find that it becomes approximately Gaussian, as long as the quantum outputs are sufficiently orthogonal. We present an analytic derivation of the large-width limit where the non-Gaussian contribution to the neuron correlations is suppressed by large width. Interestingly, we observe that now the \emph{width} is defined by the number of independent Hermitian operators in the variational ansatz, which is upper-bounded by (a polynomial of) the dimension of the Hilbert space. Thus, a large Hilbert space size will naturally bring our neural network to the large-width limit. Moreover, the orthogonality assumption in the variational ansatz could be achieved statistically using randomized assumptions. If not, the hybrid quantum-classical neural networks could still learn features even at the large width, indicating a significant difference comparing to the classical neural networks. 

\begin{figure}[h]
    \centering
    \includegraphics[width=0.5\textwidth]{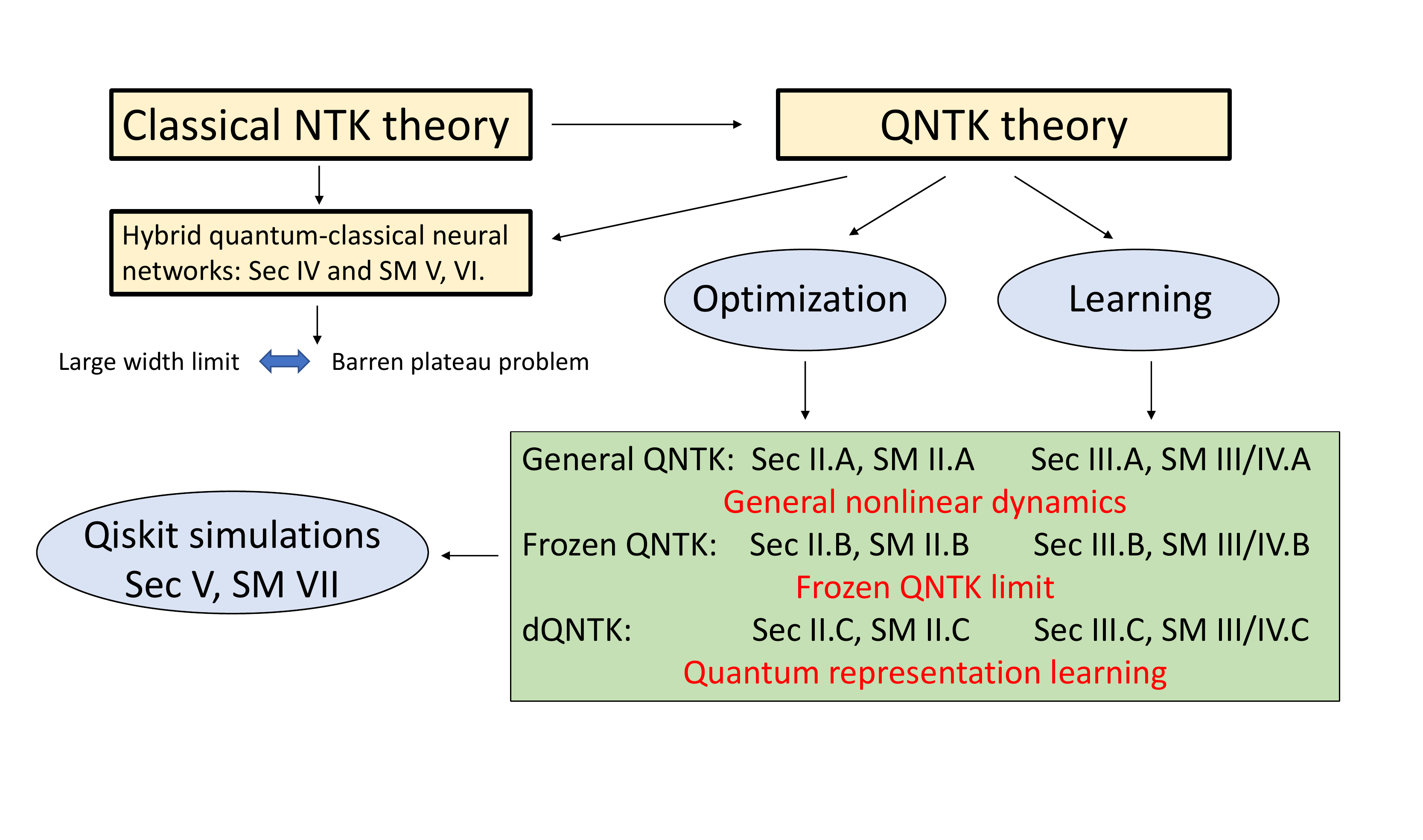}
    \caption{Structure of our paper. In Section \ref{opti} we establish the theory of QNTK in the context of optimization without data for generic variational quantum ansatz, which is the typical task in quantum simulation. In Section \ref{learn}, we establish the theory of quantum machine learning with the help of QNTK. In Section \ref{hybrid}, we define the hybrid quantum-classical neural network model, and we prove that in the large-width limit, the model is approximated by the Gaussian process. In Section \ref{numerics}, we give numerical examples to demonstrate our quantum representation theory. In Section \ref{conc}, we discuss the implication of this work, and outline open problems for future works. More technical details are given in the Supplementary Material (SM). }
    \label{fig:structure}
\end{figure}

We test the analytical derivations of our theory comparing against numerical experiments with the IBM quantum device simulator~\cite{aleksandrowicz2019qiskit}, on a classification problem in the supervised learning setting, finding good agreement with the theory. The structure of this paper and the ideas presented are summarized in Fig. \ref{fig:structure}.

\section{Theory of quantum optimization}\label{opti}

\subsection{QNTK for optimization}\label{Quantum NTK for optimization}
We start from a relatively simple example about the optimization of a quantum cost function, without a model to be learned from some data associate to it. Let a variational quantum wavefunction~\cite{peruzzo2014variational,farhi2014quantum,mcclean2016theory,kandala2017hardware,mcardle2020quantum,cerezo2021variational} be given as 
\begin{align}\label{trial_anzatz}
\left| {\phi (\theta )} \right\rangle  = U(\theta )\left| {{\Psi _0}} \right\rangle  = \left( {\prod\limits_{\ell  = 1}^L {{W_\ell }\exp } \left( {i{\theta _\ell }{X_\ell }} \right)} \right)\left| {{\Psi _0}} \right\rangle .
\end{align}
Here we have defined $L$ unitary operators of the type  $U_\ell(\theta_{\ell})=\exp(i \theta_\ell X_\ell)$, with a variational parameter $
\theta_\ell$, and a Hermitian operator $X_\ell$ associated to them. We denote the collection of all variational parameters as $\theta= \{\theta_\ell\}$ and the initial state as $\ket{\Psi_0}$. Moreover, our ansatz also includes constant gates $W_\ell$s that do not depend on the variational angles.

We introduce the following mean squared error (MSE) loss function when we wish to optimize the expectation value of a Hermitian operator $O$ to its minimal eigenvalue $O_0$, which is assumed to be known here, over the class of states $\ket{\phi(\theta})$
\begin{align}\label{optiproblem}
{\cal L}(\theta ) = \frac{1}{2}{\left( {\left\langle {\Psi_0\left| {{U^\dag }(\theta )OU(\theta )} \right|\Psi_0} \right\rangle  - {O_0}} \right)^2} \equiv \frac{1}{2}\varepsilon^2.
\end{align}
Here we have defined the \emph{residual optimization error} $\varepsilon\equiv {\left\langle {\Psi_0\left| {{U^\dag }(\theta )OU(\theta )} \right|\Psi_0} \right\rangle  - {O_0}}$. When using gradient descent to optimize Eq.~(\ref{optiproblem}), the difference equation for the dynamics of the training parameter is given by
\begin{align}\label{eq:dbar_theta}
\dbar{\theta _\ell } =  - \eta \frac{{d{\cal L}(\theta )}}{{d{\theta _\ell }}} =  - \eta \varepsilon \frac{{d\varepsilon }}{{d{\theta _\ell }}}.
\end{align}
We use the notation $\dbar o$ to denote the difference between the step $t+1$ and the step $t$ during gradient descent for the quantity $o$, $\dbar o = o(t+1)-o(t)$, associated to a learning rate $\eta$. Then we have also, to the linear order in~$\theta$,
\begin{align}\label{eq:dbar_epsilon}
\dbar \varepsilon  = \sum\limits_\ell  {\frac{{d\varepsilon }}{{d{\theta _\ell }}}\dbar{\theta _\ell }}  =  - \eta \sum\limits_\ell  {\frac{{d\varepsilon }}{{d{\theta _\ell }}}\frac{{d\varepsilon }}{{d{\theta _\ell }}}\varepsilon }.
\end{align}
The object $\sum\limits_\ell  {\frac{{d\varepsilon }}{{d{\theta _\ell }}}\frac{{d\varepsilon }}{{d{\theta _\ell }}}} $ serves to construct a toy version of the NTK in the quantum setup, in the sense that it can be seen as a 1-dimensional kernel matrix with {\it training data} $O_0$. We can make our definition of a QNTK associated to an optimization problem more precise as follows:
\begin{definition}[QNTK for optimization]
The quantum neural tangent kernel (QNTK) associated to the optimization problem of Eq.~(\ref{optiproblem}) is given by 
\begin{align}
\label{eq: toy_NTK}
    &K = \sum\limits_\ell  {\frac{{d\varepsilon }}{{d{\theta _\ell }}}\frac{{d\varepsilon }}{{d{\theta _\ell }}}}  \nonumber\\
    &=-{\left\langle {{\Psi _0}\left| {U_{ + ,\ell }^\dag \left[ {{X_\ell },U_\ell ^\dag W_\ell ^\dag U_{ - ,\ell }^\dag O{U_{ - ,\ell }}{W_\ell }{U_\ell }} \right]{U_{ + ,\ell }}} \right|{\Psi _0}} \right\rangle ^2},
\end{align}
where
\begin{align} 
{U_{ - ,\ell }} \equiv \prod \limits_{\ell ' = 1}^{\ell  - 1} {W_{\ell'}{U_{\ell '}}},~~{U_{ + ,\ell }} \equiv \prod \limits_{\ell ' = \ell +1 }^L {W_{\ell'}{U_{\ell '}}}. 
\end{align}
\end{definition}
It is easy to show that the quantity squared in Eq~(\ref{eq: toy_NTK}) is imaginary, hence $K$ is always non-negative, $K\ge 0$. A derivation of Eq.~(\ref{eq: toy_NTK}) can be found in SM.

\subsection{Frozen QNTK limit for optimization}\label{frozen_QNTK_opt}
An analytic theory of the NTK is established when the learning rate is sufficiently small. It is defined by solving the coupled difference equations Eqs.~(\ref{eq:dbar_theta}, \ref{eq:dbar_epsilon}), which we report here
\begin{align}
&\dbar{\theta _\ell } = - \eta \varepsilon \frac{{d\varepsilon }}{{d{\theta _\ell }}},\nonumber\\
&\dbar \varepsilon  = - \eta \sum\limits_\ell  {\frac{{d\varepsilon }}{{d{\theta _\ell }}}\frac{{d\varepsilon }}{{d{\theta _\ell }}}\varepsilon } = -\eta K \varepsilon.
\label{ODE}
\end{align}
In the continuum learning rate limit $\eta\rightarrow 0$, Eqs.~(\ref{ODE}) become coupled non-linear ordinary differential equations, which are hard to solve in general. Note that this system of equations stems from a quantum optimization problem and in general it is classically hard to even instantiate.

Nevertheless, in the following we build an analytic model for a quantum version of the \emph{frozen NTK} (frozen QNTK) in the regime of {\it lazy training}, where variational angles do not change too much. To be more precise, we assume that at a certain value $\theta^*$ our variational angles $\theta$ change by a small amount, $\theta^* + \delta \varphi$. A typical scenario is to do the Taylor expansion around such values $\theta^*$ during the convergence regime for instance. Here $\delta$ is a small scaling parameter. We will call the limit $\delta \to 0^+$ the \emph{frozen QNTK limit}.

In this limit, one can write ${W_\ell }{U_\ell } = {W_\ell }\exp (i\theta _\ell ^*{X_\ell })\exp (i\delta \varphi_\ell {X_\ell })$, so that the $\theta^*$ dependence is absorbed into the non-variational part of the unitary by defining ${W_\ell }(\theta _\ell ^*) \equiv {W_\ell }\exp (i\theta _\ell ^*{X_\ell })$, and we have ${W_\ell }{U_\ell } \to {W_\ell }(\theta _\ell ^*)\exp (i\delta \varphi_\ell{X_\ell })$. In what follows, we drop the $\theta^*$ notation and understand the variational angles as small parameters that change by $\delta$ around a value $\theta^*$. Then, expanding linearly for small $\delta$ we can define
\begin{definition}
[Frozen QNTK for quantum optimization] In the optimization problem Eq.~(\ref{optiproblem}) the frozen QNTK limit is
\begin{align}
&K =  -{\delta ^2}\times \nonumber\\
&\sum\limits_\ell  {{{\left\langle {{\Psi _0}\left| {W_{ + ,\ell }^\dag \left[ {{X_\ell },W_\ell ^\dag W_{ - ,\ell }^\dag O{W_{ - ,\ell }}{W_\ell }} \right]{W_{ + ,\ell }}} \right|{\Psi _0}} \right\rangle }^2}} ,
\end{align}
with
\begin{align}
{W_{ - ,\ell }} \equiv \prod\limits_{\ell ' = 1}^{\ell  - 1} {{W_{\ell '}}},~~~ {W_{ + ,\ell }} \equiv \prod\limits_{\ell ' = \ell  + 1}^L {{W_{\ell '}}}.
\end{align}
\end{definition}
In the frozen kernel limit, we can state the following result about the dependency of the residual error $\epsilon$, solving Eq.~(\ref{ODE}) linearly for small $\delta$. 
\begin{theorem}[Performance guarantee of optimization within the frozen QNTK approximation]\label{optifrozenguarantee} When using standard gradient descent for the optimization problem Eq.~(\ref{optiproblem}) within the frozen QNTK limit, the residual optimization error $\varepsilon$ decays exponentially as
\begin{align}
&\varepsilon (t) = {(1 - \eta K)^t}\varepsilon (0) = \varepsilon (0) \times \bigg( 1 + \eta {\delta ^2} \times \nonumber\\
& 
\sum\limits_\ell  {{{\left\langle {{\Psi _0}\left| {W_{ + ,\ell }^\dag \left[ {{X_\ell },W_\ell ^\dag W_{ - ,\ell }^\dag O{W_{ - ,\ell }}{W_\ell }} \right]{W_{ + ,\ell }}} \right|{\Psi _0}} \right\rangle }^2}} \bigg)^t.
\end{align}
with a convergence rate defined as
\begin{align}
&{\tau _c} =  - \log (1 - \eta K) \approx \eta K\nonumber\\
&= \eta {\delta ^2}\sum\limits_\ell  {{{\left\langle {{\Psi _0}\left| {W_{ + ,\ell }^\dag \left[ {{X_\ell },W_\ell ^\dag W_{ - ,\ell }^\dag O{W_{ - ,\ell }}{W_\ell }} \right]{W_{ + ,\ell }}} \right|{\Psi _0}} \right\rangle }^2}} \nonumber\\
&\le 2\eta {\delta ^2}L{\left\| O \right\|^2}{\max _\ell }{\left\| {{X_\ell }} \right\|^2},
\end{align}
with the $\mathbb{L}_2$ norm.
\end{theorem}
The derivation is given in the SM. An immediate consequence is that the residual error will converge to zero,
\begin{align}
\varepsilon (\infty) =0.
\end{align}

\subsection{dQNTK}
The frozen QNTK limit describes the regime of the linear approximation of non-linearities. Therefore, the frozen QNTK cannot reflect the non-linear nature of the variational quantum algorithms. In order to formulate an analytical model of the non-linearities, we now analyze the leading order correction in terms of the expansion of the learning rate $\eta$ and the size of the variational angle $\delta$. We formulate the expansion of $\dbar \varepsilon$ to the second order in $\dbar \varphi $,
\begin{align}\label{dNTK}
    \dbar\varepsilon  = \sum\limits_\ell  {\frac{{d\varepsilon }}{{d{\varphi _\ell }}}\dbar{\varphi _\ell }}  + \frac{1}{2}\sum\limits_{{\ell _1},{\ell _2}} {\frac{{d^2\varepsilon }}{{d{\varphi _{{\ell _1}}}}d \varphi_{\ell_2}}\dbar{\varphi _{{\ell _1}}}\dbar{\varphi _{{\ell _2}}}}.
\end{align}
This time $\dbar \varepsilon$ during gradient descent will follow the equation \cite{roberts2021principles}:
\begin{align}
\dbar \varepsilon  =  - \eta \sum\limits_\ell  {\frac{{d\varepsilon }}{{d{\varphi _\ell }}}\frac{{d\varepsilon }}{{d{\varphi _\ell }}}} \varepsilon  + \frac{1}{2}{\eta ^2}{\varepsilon ^2}\sum\limits_{{\ell _1},{\ell _2}} {\frac{{{d^2}\varepsilon }}{{d{\varphi _{{\ell _1}}}d{\varphi _{{\ell _2}}}}}\frac{{d\varepsilon }}{{d{\varphi _{{\ell _1}}}}}\frac{{d\varepsilon }}{{d{\varphi _{{\ell _2}}}}}}.\label{second_order}
\end{align}
With this expansion at second order, we have two contributing terms in Eq.~(\ref{dNTK}). We label the first term of Eq.~(\ref{dNTK}) quantum \emph{effective} kernel, $K^E$. We use $K^E$ to distinguish it from $K$, when only a first-order expansion is considered in the description of the dynamics. It is dynamical in the sense that it depends on the value of the training parameter $\varphi$ during the dynamics regulated by a gradient descent.
We label the variable part of the second term in Eq.~(\ref{second_order}) quantum \emph{meta-kernel} or dQNTK (differential of QNTK), 
\begin{definition}[Quantum meta-kernel for optimization]
The quantum meta-kernel associated with the optimization problem in Eq~(\ref{optiproblem}) is defined via
\begin{align}
&\mu  = \sum\limits_{{\ell _1},{\ell _2}} {\frac{{{d^2}\varepsilon }}{{d{\varphi _{{\ell _1}}}d{\varphi _{{\ell _2}}}}}\frac{{d\varepsilon }}{{d{\varphi _{{\ell _1}}}}}\frac{{d\varepsilon }}{{d{\varphi _{{\ell _2}}}}}}.
\end{align}
\end{definition}
In the limit of small changes in $\theta = \theta^* + \delta \varphi$, optimization problem Eq.~\ref{optiproblem}, the quantum meta-kernel is given at the leading order perturbation theory in $\delta$ as
\begin{widetext}
\begin{align}
&\mu = {\delta ^4}\sum\limits_{{\ell _1},{\ell _2}} \begin{array}{l}
\left\langle {{\Psi _0}\left| {W_{ + ,{\ell _1}}^\dag \left[ {{X_{{\ell _1}}},W_{{\ell _1}}^\dag W_{ - ,{\ell _1}}^\dag O{W_{ - ,{\ell _1}}}{W_{{\ell _1}}}} \right]{W_{ + ,{\ell _1}}}} \right|{\Psi _0}} \right\rangle \\
\left\langle {{\Psi _0}\left| {W_{ + ,{\ell _2}}^\dag \left[ {{X_{{\ell _2}}},W_{{\ell _2}}^\dag W_{ - ,{\ell _2}}^\dag O{W_{ - ,{\ell _2}}}{W_{{\ell _2}}}} \right]{W_{ + ,{\ell _2}}}} \right|{\Psi _0}} \right\rangle  \times 
\end{array} \nonumber\\
&\left( {\left\{ \begin{array}{l}
\left\langle {{\Psi _0}\left| {W_{ + ,{\ell _1}}^\dag \left[ {{X_{{\ell _1}}},W_{{\ell _1}}^\dag W_{{\ell _1},{\ell _2}}^\dag W_{ + ,{\ell _2}}^\dag \left[ {{X_{{\ell _2}}},W_{{\ell _2}}^\dag W_{ - ,{\ell _2}}^\dag O{W_{ - ,{\ell _2}}}{W_{{\ell _2}}}} \right]{W_{{\ell _2},{\ell _1}}}{W_{{\ell _1}}}} \right]{W_{ + ,{\ell _1}}}} \right|{\Psi _0}} \right\rangle :{\ell _1} \ge {\ell _2}\\
\left\langle {{\Psi _0}\left| {W_{ + ,{\ell _2}}^\dag \left[ {{X_{{\ell _2}}},W_{{\ell _2}}^\dag W_{{\ell _2},{\ell _1}}^\dag W_{ + ,{\ell _1}}^\dag \left[ {{X_{{\ell _1}}},W_{{\ell _1}}^\dag W_{ - ,{\ell _1}}^\dag O{W_{ - ,{\ell _1}}}{W_{{\ell _1}}}} \right]{W_{{\ell _1},{\ell _2}}}{W_{{\ell _2}}}} \right]{W_{ + ,{\ell _2}}}} \right|{\Psi _0}} \right\rangle :{\ell _1} < {\ell _2}
\end{array} \right.} \right).
\end{align}

The residual error $\varepsilon$ in the optimization problem of Eq.~(\ref{optiproblem}), can then be computed as  

\begin{align}
&\varepsilon  =\left\langle {{\Psi _0}\left| {\left( {\prod\nolimits_{\ell ' = L}^1 {W_{\ell '}^\dag } } \right)O\left( {\prod\nolimits_{\ell  = 1}^L {{W_\ell }} } \right)} \right|{\Psi _0}} \right\rangle  - {O_0}\nonumber\\
&- i\delta \sum\limits_\ell  {{\varphi _\ell }} \left\langle {{\Psi _0}\left| {W_{ + ,\ell }^\dag \left[ {{X_\ell },W_\ell ^\dag W_{ - ,\ell }^\dag O{W_{ - ,\ell }}{W_\ell }} \right]{W_{ + ,\ell }}} \right|{\Psi _0}} \right\rangle \nonumber\\
&- \frac{{{\delta ^2}}}{2}\sum\limits_{{\ell _1},{\ell _2}} {{\varphi _{{\ell _1}}}{\varphi _{{\ell _2}}} \times \left\{ \begin{array}{l}
\left\langle {{\Psi _0}\left| {W_{ + ,{\ell _1}}^\dag \left[ {{X_{{\ell _1}}},W_{{\ell _1}}^\dag W_{{\ell _1},{\ell _2}}^\dag W_{ + ,{\ell _2}}^\dag \left[ {{X_{{\ell _2}}},W_{{\ell _2}}^\dag W_{ - ,{\ell _2}}^\dag O{W_{ - ,{\ell _2}}}{W_{{\ell _2}}}} \right]{W_{{\ell _2},{\ell _1}}}{W_{{\ell _1}}}} \right]{W_{ + ,{\ell _1}}}} \right|{\Psi _0}} \right\rangle :{\ell _1} \ge {\ell _2}\\
\left\langle {{\Psi _0}\left| {W_{ + ,{\ell _2}}^\dag \left[ {{X_{{\ell _2}}},W_{{\ell _2}}^\dag W_{{\ell _2},{\ell _1}}^\dag W_{ + ,{\ell _1}}^\dag \left[ {{X_{{\ell _1}}},W_{{\ell _1}}^\dag W_{ - ,{\ell _1}}^\dag O{W_{ - ,{\ell _1}}}{W_{{\ell _1}}}} \right]{W_{{\ell _1},{\ell _2}}}{W_{{\ell _2}}}} \right]{W_{ + ,{\ell _2}}}} \right|{\Psi _0}} \right\rangle :{\ell _1} < {\ell _2}
\end{array} \right.}.
\end{align}
\end{widetext}

We are now ready to make a statement about the residual error in the limit of the dQNTK

\begin{theorem}[Performance guarantee of optimization from dQNTK] In optimization problem Eq.~(\ref{optiproblem}) at the dQNTK order, we split the residual optimization error into two pieces, the free part, and the interacting part,
\begin{align}
\varepsilon  = {\varepsilon ^F} + {\varepsilon ^I}.
\end{align}
The free part follows the exponentially decaying dynamics 
\begin{align}
{\varepsilon ^F} = {(1 - \eta K)^t}\varepsilon (0),
\end{align}
and the interacting part is given by
\begin{align}
{\varepsilon ^I}(t) =  - \eta t{\left( {1 - \eta K} \right)^{t - 1}}{K^\Delta }\varepsilon (0).
\end{align}
Here we have
\begin{widetext}
\begin{align}
&{K^\Delta } \equiv {K^E}(0) - K =\left( {\sum\limits_\ell  {\frac{{d\varepsilon }}{{d{\theta _\ell }}}\frac{{d\varepsilon }}{{d{\theta _\ell }}}} } \right)(0) - \sum\limits_\ell  {\frac{{d{\varepsilon ^F}}}{{d{\theta _\ell }}}\frac{{d{\varepsilon ^F}}}{{d{\theta _\ell }}}} \nonumber\\
&=2i{\delta ^3}\sum\limits_\ell  {\left\langle {{\Psi _0}\left| {W_{ + ,\ell }^\dag \left[ {{X_\ell },W_\ell ^\dag W_{ - ,\ell }^\dag O{W_{ - ,\ell }}{W_\ell }} \right]{W_{ + ,\ell }}} \right|{\Psi _0}} \right\rangle } \nonumber\\
&\sum\limits_{\ell '} {\begin{array}{*{20}{l}}
{\left\langle {{\Psi _0}\left| {W_{ + ,\ell '}^\dag \left[ {{X_{\ell '}},W_{\ell '}^\dag W_{\ell ',\ell }^\dag W_{ + ,\ell }^\dag \left[ {{X_\ell },W_\ell ^\dag W_{ - ,\ell }^\dag O{W_{ - ,\ell }}{W_\ell }} \right]{W_{\ell ,\ell '}}{W_{\ell '}}} \right]{W_{ + ,\ell '}}} \right|{\Psi _0}} \right\rangle :\ell ' \ge \ell }\\
{\left\langle {{\Psi _0}\left| {W_{ + ,\ell }^\dag \left[ {{X_\ell },W_\ell ^\dag W_{\ell ,\ell '}^\dag W_{ + ,\ell '}^\dag \left[ {{X_{\ell '}},W_{\ell '}^\dag W_{ - ,\ell '}^\dag O{W_{ - ,\ell '}}{W_{\ell '}}} \right]{W_{\ell ',\ell }}{W_\ell }} \right]{W_{ + ,\ell }}} \right|{\Psi _0}} \right\rangle :\ell ' < \ell }
\end{array}{\varphi _{\ell '}}(0)} .
\end{align}
\end{widetext}
Thus, the residual optimization error $\varepsilon$ will always finally approach zero,\begin{align}
\varepsilon (\infty) =0.
\end{align}
\end{theorem}
Thus, the leading order perturbative correction gives the contribution $\mathcal{O}(\delta^3)$.

\section{Theory of learning}\label{learn}
\subsection{General theory}
The results outlined in Section~\ref{opti} can be extended in the context of supervised learning from a data space $\mathcal{D}$. In particular, we are given a training set contained in the dataspace $\mathcal{A} \subset \mathcal{D}$. The data can be loaded into quantum states through a quantum feature map \cite{schuld2019quantum,havlivcek2019supervised}. 
We define the variational quantum ansatz with a single \emph{layer} by regarding the output of a quantum neural network as
\begin{align}
{z_{i;\delta }} \equiv {z_i}\left( {\theta ,{{\bf{x}}_\delta }} \right) = \left\langle {\phi \left( {{{\bf{x}}_\delta }} \right)} \right|{U^\dag }O_i U\left| {\phi \left( {{{\bf{x}}_\delta }} \right)} \right\rangle.
\end{align}
Here, we assume that $O_i$ is taken from $\mathcal{O}(\mathcal{H})$, a subset of the space of Hermitian operators of the Hilbert space $\mathcal{H}$, and the index $i$ describes the $i$-th component of the output, associated to the $i$-th operator $O_i$. The above \emph{Hermitian operator expectation value evaluation} model is a common definition of the quantum neural network. One could also measure the real and imaginary parts directly to define a complexified version of the quantum neural network, useful in the context of amplitude encoding for the ${z_{i;\delta }}$, as discussed in the Supplementary Material.
We are now in the position of introducing the loss function
\begin{align}\label{learningproblem}
{L_{\cal A}}(\theta ) = \frac{1}{2}\sum\limits_{\tilde \alpha ,i}^{} {{{\left( {{y_{i;\tilde \alpha }} - {z_{i;\tilde \alpha }}} \right)}^2}}  = \frac{1}{2}\sum\limits_{\tilde \alpha ,i} {\varepsilon _{i;\tilde \alpha }^2}.
\end{align}
Here, we call $\varepsilon_{i;\tilde{\alpha}}$ the residual training error and we assume $y_{i;\tilde{\alpha}}$ is associated with the encoded data $\phi_{i}(\mathbf{x}_{\tilde{\alpha}})$.
 Now, similarly to what described in section~\ref{Quantum NTK for optimization}, we have the gradient descent equation
\begin{align}
\dbar {z_{i;\delta }} = - \eta  \sum\limits_{\ell ,i',\tilde \alpha }^{} {{\varepsilon _{i';\tilde \alpha }}\frac{{d{z_{i;\delta }}}}{{d{\theta _\ell }}}\frac{{d{z_{i';\tilde \alpha }}}}{{d{\theta _\ell }}}},
\end{align}
with an associated kernel
\begin{align}
K_{\delta ,\tilde \alpha }^{ii'} = \sum\limits_\ell ^{} {\frac{{d{z_{i;\delta }}}}{{d{\theta _\ell }}}\frac{{d{z_{i';\tilde \alpha }}}}{{d{\theta _\ell }}}}.
\end{align}
To ease the notation, we shall define the joint index 
\begin{align}
(\delta  ,i) = \bar a,~~~~~(\tilde \alpha,i' ) = {\hat b },
\end{align}
which are running in the space $\mathcal{D}\times\mathcal{O}(\mathcal{H}) $ and $\mathcal{A} \times\mathcal{O}(\mathcal{H})$ respectively (we use $\hat{a}$ to indicate that the corresponding data component is in the sample set $\mathcal{A}$, and if we wish to make a general data point we will denote it as $\bar{a}$), and our gradient descent equations are 
\begin{align}
{\dbar z _{\bar{a} }} =  -\eta  \sum\limits_{\hat b } {{K_{\bar{a} \hat b }}{\varepsilon _{\hat b }}}.
\end{align}
It is possible to show that this kernel is always positive semidefinite and Hermitian, see Supplementary Material for a proof. Now recalling Eq.(\ref{trial_anzatz}), we are in the position to give an analytical expression for the QNTK for a supervised learning problem as follows. Details on the derivation can be found in the Supplementary Material.  
\begin{widetext}
\begin{definition}[QNTK for quantum machine learning]
The QNTK for the quantum learning model Eq.~(\ref{learningproblem}) is given by
\begin{align}
&K_{\delta ,\tilde \alpha }^{ii'} = \sum\limits_\ell  {\frac{{d{z_{i;\delta}}}}{{d{\theta _\ell }}}} \frac{{d{z_{i';\tilde \alpha}}}}{{d{\theta _\ell }}} = - \sum\limits_\ell  {\left( {\begin{array}{*{20}{l}}
{\left\langle {\phi \left( {{{\bf{x}}_\delta }} \right)\left| {U_{ + ,\ell }^\dag \left[ {{X_\ell },U_\ell ^\dag W_\ell ^\dag U_{ - ,\ell }^\dag {O_i}{U_{ - ,\ell }}{W_\ell }{U_\ell }} \right]{U_{ + ,\ell }}} \right|\phi \left( {{{\bf{x}}_\delta }} \right)} \right\rangle  \times }\\
{\left\langle {\phi \left( {{{\bf{x}}_{\tilde \alpha }}} \right)\left| {U_{ + ,\ell }^\dag \left[ {{X_\ell },U_\ell ^\dag W_\ell ^\dag U_{ - ,\ell }^\dag {O_{i'}}{U_{ - ,\ell }}{W_\ell }{U_\ell }} \right]{U_{ + ,\ell }}} \right|\phi \left( {{{\bf{x}}_{\tilde \alpha }}} \right)} \right\rangle }
\end{array}} \right)}.
\end{align}
\end{definition}
\end{widetext}

\subsection{Absence of representation learning in the frozen limit}
In the frozen QNTK case, the kernel is static, and the learning algorithm cannot learn \emph{features} from the data. In the same fashion of section~\ref{frozen_QNTK_opt}, we take \emph{the frozen QNTK limit} where the changes of variational angles $\theta$ are small. Using the previous notations we can define the QNTK in for quantum machine learning in the frozen limit, and a performance guarantee for the error on the loss function in this regime as follows.

\begin{definition}[Frozen QNTK for quantum machine learning] 
In the quantum learning model Eq.~(\ref{learningproblem}) with the frozen QNTK limit,
\begin{widetext}
\begin{align}
K_{\delta ,\tilde \alpha }^{ii'} =  - {\delta ^2}\sum\limits_\ell  {\left( {\begin{array}{*{20}{l}}
{\left\langle {\phi \left( {{{\bf{x}}_\delta }} \right)\left| {W_{ + ,\ell }^\dag \left[ {{X_\ell },W_\ell ^\dag W_{ - ,\ell }^\dag {O_i}{W_{ - ,\ell }}{W_\ell }} \right]{W_{ + ,\ell }}} \right|\phi \left( {{{\bf{x}}_\delta }} \right)} \right\rangle  \times }\\
{\left\langle {\phi \left( {{{\bf{x}}_{\tilde \alpha }}} \right)\left| {W_{ + ,\ell }^\dag \left[ {{X_\ell },W_\ell ^\dag W_{ - ,\ell }^\dag {O_{i'}}{W_{ - ,\ell }}{W_\ell }} \right]{W_{ + ,\ell }}} \right|\phi \left( {{{\bf{x}}_{\tilde \alpha }}} \right)} \right\rangle }
\end{array}} \right)}.
\end{align}
\end{widetext}
\end{definition}

\begin{theorem}[Performance guarantee of quantum machine learning in the frozen QNTK limit] In the quantum learning model Eq.(~\ref{learningproblem}) with the frozen QNTK limit, the residual optimization error decays exponentially during the gradient descent as
\begin{align}
&{\varepsilon _{{{\hat a }_1}}}(t) = \sum\limits_{{{\hat a }_2}} {{U_{{{\hat a }_1}{{\hat a }_2}}}} (t){\varepsilon _{{{\hat a }_2}}}(0),\nonumber\\
&{U_{{{\hat a }_1}{{\hat a }_2}}}(t) = {\left[ {{{\left( {1 - {\eta }K} \right)}^t}} \right]_{{{\hat a }_1}{{\hat a }_2}}}.
\end{align}
The convergence rate is defined as
\begin{align}
{\tau _c} = \left\| { - \log \left( {1 - \eta K} \right)} \right\| \approx \eta \left\| {K_{\delta ,\tilde \alpha }^{ii'}} \right\|.
\end{align}
\end{theorem}
Then we obtain for the quantum learning model Eq.~\ref{learningproblem} with the frozen QNTK limit, the asymptotic dynamics with the $\mathcal{D}\times \mathcal{O}(\mathcal{H})$ index $\bar{a}$, is given by
\begin{align}
{z_{\bar a }}(  \infty ) = {z_{\bar a }}(0) - \sum\limits_{{{\hat a }_1},{{\hat a }_2}} {{\tilde K}^{{{\hat a }_1}{{\hat a }_2}}}{K_{\bar a {{\hat a }_1}}} {\varepsilon_{{{\hat a }_2}}(0)}.
\end{align}
Here $\tilde{K}$ means that the kernel defined only restricted to the space $\mathcal{A}\times \mathcal{O}(\mathcal{H})$ (note that it is different from the kernel inverse defined for the whole space in general), and we denote the kernel inverse as 
\begin{align}
\sum\limits_{\hat a  \in {\cal A} \times {\mathcal{O}(\mathcal{H})}} {{{\tilde K}^{{{\hat a }_1}{{\hat a }_2}}}} {{\tilde K}_{{{\hat a }_2}{{\hat a }_3}}} = \delta _{{{\hat a }_3}}^{{{\hat a }_1}}.
\end{align}
Specifically, if we assume $\bar{a}$ indicates the data in the space $\mathcal{A}\times \mathcal{O}(\mathcal{H})$, we will have $\varepsilon_{\bar{a}}(\infty)=0$.
Proofs and details of these results are given in the SM. Moreover, the asymptotic value is different from the frozen QNTK case in the optimization problem, because of the existence of the difference between the training set $\mathcal{A}$ and the whole data space.

\subsection{Representation learning in the dynamical setting}
In the dynamical case, the kernel is changing during the gradient descent optimization, due to non-linearity in the unitary operations. In this case then the variational quantum circuits could naturally serve as architectures of representation learning in the classical sense. 

We generalize the leading order perturbation theory of optimization naturally to the learning case, and we state the main theorems here. First, we have
\begin{theorem}[Performance guarantee of quantum machine learning in the dQNTK limit] In the quantum learning model Eq.~(\ref{learningproblem}) at the dQNTK order, the training error is given by two contributions, a free and interacting part, as follows
\begin{align}
{\varepsilon _{\hat a }}(t) = \varepsilon _{\hat a }^F(t) + \varepsilon _{\hat a }^I(t),
\end{align}
where
\begin{align}
&\varepsilon _{\hat a }^F(t) = \sum\limits_{{{\hat a }_1}} {{U_{\hat a {{\hat a }_1}}}} (t){\varepsilon _{{{\hat a }_1}}}(0),\nonumber\\
&{U_{{{\hat a }_1}{{\hat a }_2}}}(t) = {\left[ {{{\left( {1 - \eta  K} \right)}^t}} \right]_{{{\hat a }_1}{{\hat a }_2}}},
\end{align}
and
\begin{align}
\varepsilon _{\hat a }^I(t) = {\left( { - \eta \sum\limits_{s = 0}^{t - 1} {{{\left( {1 - \eta  K} \right)}^{t - 1 - s}}} {K^\Delta }{{\left( {1 - \eta  K} \right)}^s}\varepsilon (0)} \right)_{\hat a }}.
\end{align}
Here $K$ is the frozen (linear) part of the QNTK. Using a matrix notation for the compact indices $\hat{a}$, in the space $\mathcal{A}\times \mathcal{O}(\mathcal{H})$, we have
\begin{align}
\left\| {{\varepsilon ^I}(t)} \right\| \le \eta t{\left\| {1 - \eta  K} \right\|^{t - 1}}\left\| {{K^\Delta }} \right\|\left\| {\varepsilon (0)} \right\|.
\end{align}
where $K^\Delta$ is defined as 
\begin{align}
K_{\delta ,\tilde \alpha }^{\Delta ,ii'} = i{\delta ^3}\sum\limits_{\ell ,\ell '} {{\varphi _{\ell '}}(0)G_{\ell ',\ell }^{\delta ,i}\Theta _\ell ^{\tilde \alpha ,i'}}  + i{\delta ^3}\sum\limits_{\ell ,\ell '} {{\varphi _{\ell '}}(0)G_{\ell ',\ell }^{\tilde \alpha ,i'}\Theta _\ell ^{\delta ,i}},
\end{align}
and
\begin{widetext}
\begin{align}
&G_{{\ell _1},{\ell _2}}^{\delta ,i} \equiv {G_{{\ell _1},{\ell _2}}}(\phi ({{\bf{x}}_\delta }),{O_i}) = \nonumber\\
&\left( {\left\{ {\begin{array}{*{20}{l}}
{\left\langle {\phi ({{\bf{x}}_\delta })\left| {W_{ + ,{\ell _1}}^\dag \left[ {{X_{{\ell _1}}},W_{{\ell _1}}^\dag W_{{\ell _1},{\ell _2}}^\dag W_{ + ,{\ell _2}}^\dag \left[ {{X_{{\ell _2}}},W_{{\ell _2}}^\dag W_{ - ,{\ell _2}}^\dag {O_i}{W_{ - ,{\ell _2}}}{W_{{\ell _2}}}} \right]{W_{{\ell _2},{\ell _1}}}{W_{{\ell _1}}}} \right]{W_{ + ,{\ell _1}}}} \right|\phi ({{\bf{x}}_\delta })} \right\rangle :{\ell _1} \ge {\ell _2}}\\
{\left\langle {\phi ({{\bf{x}}_\delta })\left| {W_{ + ,{\ell _2}}^\dag \left[ {{X_{{\ell _2}}},W_{{\ell _2}}^\dag W_{{\ell _2},{\ell _1}}^\dag W_{ + ,{\ell _1}}^\dag \left[ {{X_{{\ell _1}}},W_{{\ell _1}}^\dag W_{ - ,{\ell _1}}^\dag {O_i}{W_{ - ,{\ell _1}}}{W_{{\ell _1}}}} \right]{W_{{\ell _1},{\ell _2}}}{W_{{\ell _2}}}} \right]{W_{ + ,{\ell _2}}}} \right|\phi ({{\bf{x}}_\delta })} \right\rangle :{\ell _1} < {\ell _2}}
\end{array}} \right.} \right),\nonumber\\
&\Theta _\ell ^{\delta ,i} \equiv {\Theta _\ell }(\phi ({{\bf{x}}_\delta }),{O_i})= \left\langle {\phi ({{\bf{x}}_\delta })\left| {W_{ + ,\ell }^\dag \left[ {{X_\ell },W_\ell ^\dag W_{ - ,\ell }^\dag {O_i}{W_{ - ,\ell }}{W_\ell }} \right]{W_{ + ,\ell }}} \right|\phi ({{\bf{x}}_\delta })} \right\rangle.
\end{align}
\end{widetext}
\end{theorem}
\begin{widetext}
For the quantum learning model Eq.~\ref{learningproblem} at the dQNTK order, the dynamics given by gradient descent on a general data point is given by
\begin{align}
&{z_{\bar a  }}(\infty ) = {z_{\bar a  }}(0) - \sum\limits_{{{\hat a  }_1},{{\hat a  }_2}} {{K_{\bar a  {{\hat a  }_1}}}} {{\tilde K}^{{{\hat a  }_1}{{\hat a  }_2}}}{\varepsilon _{{{\hat a  }_2}}}(0)\nonumber\\
&+ \sum\limits_{{{\hat a  }_1},{{\hat a  }_2},{{\hat a  }_3},{{\hat a  }_4}} {\left[ {{\mu  _{{{\hat a  }_1}\bar a  {{\hat a  }_2}}} - \sum\limits_{{{\hat a  }_5},{{\hat a  }_6}} {{K_{\bar a  {{\hat a  }_5}}}} {{\tilde K}^{{{\hat a  }_5}{{\hat a  }_6}}}{\mu _{{{\hat a  }_1}{{\hat a  }_6}{{\hat a  }_2}}}} \right]} Z_A^{{{\hat a  }_1}{{\hat a  }_2}{{\hat a  }_3}{{\hat a  }_4}}{\varepsilon _{{{\hat a  }_3}}}(0){\varepsilon _{{{\hat a  }_4}}}(0)\nonumber\\
&+ \sum\limits_{{{\hat a  }_1},{{\hat a  }_2},{{\hat a  }_3},{{\hat a  }_4}} {\left[ {{\mu  _{\bar a  {{\hat a  }_1}{{\hat a  }_2}}} - \sum\limits_{{{\hat a  }_5},{{\hat a  }_6}} {{K_{\bar a  {{\hat a  }_5}}}} {{\tilde K}^{{{\hat a  }_5}{{\hat a  }_6}}}{\mu  _{{{\hat a  }_6}{{\hat a  }_1}{{\hat a  }_2}}}} \right]} Z_B^{{{\hat a  }_1}{{\hat a  }_2}{{\hat a  }_3}{{\hat a  }_4}}{\varepsilon _{{{\hat a  }_3}}}(0){\varepsilon _{{{\hat a  }_4}}}(0),
\end{align}
\end{widetext}
where $Z_{A,B}$s are called the quantum algorithm projectors (see \cite{roberts2021ai,summer} for their original framework),
\begin{align}
&Z_A^{{{\hat a  }_1}{{\hat a  }_2}{{\hat a  }_3}{{\hat a  }_4}}  \nonumber\\
&\equiv{{\tilde K}^{{{\hat a  }_1}{{\hat a  }_3}}}{{\tilde K}^{{{\hat a  }_2}{{\hat a  }_4}}} - \sum\limits_{{{\hat a  }_5}} {{{\tilde K}^{{{\hat a  }_2}{{\hat a  }_5}}}} X_\parallel ^{{{\hat a  }_1}{{\hat a  }_5}{{\hat a  }_3}{{\hat a  }_4}},\nonumber\\
&Z_B^{{{\hat a  }_1}{{\hat a  }_2}{{\hat a  }_3}{{\hat a  }_4}}  \nonumber\\
&\equiv{{\tilde K}^{{{\hat a  }_1}{{\hat a  }_3}}}{{\tilde K}^{{{\hat a  }_2}{{\hat a  }_4}}} - \sum\limits_{{{\hat a  }_5}} {{{\tilde K}^{{{\hat a  }_2}{{\hat a  }_5}}}} X_\parallel ^{{{\hat a  }_1}{{\hat a  }_5}{{\hat a  }_3}{{\hat a  }_4}} + \frac{\eta  }{2}X_\parallel ^{{{\hat a  }_1}{{\hat a  }_2}{{\hat a  }_3}{{\hat a  }_4}},
\end{align}
and $X_{\parallel}$ is defined as
\begin{align}
X_\parallel ^{{{\hat a  }_1}{{\hat a  }_2}{{\hat a  }_3}{{\hat a  }_4}} = \sum\limits_{s = 0}^\infty  {{{\left[ {{{(1 - {\eta  }K)}^s}} \right]}_{{{\hat a  }_1}{{\hat a  }_3}}}} {\left[ {{{(1 - {\eta  } K)}^s}} \right]_{{{\hat a  }_2}{{\hat a  }_4}}},
\end{align}
or
\begin{align}
&\delta _{{{\hat a  }_5}}^{{{\hat a  }_1}}\delta _{{{\hat a  }_6}}^{{{\hat a  }_2}} = \sum\limits_{{{\hat a  }_3},{{\hat a  }_4}} {X_\parallel ^{{{\hat a  }_1}{{\hat a  }_2}{{\hat a  }_3}{{\hat a  }_4}}} \times\nonumber\\
&\left( {{{\tilde K}_{{{\hat a  }_3}{{\hat a  }_5}}}{\delta _{{{\hat a  }_4}{{\hat a  }_6}}} + {\delta _{{{\hat a  }_3}{{\hat a  }_5}}}{{\tilde K}_{{{\hat a  }_4}{{\hat a  }_6}}} - {\eta  }{{\tilde K}_{{{\hat a  }_3}{{\hat a  }_5}}}{{\tilde K}_{{{\hat a  }_4}{{\hat a  }_6}}}} \right).
\end{align}
Finally, $\mu$ is the quantum meta-kernel in the quantum machine learning context,
\begin{align}
&\mu _{{\delta _0}{\delta _1}{\delta _2}}^{{i_0}{i_1}{i_2}} = {\mu _{{{\bar a}_0}{{\bar a}_1}{{\bar a}_2}}} = {\left. {\sum\limits_{{\ell _1},{\ell _2}} {\frac{{{d^2}{z_{{i_0};{\delta _0}}}}}{{d{\varphi _{{\ell _1}}}d{\varphi _{{\ell _2}}}}}} \left( {\frac{{d{z_{{i_1};{\delta _1}}}}}{{d{\varphi _{{\ell _1}}}}}\frac{{d{z_{{i_2};{\delta _2}}}}}{{d{\varphi _{{\ell _2}}}}}} \right)} \right|_{\varphi  = 0}}\nonumber\\
&={\delta ^4}\sum\limits_{{\ell _1},{\ell _2}} {\Theta _{{\ell _1}}^{{\delta _1},{i_1}}\Theta _{{\ell _2}}^{{\delta _2},{i_2}}G_{{\ell _1},{\ell _2}}^{{\delta _0},{i_0}}}.
\end{align}
Specifically, if we assume that $\bar{a}$ is from $\mathcal{A}\times \mathcal{O}(\mathcal{H})$, we will get $\varepsilon_{\bar{a}} (\infty)=0$.
More details of $\mu$ are given in SM. The existence of quantum algorithm projectors shows the \emph{quantum algorithm dependence} of the variational quantum circuits, which indicates powerful representation learning potential because of non-linearity.

\section{Hybrid Quantum-classical network and the large-width limit}\label{hybrid}
In this section we define a setting in which one can speak of a quantum analog of the large-width limit for NTKs. In such a limit, we expect that the dynamics linearizes during the whole training process, similar to what happens in the frozen regime of lazy training, and the correlation function of the outputs neuraons becomes Gaussian. The classical NTK theory requires a random initialization of weights and bias and takes the large-width limit of neural network architectures. In the quantum setup, the random initialization is a random choice of trainable ans\"{a}tze.

\begin{figure}[h]
    \centering
    \includegraphics[width=0.5\textwidth]{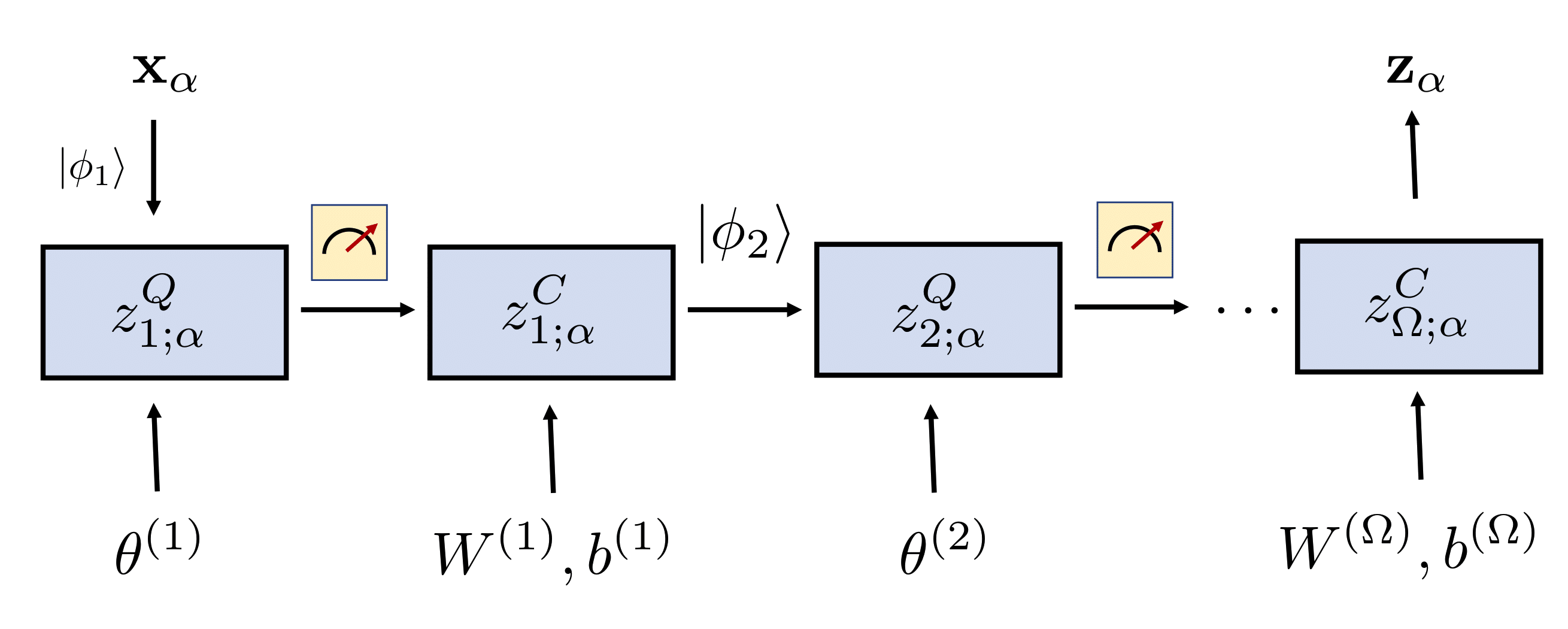}
    \caption{The hybrid quantum-classical neural network considered here. We repetitively apply quantum and classical neural networks in our architecture, with feature map encoding $\ket{\psi}$ and quantum measurements, mapping the data point $\mathbf{x}_\alpha$ to the prediction $\mathbf{z}_\alpha$. }
    \label{fig:hybrid}
\end{figure}

To see it more clearly, we consider a hybrid quantum classical neural network model \cite{otterbach2017unsupervised,farhi2018classification}. Starting from a quantum neural network, we measure the output neurons from the quantum architecture and dress them with a single-layer classical neural network. The output of the classical neural network could be then re-encoded into a quantum register via another quantum feature map. A single quantum to classical step can be called one \emph{hybrid layer}, and then one could construct multiple hybrid layers connected by feature map encoding, see Fig. \ref{fig:hybrid} for an illustration.

For the quantum part of the circuit, we use the same structure of quantum neural networks with Hermitian operator expectation values. Mathematically, the model is defined as
\begin{align}
&z_{1;\alpha ;{j_1}}^Q = \left\langle {{\phi _1}\left( {{{\bf{x}}_\alpha }} \right)\left| {{U^{\dag ,1}}\left( {{\theta ^1}} \right)O_{{j_1}}^1{U^1}\left( {{\theta ^1}} \right)} \right|{\phi _1}\left( {{{\bf{x}}_\alpha }} \right)} \right\rangle  ,\label{eq1}\\
&z_{\omega ;\alpha ;{j_\omega }}^Q = \left\langle {{\phi _\omega }\left( {{{\bf{w}}_{\omega  - 1;\alpha }}} \right)} \right|{U^{\dag ,\omega }}({\theta ^\omega }){O^{\omega}_{{j_\omega }}}{U^\omega }({\theta ^\omega })\left| {{\phi _\omega }\left( {{{\bf{w}}_{\omega  - 1;\alpha }}} \right)} \right\rangle  ,\label{eq2}\\
&w_{\omega ;\alpha ;j_\omega ^C} = \sigma _{j_\omega ^C}^\omega \left( {\sum\limits_{{j_\omega } = 1}^{\dim {{\mathcal{O^\omega}(\mathcal{H^\omega})} }} {W_{j_\omega ^C,{j_\omega }}^\omega z_{\omega ;\alpha ;{j_\omega }}^Q}  + b_{j_\omega ^C}^\omega } \right) \equiv  \nonumber\\
&\sigma _{j_\omega ^C}^\omega \left(z^C_{\omega;\alpha;j_{\omega}^C } \right).\label{eq3}\nonumber\\
& 
\end{align}
Here, Eq.~\ref{eq1} initializes the quantum neural network, mapping the data $\mathbf{x}_\alpha$ to components $j_1$, labeling the index in the space of Hermitian operator we use $\mathcal{O}^1(\mathcal{H}^1)$. The variational ansatz is similar to what we have discussed before, but they might be different in different layers. We use the label $\omega$ to denote the order of hybrid layers, ranging from 1 to the total number of hybrid layers. We introduce the quantum ansatz ${U^\omega }({\theta ^\omega }) = \prod\limits_{{\ell _\omega } = 1}^{{L_\omega }} W_{\ell_\omega}^\omega{\exp } \left( {i\theta _{{\ell _\omega }}^\omega X_{{\ell _\omega }}^\omega } \right)$, the feature map ${\phi _\omega }$, and the operator space $\mathcal{O}^\omega(\mathcal{H}^\omega)$ index $j_{\omega}$. Eq.~\ref{eq2} introduces the recursive encoding from the classical neural network data ${\bf{w}}_{\omega-1;\alpha } = {(w_{\omega-1 ;\alpha })_{j_{\omega-1} ^C}}$ to the space $\mathcal{O}^\omega (\mathcal{H}^\omega)$, where the classical data vector $w_{\omega ;\alpha ;j_\omega ^C}$ is obtained through a single-layer classical neural network with the non-linear activation $\sigma_{j_\omega^C}^\omega$, weight matrix $W_{j_\omega ^C,{j_\omega }}^\omega$, and bias vector $b_{j_\omega ^C}^\omega $ with the classical index ${j_\omega^C}$, and the preactivation $z^C_{\omega;\alpha;j_{\omega}^C }$. When we intialize the hybrid network, the classical weights and biases are statistically Gaussian following the LeCun parametrization \footnote{This is somewhat different from the so-called \emph{NTK parametrization} in some literature. See details in SM. Moreover, here we assume the weights and biases are real.},
\begin{align}
&\mathbb{E}\left( {W_{j_{1,\omega }^C,{j_{1,\omega }}}^\omega W_{j_{2,\omega }^C,{j_{2,\omega }}}^\omega } \right) = {\delta _{j_{1,\omega }^C,j_{2,\omega }^C}}{\delta _{{j_{1,\omega }},{j_{2,\omega }}}}\frac{{C_W^\omega }}{{\dim {{\mathcal{O}}^\omega(\mathcal{H}^\omega) }}},\nonumber\\
&\mathbb{E}\left( {b_{j_{1,\omega }^C}^\omega b_{j_{2,\omega }^C}^\omega } \right) = {\delta _{j_{1,\omega }^C,j_{2,\omega }^C}}C_b^\omega .
\end{align}
Note that in this case, the role of \emph{width} in the large-width theory is replaced the dimension of the operator space, $\dim ({{\cal O}^\omega }({{\cal H}^\omega }))$. The value of the dimension (width) could be arbitrary in principle, but it is upper bounded by the square of the dimension of the Hilbert space, $\dim ({{\cal O}^\omega }({{\cal H}^\omega })) \le \dim (\mathcal{H}^\omega)^2$, in the qubit system. 

If we now assume that our quantum training parameters $\theta^{\omega}$ are chosen from ensembles (or the variational ans\"atze themselves are from some ensembles), similar to the classical assumption. Denoting the expectation value from quantum ensembles as $\mathbb{E}$, we will show the following statement,
\begin{theorem}[Non-Gaussianity from large width] The four-point function of classical preactivations is nearly Gaussian if $\dim ({{\cal O}^\omega }({{\cal H}^\omega })) $ is large, 
\begin{align}
&{\mathbb{E}_{{\rm{conn}}}}\left( {z_{\omega ;{\alpha _1};j_{1,\omega }^C}^Cz_{\omega ;{\alpha _2};j_{2,\omega }^C}^Cz_{\omega ;{\alpha _3};j_{3,\omega }^C}^Cz_{\omega ;{\alpha _4};j_{4,\omega }^C}^C} \right) \nonumber\\
&= \mathcal{O}(\frac{1}{\dim ({{\cal O}^\omega }({{\cal H}^\omega }))}),
\end{align}
as long as,
\begin{align} \label{ortho}
&{\mathbb{E}_{\rm{conn}}}\left( {z_{\omega ;{\alpha _1 };{j_{1,\omega }}}^Qz_{\omega ;{\alpha _2 };{j_{1,\omega }}}^Qz_{\omega ;{\alpha _3};{j_{2,\omega }}}^Qz_{\omega ;{\alpha _4};{j_{2,\omega }}}^Q} \right) \nonumber\\
&= \mathcal{O}(1) \times {\delta _{{j_{1,\omega }},{j_{2,\omega }}}},
\end{align}
and their permutations for all $\omega$s. Here the notation $\mathbb{E}_{\rm{conn}}$ means the connected Gaussian correlators subtracting Wick contractions. 
\end{theorem}
More details are given in SM. The orthogonal condition Eq. \ref{ortho} can be naturally achieved by randomized architectures, for instance, Haar randomness and $k$-designs. We interpret the result as:
\begin{itemize}
\item In this hybrid case, the role of \emph{width} in the neural network is upper bounded by the square dimension of the $w$-th Hilbert space. Thus, if we scale up the number of qubits, we are naturally in the large-width limit. However, if our variational ansatz is sparse enough such that the operator space dimension $\dim \mathcal{O}(\mathcal{H})$ is small, then we will have significant finite width effects. 
\item The condition Eq.~(\ref{ortho}) for quantum outputs is naturally satisfied by random architectures. If we assume that our variational ansatz is highly random, we are expected to have similar Gaussian process behaviors as the large-width limit of classical neural networks. However, the same assumption will generically lead to the barren plateau problem \cite{mcclean2018barren}, where the derivatives of the loss function will move slowly when we scale up our operator space dimension. Our result shows a possible connection between the large-width limit and the barren plateau problem. 
\item Moreover, the orthogonal condition in Eq. \ref{ortho} we impose does not mean that we have to set the ansatz to be highly random. It could also be potentially satisfied by fixed ans\"atze, for instance, with some error-correction types of orthogonal conditions. If the condition is generally not satisfied, the variational architecture we study could have highly non-Gaussian and representation learning features, although it might be theoretically hard to understand \footnote{Similar analysis could be done on dynamics, see SM for further comments.}. 
\end{itemize}

\section{Numerical results}\label{numerics}
\begin{figure}[h]
    \centering
    \includegraphics[width=0.5\textwidth]{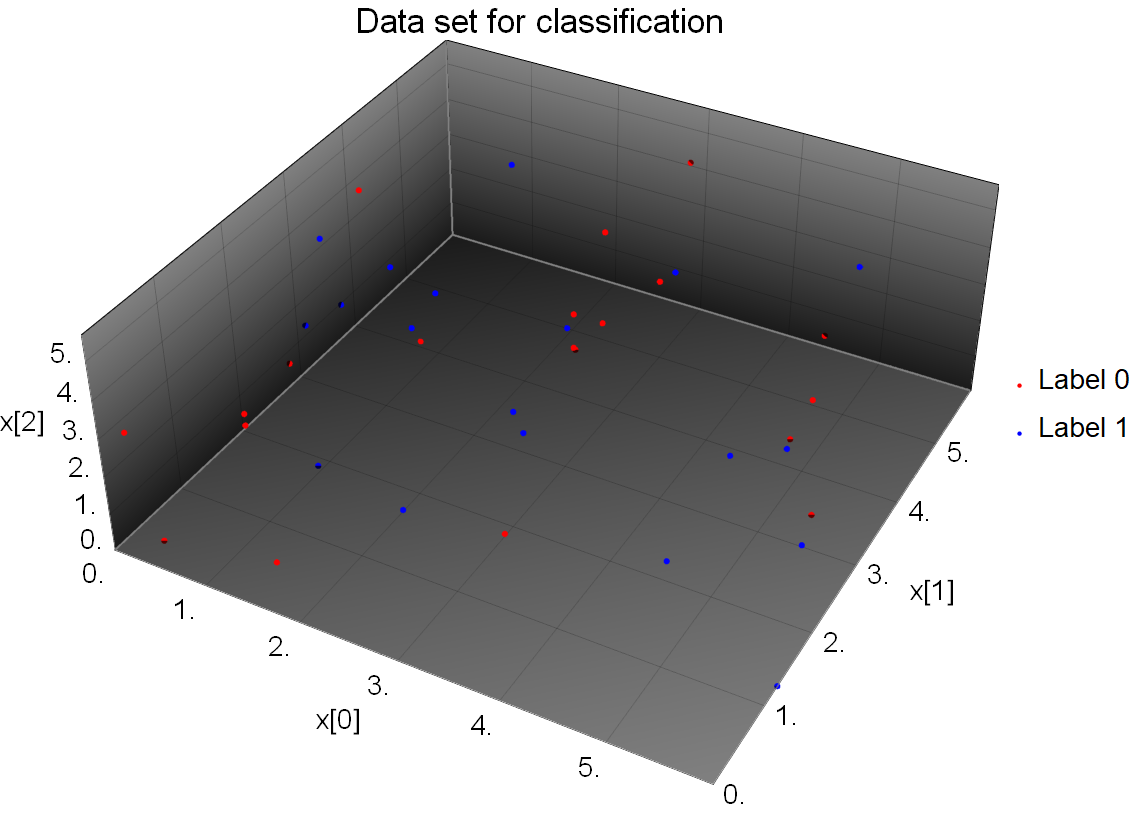}
    \caption{A 3D illustration of our data set for training. Here we use three inputs $x[0,1,2]$ and label them with 0 or 1, colored by red or blue respectively. The data set is generated using \texttt{ad\_hoc\_data}.}
    \label{fig:data}
\end{figure}

In this section, we test our QNTK theory in practice, using the Qiskit software library~\cite{aleksandrowicz2019qiskit} to simulate the implementation of a  paradigmatic quantum machine learning task on quantum processors, both in noiseless and noisy cases. We consider a variational classification problem in supervised learning with three qubits. The data set is generated with the \texttt{ad\_hoc\_data} functionality as provided in \texttt{qiskit.ml.datasets} within the Qiskit Machine Learning module~\cite{ibmqml}, see Fig. \ref{fig:data} for an illustration. 

Our numerical experiments are performed first using the noiseless \texttt{statevector{\_}simulator} backend, then including both statistical ($n_{\text{shots}}=8192$) and simulated hardware noise with the Qiskit \texttt{qasm{\_}simulator}. A simplified model of device noise, featuring the qubit relaxation and dephasing, single-qubit and 2-qubit gate errors and readout inaccuracies, is constructed with the \texttt{NoiseModel.from{\_}backend()} Qiskit method and parametrized using our calibration data from the \texttt{ibmq{\_}bogota} superconducting processor (accessed on Oct, 15 2021).

\begin{figure}[h]
    \centering
    \includegraphics[width=0.5\textwidth]{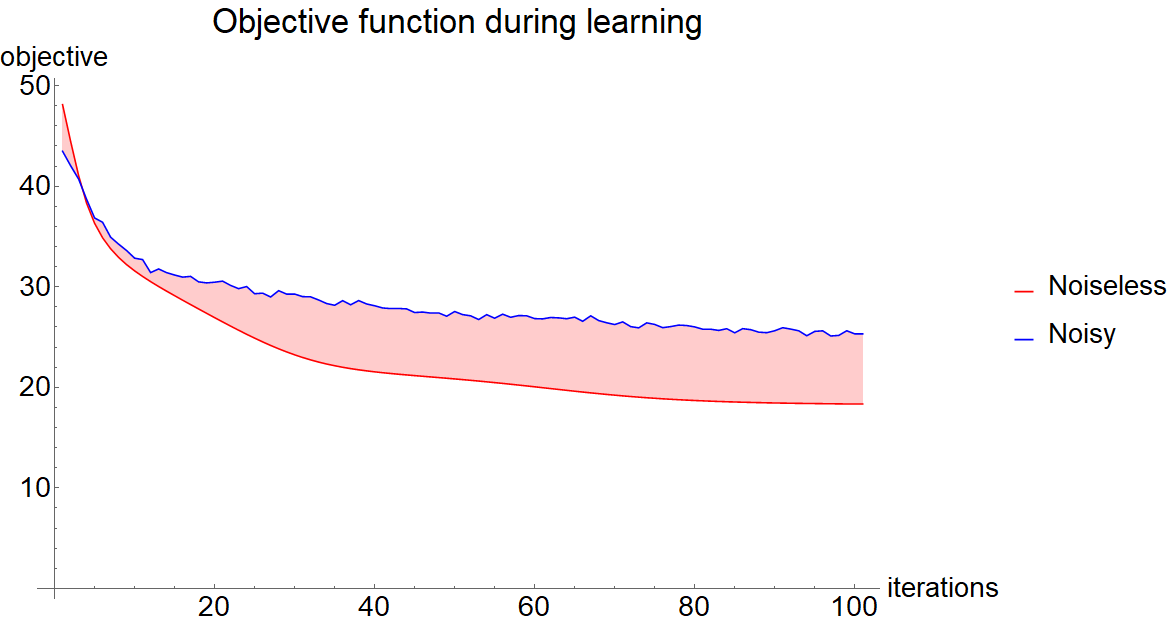}
    \caption{The convergence of objective function during gradient descent. Here we compare the ideal and noisy cases, labeled by red or blue respectively.} 
    \label{fig:convergence}
\end{figure}

We implement supervised learning using a Qiskit Machine Learning \texttt{NeuralNetworkClassifier} with a squared error loss, obtaining reasonable convergence with gradient descent algorithms (See Fig. \ref{fig:convergence}). The underlying variational quantum classifier is based on the \texttt{TwoLayerQNN} design, with a 3-input \texttt{ZZFeatureMap} and a  \texttt{RealAmplitudes} trainable ansatz with 3 repetitions and 12 parameters. Further details on numerical simulations are given in SM. Note that we do not demand a perfect convergence around the global minimum, since the QNTK theory only cares about the derivatives of the residual learning errors, which is invariant by shifting a constant or changing the initial condition when solving the training dynamics. In the classical theory of NTK, in the infinite width case, for instance, the multilayer perceptron (MLP) model is both overparametrized and generalized, and the answer would give the global minimum. Including finite width corrections, there might be multiple local minima, and it is a feature of representation learning. Moreover, we use the error mitigation protocol by applying $\texttt{CompleteMeasFitter}$ from $\texttt{qiskit.ignis.mitigation.measurement}$ to mitigate readout noise. 

\begin{figure}[h]
    \centering
    \includegraphics[width=0.5\textwidth]{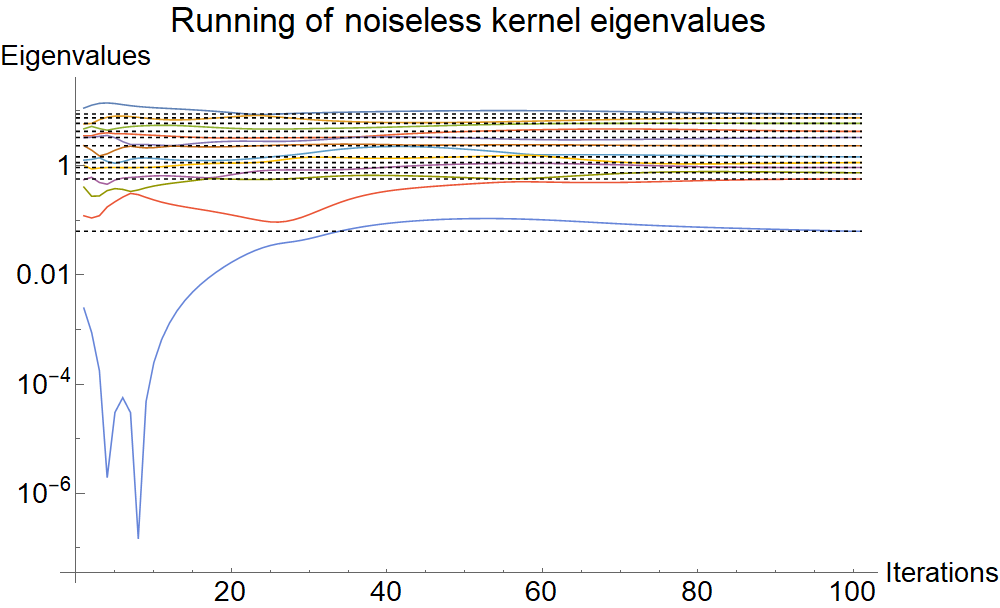}
    \includegraphics[width=0.5\textwidth]{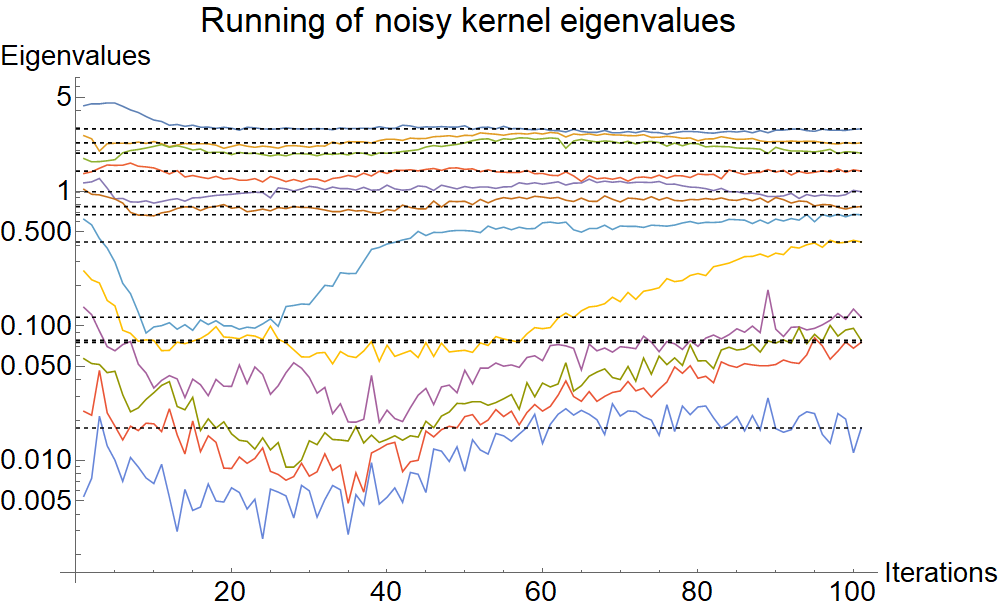}
    \caption{Kernel eigenvalues during the gradient descent dynamics. Up: noiseless simulation. Down: noisy simulation including a model of device errors. The solid curves are 12 nonzero eigenvalues of the QNTK, while the dashed lines are theoretical predictions of the frozen QNTK at the late time.}
    \label{fig:eigens}
\end{figure}

In Fig. \ref{fig:eigens}, we compute the QNTK eigenvalues for both the noiseless and noisy simulations, comparing them with theoretical predictions. Since we are in the underparametrized regime, the number of non-zero eigenvalues of the QNTK is the same as the number of variational angles, which is 12 in our experiments. We find agreement between those two in the late time, which shows the power of predictability using the QNTK theory.  

\section{Conclusions and open problems}\label{conc}
The results presented here establish a general framework of a QNTK theory, deriving analytical treatment of the optimization and learning dynamics in that regime. 
We outline the following open problems for future study. 
\begin{itemize}
    \item Our research gives practical guidance to the design variational quantum algorithms. One could compute the QNTK, or the kernel itself in the quantum kernel method \cite{havlivcek2019supervised}. If their eigenvalues are large, it will indicate faster convergence. It will be interesting to compare those results with other theoretical criteria about the quality of the variational quantum algorithms \cite{abbas2021power,haug2021capacity}.
    \item It would be interesting if one could investigate when the frozen QNTK limit is useful in other contexts. In Trotter product formulas, for instance, 
    \begin{align}
    \lim _{n \rightarrow \infty}\left(e^{i a X / n} e^{i b Z / n}\right)^{n},
    \end{align}
    to implement the gate $U=e^{i(a X+b Z)}$. Thus, small variational angles might widely appear in real cases of quantum architectures, even beyond the regime of lazy training around convergence. 
    \item Connection to the barren plateau problem. Our work suggests a possible connection between the barren plateau problem in variational quantum algorithms and the large-width limit in classical neural networks, by observing the following similarity between the LeCun parametrization above 
    \begin{align}
    \mathbb{E}\left( {{W}_{j_{1}^{C}{{j}_{1}}}}{{W}_{j_{2}^{C}{{j}_{2}}}} \right)={{\delta }_{j_{1}^{C},j_{2}^{C}}}{{\delta }_{{{j}_{1}},{{j}_{2}}}}\frac{{{C}_{W}}}{\text{width}},
    \end{align}
    and the 1-design random formula \cite{Roberts:2016hpo}
    \begin{align}
        \mathbb{E}({{U}_{ij}}U_{kl}^{\dagger })=\frac{{{\delta }_{il}}{{\delta }_{jk}}}{\dim \mathcal{H}}.
    \end{align}
    \item It will be interesting to explore the robustness of the QNTK theory against noise. Specifically, we have obtained an exponential convergence when the QNTK is frozen. 
    \item Finally, it will be interesting to dive deeper when the perturbative analysis fails in the non-linear regime. One could draw phase diagrams of quantum machine learning about some \emph{order parameters}, for instance, the learning rate \cite{lewkowycz2020large}. Those studies will deepen our theoretical understanding of quantum machine learning.   
\end{itemize}

\section*{Acknowledgement}

We thank Bryan Clark, Cliff Cheung, Jens Eisert, Thomas Faulkner, Guy Gur-Ari, Yoni Kahn, Risi Kondor, David Meltzer, Ryan Mishmash, John Preskill, Dan A. Roberts, David Simmons-Duffin, Kristan Temme, Lian-Tao Wang, Rebecca Willett, and Changchun Zhong for useful discussions. Specifically, JL would also acknowledge conversations with Dan A. Roberts on clarifying his book and lecture \cite{roberts2021principles,summer}. JL is supported in part by International Business Machines (IBM) Quantum through the Chicago Quantum Exchange, and the Pritzker School of Molecular Engineering at the University of Chicago through AFOSR MURI (FA9550-21-1-0209).
LJ acknowledges support from the ARO (W911NF-18-1-0020, W911NF-18-1-0212), ARO MURI (W911NF-16-1-0349), AFOSR MURI (FA9550-19-1-0399, FA9550-21-1-0209), DoE Q-NEXT, NSF (EFMA-1640959, OMA-1936118, EEC-1941583), NTT Research, and the Packard Foundation (2013-39273).

\emph{Note Added:} A recent independent paper  \cite{shirai2021quantum} that addresses the same topic has been posted publicly on the arXiv four days prior to the publication of this manuscript.

\bibliographystyle{apsrev4-1}
\bibliography{representation.bib}

\begin{thebibliography}{46}%
\makeatletter
\providecommand \@ifxundefined [1]{%
 \@ifx{#1\undefined}
}%
\providecommand \@ifnum [1]{%
 \ifnum #1\expandafter \@firstoftwo
 \else \expandafter \@secondoftwo
 \fi
}%
\providecommand \@ifx [1]{%
 \ifx #1\expandafter \@firstoftwo
 \else \expandafter \@secondoftwo
 \fi
}%
\providecommand \natexlab [1]{#1}%
\providecommand \enquote  [1]{``#1''}%
\providecommand \bibnamefont  [1]{#1}%
\providecommand \bibfnamefont [1]{#1}%
\providecommand \citenamefont [1]{#1}%
\providecommand \href@noop [0]{\@secondoftwo}%
\providecommand \href [0]{\begingroup \@sanitize@url \@href}%
\providecommand \@href[1]{\@@startlink{#1}\@@href}%
\providecommand \@@href[1]{\endgroup#1\@@endlink}%
\providecommand \@sanitize@url [0]{\catcode `\\12\catcode `\$12\catcode
  `\&12\catcode `\#12\catcode `\^12\catcode `\_12\catcode `\%12\relax}%
\providecommand \@@startlink[1]{}%
\providecommand \@@endlink[0]{}%
\providecommand \url  [0]{\begingroup\@sanitize@url \@url }%
\providecommand \@url [1]{\endgroup\@href {#1}{\urlprefix }}%
\providecommand \urlprefix  [0]{URL }%
\providecommand \Eprint [0]{\href }%
\providecommand \doibase [0]{http://dx.doi.org/}%
\providecommand \selectlanguage [0]{\@gobble}%
\providecommand \bibinfo  [0]{\@secondoftwo}%
\providecommand \bibfield  [0]{\@secondoftwo}%
\providecommand \translation [1]{[#1]}%
\providecommand \BibitemOpen [0]{}%
\providecommand \bibitemStop [0]{}%
\providecommand \bibitemNoStop [0]{.\EOS\space}%
\providecommand \EOS [0]{\spacefactor3000\relax}%
\providecommand \BibitemShut  [1]{\csname bibitem#1\endcsname}%
\let\auto@bib@innerbib\@empty
\bibitem [{\citenamefont {Harrow}\ \emph {et~al.}(2009)\citenamefont {Harrow},
  \citenamefont {Hassidim},\ and\ \citenamefont {Lloyd}}]{harrow2009quantum}%
  \BibitemOpen
  \bibfield  {author} {\bibinfo {author} {\bibfnamefont {A.~W.}\ \bibnamefont
  {Harrow}}, \bibinfo {author} {\bibfnamefont {A.}~\bibnamefont {Hassidim}}, \
  and\ \bibinfo {author} {\bibfnamefont {S.}~\bibnamefont {Lloyd}},\
  }\href@noop {} {\bibfield  {journal} {\bibinfo  {journal} {Physical review
  letters}\ }\textbf {\bibinfo {volume} {103}},\ \bibinfo {pages} {150502}
  (\bibinfo {year} {2009})}\BibitemShut {NoStop}%
\bibitem [{\citenamefont {Wiebe}\ \emph {et~al.}(2012)\citenamefont {Wiebe},
  \citenamefont {Braun},\ and\ \citenamefont {Lloyd}}]{wiebe2012quantum}%
  \BibitemOpen
  \bibfield  {author} {\bibinfo {author} {\bibfnamefont {N.}~\bibnamefont
  {Wiebe}}, \bibinfo {author} {\bibfnamefont {D.}~\bibnamefont {Braun}}, \ and\
  \bibinfo {author} {\bibfnamefont {S.}~\bibnamefont {Lloyd}},\ }\href@noop {}
  {\bibfield  {journal} {\bibinfo  {journal} {Physical review letters}\
  }\textbf {\bibinfo {volume} {109}},\ \bibinfo {pages} {050505} (\bibinfo
  {year} {2012})}\BibitemShut {NoStop}%
\bibitem [{\citenamefont {Lloyd}\ \emph {et~al.}(2014)\citenamefont {Lloyd},
  \citenamefont {Mohseni},\ and\ \citenamefont
  {Rebentrost}}]{lloyd2014quantum}%
  \BibitemOpen
  \bibfield  {author} {\bibinfo {author} {\bibfnamefont {S.}~\bibnamefont
  {Lloyd}}, \bibinfo {author} {\bibfnamefont {M.}~\bibnamefont {Mohseni}}, \
  and\ \bibinfo {author} {\bibfnamefont {P.}~\bibnamefont {Rebentrost}},\
  }\href@noop {} {\bibfield  {journal} {\bibinfo  {journal} {Nature Physics}\
  }\textbf {\bibinfo {volume} {10}},\ \bibinfo {pages} {631} (\bibinfo {year}
  {2014})}\BibitemShut {NoStop}%
\bibitem [{\citenamefont {Wittek}(2014)}]{wittek2014quantum}%
  \BibitemOpen
  \bibfield  {author} {\bibinfo {author} {\bibfnamefont {P.}~\bibnamefont
  {Wittek}},\ }\href@noop {} {\emph {\bibinfo {title} {Quantum machine
  learning: what quantum computing means to data mining}}}\ (\bibinfo
  {publisher} {Academic Press},\ \bibinfo {year} {2014})\BibitemShut {NoStop}%
\bibitem [{\citenamefont {Wiebe}\ \emph {et~al.}(2014)\citenamefont {Wiebe},
  \citenamefont {Kapoor},\ and\ \citenamefont {Svore}}]{wiebe2014quantum}%
  \BibitemOpen
  \bibfield  {author} {\bibinfo {author} {\bibfnamefont {N.}~\bibnamefont
  {Wiebe}}, \bibinfo {author} {\bibfnamefont {A.}~\bibnamefont {Kapoor}}, \
  and\ \bibinfo {author} {\bibfnamefont {K.~M.}\ \bibnamefont {Svore}},\
  }\href@noop {} {\bibfield  {journal} {\bibinfo  {journal} {arXiv preprint
  arXiv:1412.3489}\ } (\bibinfo {year} {2014})}\BibitemShut {NoStop}%
\bibitem [{\citenamefont {Rebentrost}\ \emph {et~al.}(2014)\citenamefont
  {Rebentrost}, \citenamefont {Mohseni},\ and\ \citenamefont
  {Lloyd}}]{rebentrost2014quantum}%
  \BibitemOpen
  \bibfield  {author} {\bibinfo {author} {\bibfnamefont {P.}~\bibnamefont
  {Rebentrost}}, \bibinfo {author} {\bibfnamefont {M.}~\bibnamefont {Mohseni}},
  \ and\ \bibinfo {author} {\bibfnamefont {S.}~\bibnamefont {Lloyd}},\
  }\href@noop {} {\bibfield  {journal} {\bibinfo  {journal} {Physical review
  letters}\ }\textbf {\bibinfo {volume} {113}},\ \bibinfo {pages} {130503}
  (\bibinfo {year} {2014})}\BibitemShut {NoStop}%
\bibitem [{\citenamefont {Biamonte}\ \emph {et~al.}(2017)\citenamefont
  {Biamonte}, \citenamefont {Wittek}, \citenamefont {Pancotti}, \citenamefont
  {Rebentrost}, \citenamefont {Wiebe},\ and\ \citenamefont
  {Lloyd}}]{biamonte2017quantum}%
  \BibitemOpen
  \bibfield  {author} {\bibinfo {author} {\bibfnamefont {J.}~\bibnamefont
  {Biamonte}}, \bibinfo {author} {\bibfnamefont {P.}~\bibnamefont {Wittek}},
  \bibinfo {author} {\bibfnamefont {N.}~\bibnamefont {Pancotti}}, \bibinfo
  {author} {\bibfnamefont {P.}~\bibnamefont {Rebentrost}}, \bibinfo {author}
  {\bibfnamefont {N.}~\bibnamefont {Wiebe}}, \ and\ \bibinfo {author}
  {\bibfnamefont {S.}~\bibnamefont {Lloyd}},\ }\href@noop {} {\bibfield
  {journal} {\bibinfo  {journal} {Nature}\ }\textbf {\bibinfo {volume} {549}},\
  \bibinfo {pages} {195} (\bibinfo {year} {2017})}\BibitemShut {NoStop}%
\bibitem [{\citenamefont {McClean}\ \emph {et~al.}(2018)\citenamefont
  {McClean}, \citenamefont {Boixo}, \citenamefont {Smelyanskiy}, \citenamefont
  {Babbush},\ and\ \citenamefont {Neven}}]{mcclean2018barren}%
  \BibitemOpen
  \bibfield  {author} {\bibinfo {author} {\bibfnamefont {J.~R.}\ \bibnamefont
  {McClean}}, \bibinfo {author} {\bibfnamefont {S.}~\bibnamefont {Boixo}},
  \bibinfo {author} {\bibfnamefont {V.~N.}\ \bibnamefont {Smelyanskiy}},
  \bibinfo {author} {\bibfnamefont {R.}~\bibnamefont {Babbush}}, \ and\
  \bibinfo {author} {\bibfnamefont {H.}~\bibnamefont {Neven}},\ }\href@noop {}
  {\bibfield  {journal} {\bibinfo  {journal} {Nature communications}\ }\textbf
  {\bibinfo {volume} {9}},\ \bibinfo {pages} {1} (\bibinfo {year}
  {2018})}\BibitemShut {NoStop}%
\bibitem [{\citenamefont {Schuld}\ and\ \citenamefont
  {Killoran}(2019)}]{schuld2019quantum}%
  \BibitemOpen
  \bibfield  {author} {\bibinfo {author} {\bibfnamefont {M.}~\bibnamefont
  {Schuld}}\ and\ \bibinfo {author} {\bibfnamefont {N.}~\bibnamefont
  {Killoran}},\ }\href@noop {} {\bibfield  {journal} {\bibinfo  {journal}
  {Physical review letters}\ }\textbf {\bibinfo {volume} {122}},\ \bibinfo
  {pages} {040504} (\bibinfo {year} {2019})}\BibitemShut {NoStop}%
\bibitem [{\citenamefont {Tang}(2019)}]{tang2019quantum}%
  \BibitemOpen
  \bibfield  {author} {\bibinfo {author} {\bibfnamefont {E.}~\bibnamefont
  {Tang}},\ }in\ \href@noop {} {\emph {\bibinfo {booktitle} {Proceedings of the
  51st Annual ACM SIGACT Symposium on Theory of Computing}}}\ (\bibinfo {year}
  {2019})\ pp.\ \bibinfo {pages} {217--228}\BibitemShut {NoStop}%
\bibitem [{\citenamefont {Havl{\'\i}{\v{c}}ek}\ \emph
  {et~al.}(2019)\citenamefont {Havl{\'\i}{\v{c}}ek}, \citenamefont
  {C{\'o}rcoles}, \citenamefont {Temme}, \citenamefont {Harrow}, \citenamefont
  {Kandala}, \citenamefont {Chow},\ and\ \citenamefont
  {Gambetta}}]{havlivcek2019supervised}%
  \BibitemOpen
  \bibfield  {author} {\bibinfo {author} {\bibfnamefont {V.}~\bibnamefont
  {Havl{\'\i}{\v{c}}ek}}, \bibinfo {author} {\bibfnamefont {A.~D.}\
  \bibnamefont {C{\'o}rcoles}}, \bibinfo {author} {\bibfnamefont
  {K.}~\bibnamefont {Temme}}, \bibinfo {author} {\bibfnamefont {A.~W.}\
  \bibnamefont {Harrow}}, \bibinfo {author} {\bibfnamefont {A.}~\bibnamefont
  {Kandala}}, \bibinfo {author} {\bibfnamefont {J.~M.}\ \bibnamefont {Chow}}, \
  and\ \bibinfo {author} {\bibfnamefont {J.~M.}\ \bibnamefont {Gambetta}},\
  }\href@noop {} {\bibfield  {journal} {\bibinfo  {journal} {Nature}\ }\textbf
  {\bibinfo {volume} {567}},\ \bibinfo {pages} {209} (\bibinfo {year}
  {2019})}\BibitemShut {NoStop}%
\bibitem [{\citenamefont {Huang}\ \emph
  {et~al.}(2021{\natexlab{a}})\citenamefont {Huang}, \citenamefont {Broughton},
  \citenamefont {Mohseni}, \citenamefont {Babbush}, \citenamefont {Boixo},
  \citenamefont {Neven},\ and\ \citenamefont {McClean}}]{huang2021power}%
  \BibitemOpen
  \bibfield  {author} {\bibinfo {author} {\bibfnamefont {H.-Y.}\ \bibnamefont
  {Huang}}, \bibinfo {author} {\bibfnamefont {M.}~\bibnamefont {Broughton}},
  \bibinfo {author} {\bibfnamefont {M.}~\bibnamefont {Mohseni}}, \bibinfo
  {author} {\bibfnamefont {R.}~\bibnamefont {Babbush}}, \bibinfo {author}
  {\bibfnamefont {S.}~\bibnamefont {Boixo}}, \bibinfo {author} {\bibfnamefont
  {H.}~\bibnamefont {Neven}}, \ and\ \bibinfo {author} {\bibfnamefont {J.~R.}\
  \bibnamefont {McClean}},\ }\href@noop {} {\bibfield  {journal} {\bibinfo
  {journal} {Nature communications}\ }\textbf {\bibinfo {volume} {12}},\
  \bibinfo {pages} {1} (\bibinfo {year} {2021}{\natexlab{a}})}\BibitemShut
  {NoStop}%
\bibitem [{\citenamefont {Liu}\ \emph {et~al.}(2021)\citenamefont {Liu},
  \citenamefont {Arunachalam},\ and\ \citenamefont {Temme}}]{liu2021rigorous}%
  \BibitemOpen
  \bibfield  {author} {\bibinfo {author} {\bibfnamefont {Y.}~\bibnamefont
  {Liu}}, \bibinfo {author} {\bibfnamefont {S.}~\bibnamefont {Arunachalam}}, \
  and\ \bibinfo {author} {\bibfnamefont {K.}~\bibnamefont {Temme}},\
  }\href@noop {} {\bibfield  {journal} {\bibinfo  {journal} {Nature Physics}\
  ,\ \bibinfo {pages} {1}} (\bibinfo {year} {2021})}\BibitemShut {NoStop}%
\bibitem [{\citenamefont {Huang}\ \emph
  {et~al.}(2021{\natexlab{b}})\citenamefont {Huang}, \citenamefont {Kueng},\
  and\ \citenamefont {Preskill}}]{huang2021information}%
  \BibitemOpen
  \bibfield  {author} {\bibinfo {author} {\bibfnamefont {H.-Y.}\ \bibnamefont
  {Huang}}, \bibinfo {author} {\bibfnamefont {R.}~\bibnamefont {Kueng}}, \ and\
  \bibinfo {author} {\bibfnamefont {J.}~\bibnamefont {Preskill}},\ }\href@noop
  {} {\bibfield  {journal} {\bibinfo  {journal} {Physical Review Letters}\
  }\textbf {\bibinfo {volume} {126}},\ \bibinfo {pages} {190505} (\bibinfo
  {year} {2021}{\natexlab{b}})}\BibitemShut {NoStop}%
\bibitem [{\citenamefont {Lee}\ \emph {et~al.}(2017)\citenamefont {Lee},
  \citenamefont {Bahri}, \citenamefont {Novak}, \citenamefont {Schoenholz},
  \citenamefont {Pennington},\ and\ \citenamefont
  {Sohl-Dickstein}}]{lee2017deep}%
  \BibitemOpen
  \bibfield  {author} {\bibinfo {author} {\bibfnamefont {J.}~\bibnamefont
  {Lee}}, \bibinfo {author} {\bibfnamefont {Y.}~\bibnamefont {Bahri}}, \bibinfo
  {author} {\bibfnamefont {R.}~\bibnamefont {Novak}}, \bibinfo {author}
  {\bibfnamefont {S.~S.}\ \bibnamefont {Schoenholz}}, \bibinfo {author}
  {\bibfnamefont {J.}~\bibnamefont {Pennington}}, \ and\ \bibinfo {author}
  {\bibfnamefont {J.}~\bibnamefont {Sohl-Dickstein}},\ }\href@noop {}
  {\bibfield  {journal} {\bibinfo  {journal} {arXiv preprint arXiv:1711.00165}\
  } (\bibinfo {year} {2017})}\BibitemShut {NoStop}%
\bibitem [{\citenamefont {Jacot}\ \emph {et~al.}(2018)\citenamefont {Jacot},
  \citenamefont {Gabriel},\ and\ \citenamefont {Hongler}}]{jacot2018neural}%
  \BibitemOpen
  \bibfield  {author} {\bibinfo {author} {\bibfnamefont {A.}~\bibnamefont
  {Jacot}}, \bibinfo {author} {\bibfnamefont {F.}~\bibnamefont {Gabriel}}, \
  and\ \bibinfo {author} {\bibfnamefont {C.}~\bibnamefont {Hongler}},\
  }\href@noop {} {\bibfield  {journal} {\bibinfo  {journal} {arXiv preprint
  arXiv:1806.07572}\ } (\bibinfo {year} {2018})}\BibitemShut {NoStop}%
\bibitem [{\citenamefont {Lee}\ \emph {et~al.}(2019)\citenamefont {Lee},
  \citenamefont {Xiao}, \citenamefont {Schoenholz}, \citenamefont {Bahri},
  \citenamefont {Novak}, \citenamefont {Sohl-Dickstein},\ and\ \citenamefont
  {Pennington}}]{lee2019wide}%
  \BibitemOpen
  \bibfield  {author} {\bibinfo {author} {\bibfnamefont {J.}~\bibnamefont
  {Lee}}, \bibinfo {author} {\bibfnamefont {L.}~\bibnamefont {Xiao}}, \bibinfo
  {author} {\bibfnamefont {S.}~\bibnamefont {Schoenholz}}, \bibinfo {author}
  {\bibfnamefont {Y.}~\bibnamefont {Bahri}}, \bibinfo {author} {\bibfnamefont
  {R.}~\bibnamefont {Novak}}, \bibinfo {author} {\bibfnamefont
  {J.}~\bibnamefont {Sohl-Dickstein}}, \ and\ \bibinfo {author} {\bibfnamefont
  {J.}~\bibnamefont {Pennington}},\ }\href@noop {} {\bibfield  {journal}
  {\bibinfo  {journal} {Advances in neural information processing systems}\
  }\textbf {\bibinfo {volume} {32}},\ \bibinfo {pages} {8572} (\bibinfo {year}
  {2019})}\BibitemShut {NoStop}%
\bibitem [{\citenamefont {Arora}\ \emph {et~al.}(2019)\citenamefont {Arora},
  \citenamefont {Du}, \citenamefont {Hu}, \citenamefont {Li}, \citenamefont
  {Salakhutdinov},\ and\ \citenamefont {Wang}}]{arora2019exact}%
  \BibitemOpen
  \bibfield  {author} {\bibinfo {author} {\bibfnamefont {S.}~\bibnamefont
  {Arora}}, \bibinfo {author} {\bibfnamefont {S.~S.}\ \bibnamefont {Du}},
  \bibinfo {author} {\bibfnamefont {W.}~\bibnamefont {Hu}}, \bibinfo {author}
  {\bibfnamefont {Z.}~\bibnamefont {Li}}, \bibinfo {author} {\bibfnamefont
  {R.}~\bibnamefont {Salakhutdinov}}, \ and\ \bibinfo {author} {\bibfnamefont
  {R.}~\bibnamefont {Wang}},\ }\href@noop {} {\bibfield  {journal} {\bibinfo
  {journal} {arXiv preprint arXiv:1904.11955}\ } (\bibinfo {year}
  {2019})}\BibitemShut {NoStop}%
\bibitem [{\citenamefont {Sohl-Dickstein}\ \emph {et~al.}(2020)\citenamefont
  {Sohl-Dickstein}, \citenamefont {Novak}, \citenamefont {Schoenholz},\ and\
  \citenamefont {Lee}}]{sohl2020infinite}%
  \BibitemOpen
  \bibfield  {author} {\bibinfo {author} {\bibfnamefont {J.}~\bibnamefont
  {Sohl-Dickstein}}, \bibinfo {author} {\bibfnamefont {R.}~\bibnamefont
  {Novak}}, \bibinfo {author} {\bibfnamefont {S.~S.}\ \bibnamefont
  {Schoenholz}}, \ and\ \bibinfo {author} {\bibfnamefont {J.}~\bibnamefont
  {Lee}},\ }\href@noop {} {\bibfield  {journal} {\bibinfo  {journal} {arXiv
  preprint arXiv:2001.07301}\ } (\bibinfo {year} {2020})}\BibitemShut {NoStop}%
\bibitem [{\citenamefont {Yang}\ and\ \citenamefont
  {Hu}(2020)}]{yang2020feature}%
  \BibitemOpen
  \bibfield  {author} {\bibinfo {author} {\bibfnamefont {G.}~\bibnamefont
  {Yang}}\ and\ \bibinfo {author} {\bibfnamefont {E.~J.}\ \bibnamefont {Hu}},\
  }\href@noop {} {\bibfield  {journal} {\bibinfo  {journal} {arXiv preprint
  arXiv:2011.14522}\ } (\bibinfo {year} {2020})}\BibitemShut {NoStop}%
\bibitem [{\citenamefont {Yaida}(2020)}]{yaida2020non}%
  \BibitemOpen
  \bibfield  {author} {\bibinfo {author} {\bibfnamefont {S.}~\bibnamefont
  {Yaida}},\ }in\ \href@noop {} {\emph {\bibinfo {booktitle} {Mathematical and
  Scientific Machine Learning}}}\ (\bibinfo {organization} {PMLR},\ \bibinfo
  {year} {2020})\ pp.\ \bibinfo {pages} {165--192}\BibitemShut {NoStop}%
\bibitem [{\citenamefont {Dyer}\ and\ \citenamefont
  {Gur-Ari}(2019)}]{dyer2019asymptotics}%
  \BibitemOpen
  \bibfield  {author} {\bibinfo {author} {\bibfnamefont {E.}~\bibnamefont
  {Dyer}}\ and\ \bibinfo {author} {\bibfnamefont {G.}~\bibnamefont {Gur-Ari}},\
  }\href@noop {} {\bibfield  {journal} {\bibinfo  {journal} {arXiv preprint
  arXiv:1909.11304}\ } (\bibinfo {year} {2019})}\BibitemShut {NoStop}%
\bibitem [{\citenamefont {Halverson}\ \emph {et~al.}(2021)\citenamefont
  {Halverson}, \citenamefont {Maiti},\ and\ \citenamefont
  {Stoner}}]{halverson2021neural}%
  \BibitemOpen
  \bibfield  {author} {\bibinfo {author} {\bibfnamefont {J.}~\bibnamefont
  {Halverson}}, \bibinfo {author} {\bibfnamefont {A.}~\bibnamefont {Maiti}}, \
  and\ \bibinfo {author} {\bibfnamefont {K.}~\bibnamefont {Stoner}},\
  }\href@noop {} {\bibfield  {journal} {\bibinfo  {journal} {Machine Learning:
  Science and Technology}\ }\textbf {\bibinfo {volume} {2}},\ \bibinfo {pages}
  {035002} (\bibinfo {year} {2021})}\BibitemShut {NoStop}%
\bibitem [{\citenamefont {Roberts}(2021)}]{roberts2021ai}%
  \BibitemOpen
  \bibfield  {author} {\bibinfo {author} {\bibfnamefont {D.~A.}\ \bibnamefont
  {Roberts}},\ }\href@noop {} {\bibfield  {journal} {\bibinfo  {journal} {arXiv
  preprint arXiv:2104.00008}\ } (\bibinfo {year} {2021})}\BibitemShut {NoStop}%
\bibitem [{\citenamefont {Roberts}\ \emph {et~al.}(2021)\citenamefont
  {Roberts}, \citenamefont {Yaida},\ and\ \citenamefont
  {Hanin}}]{roberts2021principles}%
  \BibitemOpen
  \bibfield  {author} {\bibinfo {author} {\bibfnamefont {D.~A.}\ \bibnamefont
  {Roberts}}, \bibinfo {author} {\bibfnamefont {S.}~\bibnamefont {Yaida}}, \
  and\ \bibinfo {author} {\bibfnamefont {B.}~\bibnamefont {Hanin}},\
  }\href@noop {} {\bibfield  {journal} {\bibinfo  {journal} {arXiv preprint
  arXiv:2106.10165}\ } (\bibinfo {year} {2021})}\BibitemShut {NoStop}%
\bibitem [{\citenamefont {Liu}(2021)}]{Liu:2021ohs}%
  \BibitemOpen
  \bibfield  {author} {\bibinfo {author} {\bibfnamefont {J.}~\bibnamefont
  {Liu}},\ }\emph {\bibinfo {title} {{Does Richard Feynman Dream of Electric
  Sheep? Topics on Quantum Field Theory, Quantum Computing, and Computer
  Science}}},\ \href {\doibase 10.7907/adtc-ss13} {Ph.D. thesis},\ \bibinfo
  {school} {Caltech} (\bibinfo {year} {2021})\BibitemShut {NoStop}%
\bibitem [{\citenamefont {Nakaji}\ \emph {et~al.}(2021)\citenamefont {Nakaji},
  \citenamefont {Tezuka},\ and\ \citenamefont
  {Yamamoto}}]{nakaji2021quantumenhanced}%
  \BibitemOpen
  \bibfield  {author} {\bibinfo {author} {\bibfnamefont {K.}~\bibnamefont
  {Nakaji}}, \bibinfo {author} {\bibfnamefont {H.}~\bibnamefont {Tezuka}}, \
  and\ \bibinfo {author} {\bibfnamefont {N.}~\bibnamefont {Yamamoto}},\
  }\href@noop {} {\enquote {\bibinfo {title} {Quantum-enhanced neural networks
  in the neural tangent kernel framework},}\ } (\bibinfo {year} {2021}),\
  \Eprint {http://arxiv.org/abs/2109.03786} {arXiv:2109.03786 [quant-ph]}
  \BibitemShut {NoStop}%
\bibitem [{\citenamefont {Roberts}\ and\ \citenamefont {Yaida}(2021)}]{summer}%
  \BibitemOpen
  \bibfield  {author} {\bibinfo {author} {\bibfnamefont {D.~A.}\ \bibnamefont
  {Roberts}}\ and\ \bibinfo {author} {\bibfnamefont {S.}~\bibnamefont
  {Yaida}},\ }\href@noop {} {\bibfield  {journal} {\bibinfo  {journal} {Deep
  Learning Theory Summer School at Princeton}\ } (\bibinfo {year}
  {2021})}\BibitemShut {NoStop}%
\bibitem [{\citenamefont {Chizat}\ \emph {et~al.}(2018)\citenamefont {Chizat},
  \citenamefont {Oyallon},\ and\ \citenamefont {Bach}}]{chizat2018lazy}%
  \BibitemOpen
  \bibfield  {author} {\bibinfo {author} {\bibfnamefont {L.}~\bibnamefont
  {Chizat}}, \bibinfo {author} {\bibfnamefont {E.}~\bibnamefont {Oyallon}}, \
  and\ \bibinfo {author} {\bibfnamefont {F.}~\bibnamefont {Bach}},\ }\href@noop
  {} {\bibfield  {journal} {\bibinfo  {journal} {arXiv preprint
  arXiv:1812.07956}\ } (\bibinfo {year} {2018})}\BibitemShut {NoStop}%
\bibitem [{\citenamefont {Aleksandrowicz}\ \emph {et~al.}(2019)\citenamefont
  {Aleksandrowicz} \emph {et~al.}}]{aleksandrowicz2019qiskit}%
  \BibitemOpen
  \bibfield  {author} {\bibinfo {author} {\bibfnamefont {G.}~\bibnamefont
  {Aleksandrowicz}} \emph {et~al.},\ }\href@noop {} {\enquote {\bibinfo {title}
  {Qiskit: An open-source framework for quantum computing},}\ } (\bibinfo
  {year} {2019})\BibitemShut {NoStop}%
\bibitem [{\citenamefont {Peruzzo}\ \emph {et~al.}(2014)\citenamefont
  {Peruzzo}, \citenamefont {McClean}, \citenamefont {Shadbolt}, \citenamefont
  {Yung}, \citenamefont {Zhou}, \citenamefont {Love}, \citenamefont
  {Aspuru-Guzik},\ and\ \citenamefont {O’brien}}]{peruzzo2014variational}%
  \BibitemOpen
  \bibfield  {author} {\bibinfo {author} {\bibfnamefont {A.}~\bibnamefont
  {Peruzzo}}, \bibinfo {author} {\bibfnamefont {J.}~\bibnamefont {McClean}},
  \bibinfo {author} {\bibfnamefont {P.}~\bibnamefont {Shadbolt}}, \bibinfo
  {author} {\bibfnamefont {M.-H.}\ \bibnamefont {Yung}}, \bibinfo {author}
  {\bibfnamefont {X.-Q.}\ \bibnamefont {Zhou}}, \bibinfo {author}
  {\bibfnamefont {P.~J.}\ \bibnamefont {Love}}, \bibinfo {author}
  {\bibfnamefont {A.}~\bibnamefont {Aspuru-Guzik}}, \ and\ \bibinfo {author}
  {\bibfnamefont {J.~L.}\ \bibnamefont {O’brien}},\ }\href@noop {} {\bibfield
   {journal} {\bibinfo  {journal} {Nature communications}\ }\textbf {\bibinfo
  {volume} {5}},\ \bibinfo {pages} {1} (\bibinfo {year} {2014})}\BibitemShut
  {NoStop}%
\bibitem [{\citenamefont {Farhi}\ \emph {et~al.}(2014)\citenamefont {Farhi},
  \citenamefont {Goldstone},\ and\ \citenamefont {Gutmann}}]{farhi2014quantum}%
  \BibitemOpen
  \bibfield  {author} {\bibinfo {author} {\bibfnamefont {E.}~\bibnamefont
  {Farhi}}, \bibinfo {author} {\bibfnamefont {J.}~\bibnamefont {Goldstone}}, \
  and\ \bibinfo {author} {\bibfnamefont {S.}~\bibnamefont {Gutmann}},\
  }\href@noop {} {\bibfield  {journal} {\bibinfo  {journal} {arXiv preprint
  arXiv:1411.4028}\ } (\bibinfo {year} {2014})}\BibitemShut {NoStop}%
\bibitem [{\citenamefont {McClean}\ \emph {et~al.}(2016)\citenamefont
  {McClean}, \citenamefont {Romero}, \citenamefont {Babbush},\ and\
  \citenamefont {Aspuru-Guzik}}]{mcclean2016theory}%
  \BibitemOpen
  \bibfield  {author} {\bibinfo {author} {\bibfnamefont {J.~R.}\ \bibnamefont
  {McClean}}, \bibinfo {author} {\bibfnamefont {J.}~\bibnamefont {Romero}},
  \bibinfo {author} {\bibfnamefont {R.}~\bibnamefont {Babbush}}, \ and\
  \bibinfo {author} {\bibfnamefont {A.}~\bibnamefont {Aspuru-Guzik}},\
  }\href@noop {} {\bibfield  {journal} {\bibinfo  {journal} {New Journal of
  Physics}\ }\textbf {\bibinfo {volume} {18}},\ \bibinfo {pages} {023023}
  (\bibinfo {year} {2016})}\BibitemShut {NoStop}%
\bibitem [{\citenamefont {Kandala}\ \emph {et~al.}(2017)\citenamefont
  {Kandala}, \citenamefont {Mezzacapo}, \citenamefont {Temme}, \citenamefont
  {Takita}, \citenamefont {Brink}, \citenamefont {Chow},\ and\ \citenamefont
  {Gambetta}}]{kandala2017hardware}%
  \BibitemOpen
  \bibfield  {author} {\bibinfo {author} {\bibfnamefont {A.}~\bibnamefont
  {Kandala}}, \bibinfo {author} {\bibfnamefont {A.}~\bibnamefont {Mezzacapo}},
  \bibinfo {author} {\bibfnamefont {K.}~\bibnamefont {Temme}}, \bibinfo
  {author} {\bibfnamefont {M.}~\bibnamefont {Takita}}, \bibinfo {author}
  {\bibfnamefont {M.}~\bibnamefont {Brink}}, \bibinfo {author} {\bibfnamefont
  {J.~M.}\ \bibnamefont {Chow}}, \ and\ \bibinfo {author} {\bibfnamefont
  {J.~M.}\ \bibnamefont {Gambetta}},\ }\href@noop {} {\bibfield  {journal}
  {\bibinfo  {journal} {Nature}\ }\textbf {\bibinfo {volume} {549}},\ \bibinfo
  {pages} {242} (\bibinfo {year} {2017})}\BibitemShut {NoStop}%
\bibitem [{\citenamefont {McArdle}\ \emph {et~al.}(2020)\citenamefont
  {McArdle}, \citenamefont {Endo}, \citenamefont {Aspuru-Guzik}, \citenamefont
  {Benjamin},\ and\ \citenamefont {Yuan}}]{mcardle2020quantum}%
  \BibitemOpen
  \bibfield  {author} {\bibinfo {author} {\bibfnamefont {S.}~\bibnamefont
  {McArdle}}, \bibinfo {author} {\bibfnamefont {S.}~\bibnamefont {Endo}},
  \bibinfo {author} {\bibfnamefont {A.}~\bibnamefont {Aspuru-Guzik}}, \bibinfo
  {author} {\bibfnamefont {S.~C.}\ \bibnamefont {Benjamin}}, \ and\ \bibinfo
  {author} {\bibfnamefont {X.}~\bibnamefont {Yuan}},\ }\href@noop {} {\bibfield
   {journal} {\bibinfo  {journal} {Reviews of Modern Physics}\ }\textbf
  {\bibinfo {volume} {92}},\ \bibinfo {pages} {015003} (\bibinfo {year}
  {2020})}\BibitemShut {NoStop}%
\bibitem [{\citenamefont {Cerezo}\ \emph {et~al.}(2021)\citenamefont {Cerezo},
  \citenamefont {Arrasmith}, \citenamefont {Babbush}, \citenamefont {Benjamin},
  \citenamefont {Endo}, \citenamefont {Fujii}, \citenamefont {McClean},
  \citenamefont {Mitarai}, \citenamefont {Yuan}, \citenamefont {Cincio} \emph
  {et~al.}}]{cerezo2021variational}%
  \BibitemOpen
  \bibfield  {author} {\bibinfo {author} {\bibfnamefont {M.}~\bibnamefont
  {Cerezo}}, \bibinfo {author} {\bibfnamefont {A.}~\bibnamefont {Arrasmith}},
  \bibinfo {author} {\bibfnamefont {R.}~\bibnamefont {Babbush}}, \bibinfo
  {author} {\bibfnamefont {S.~C.}\ \bibnamefont {Benjamin}}, \bibinfo {author}
  {\bibfnamefont {S.}~\bibnamefont {Endo}}, \bibinfo {author} {\bibfnamefont
  {K.}~\bibnamefont {Fujii}}, \bibinfo {author} {\bibfnamefont {J.~R.}\
  \bibnamefont {McClean}}, \bibinfo {author} {\bibfnamefont {K.}~\bibnamefont
  {Mitarai}}, \bibinfo {author} {\bibfnamefont {X.}~\bibnamefont {Yuan}},
  \bibinfo {author} {\bibfnamefont {L.}~\bibnamefont {Cincio}},  \emph
  {et~al.},\ }\href@noop {} {\bibfield  {journal} {\bibinfo  {journal} {Nature
  Reviews Physics}\ ,\ \bibinfo {pages} {1}} (\bibinfo {year}
  {2021})}\BibitemShut {NoStop}%
\bibitem [{\citenamefont {Otterbach}\ \emph {et~al.}(2017)\citenamefont
  {Otterbach}, \citenamefont {Manenti}, \citenamefont {Alidoust}, \citenamefont
  {Bestwick}, \citenamefont {Block}, \citenamefont {Bloom}, \citenamefont
  {Caldwell}, \citenamefont {Didier}, \citenamefont {Fried}, \citenamefont
  {Hong} \emph {et~al.}}]{otterbach2017unsupervised}%
  \BibitemOpen
  \bibfield  {author} {\bibinfo {author} {\bibfnamefont {J.}~\bibnamefont
  {Otterbach}}, \bibinfo {author} {\bibfnamefont {R.}~\bibnamefont {Manenti}},
  \bibinfo {author} {\bibfnamefont {N.}~\bibnamefont {Alidoust}}, \bibinfo
  {author} {\bibfnamefont {A.}~\bibnamefont {Bestwick}}, \bibinfo {author}
  {\bibfnamefont {M.}~\bibnamefont {Block}}, \bibinfo {author} {\bibfnamefont
  {B.}~\bibnamefont {Bloom}}, \bibinfo {author} {\bibfnamefont
  {S.}~\bibnamefont {Caldwell}}, \bibinfo {author} {\bibfnamefont
  {N.}~\bibnamefont {Didier}}, \bibinfo {author} {\bibfnamefont {E.~S.}\
  \bibnamefont {Fried}}, \bibinfo {author} {\bibfnamefont {S.}~\bibnamefont
  {Hong}},  \emph {et~al.},\ }\href@noop {} {\bibfield  {journal} {\bibinfo
  {journal} {arXiv preprint arXiv:1712.05771}\ } (\bibinfo {year}
  {2017})}\BibitemShut {NoStop}%
\bibitem [{\citenamefont {Farhi}\ and\ \citenamefont
  {Neven}(2018)}]{farhi2018classification}%
  \BibitemOpen
  \bibfield  {author} {\bibinfo {author} {\bibfnamefont {E.}~\bibnamefont
  {Farhi}}\ and\ \bibinfo {author} {\bibfnamefont {H.}~\bibnamefont {Neven}},\
  }\href@noop {} {\bibfield  {journal} {\bibinfo  {journal} {arXiv preprint
  arXiv:1802.06002}\ } (\bibinfo {year} {2018})}\BibitemShut {NoStop}%
\bibitem [{Note1()}]{Note1}%
  \BibitemOpen
  \bibinfo {note} {This is somewhat different from the so-called \protect \emph
  {NTK parametrization} in some literature. See details in SM. Moreover, here
  we assume the weights and biases are real.}\BibitemShut {Stop}%
\bibitem [{Note2()}]{Note2}%
  \BibitemOpen
  \bibinfo {note} {Similar analysis could be done on dynamics, see SM for
  further comments.}\BibitemShut {Stop}%
\bibitem [{ibm(2021)}]{ibmqml}%
  \BibitemOpen
  \href@noop {} {\enquote {\bibinfo {title} {Qiskit machine learning},}\
  }\bibinfo {howpublished}
  {\url{https://qiskit.org/documentation/machine-learning/}} (\bibinfo {year}
  {2021})\BibitemShut {NoStop}%
\bibitem [{\citenamefont {Abbas}\ \emph {et~al.}(2021)\citenamefont {Abbas},
  \citenamefont {Sutter}, \citenamefont {Zoufal}, \citenamefont {Lucchi},
  \citenamefont {Figalli},\ and\ \citenamefont {Woerner}}]{abbas2021power}%
  \BibitemOpen
  \bibfield  {author} {\bibinfo {author} {\bibfnamefont {A.}~\bibnamefont
  {Abbas}}, \bibinfo {author} {\bibfnamefont {D.}~\bibnamefont {Sutter}},
  \bibinfo {author} {\bibfnamefont {C.}~\bibnamefont {Zoufal}}, \bibinfo
  {author} {\bibfnamefont {A.}~\bibnamefont {Lucchi}}, \bibinfo {author}
  {\bibfnamefont {A.}~\bibnamefont {Figalli}}, \ and\ \bibinfo {author}
  {\bibfnamefont {S.}~\bibnamefont {Woerner}},\ }\href@noop {} {\bibfield
  {journal} {\bibinfo  {journal} {Nature Computational Science}\ }\textbf
  {\bibinfo {volume} {1}},\ \bibinfo {pages} {403} (\bibinfo {year}
  {2021})}\BibitemShut {NoStop}%
\bibitem [{\citenamefont {Haug}\ \emph {et~al.}(2021)\citenamefont {Haug},
  \citenamefont {Bharti},\ and\ \citenamefont {Kim}}]{haug2021capacity}%
  \BibitemOpen
  \bibfield  {author} {\bibinfo {author} {\bibfnamefont {T.}~\bibnamefont
  {Haug}}, \bibinfo {author} {\bibfnamefont {K.}~\bibnamefont {Bharti}}, \ and\
  \bibinfo {author} {\bibfnamefont {M.}~\bibnamefont {Kim}},\ }\href@noop {}
  {\bibfield  {journal} {\bibinfo  {journal} {arXiv preprint arXiv:2102.01659}\
  } (\bibinfo {year} {2021})}\BibitemShut {NoStop}%
\bibitem [{\citenamefont {Roberts}\ and\ \citenamefont
  {Yoshida}(2017)}]{Roberts:2016hpo}%
  \BibitemOpen
  \bibfield  {author} {\bibinfo {author} {\bibfnamefont {D.~A.}\ \bibnamefont
  {Roberts}}\ and\ \bibinfo {author} {\bibfnamefont {B.}~\bibnamefont
  {Yoshida}},\ }\href {\doibase 10.1007/JHEP04(2017)121} {\bibfield  {journal}
  {\bibinfo  {journal} {JHEP}\ }\textbf {\bibinfo {volume} {04}},\ \bibinfo
  {pages} {121} (\bibinfo {year} {2017})},\ \Eprint
  {http://arxiv.org/abs/1610.04903} {arXiv:1610.04903 [quant-ph]} \BibitemShut
  {NoStop}%
\bibitem [{\citenamefont {Lewkowycz}\ \emph {et~al.}(2020)\citenamefont
  {Lewkowycz}, \citenamefont {Bahri}, \citenamefont {Dyer}, \citenamefont
  {Sohl-Dickstein},\ and\ \citenamefont {Gur-Ari}}]{lewkowycz2020large}%
  \BibitemOpen
  \bibfield  {author} {\bibinfo {author} {\bibfnamefont {A.}~\bibnamefont
  {Lewkowycz}}, \bibinfo {author} {\bibfnamefont {Y.}~\bibnamefont {Bahri}},
  \bibinfo {author} {\bibfnamefont {E.}~\bibnamefont {Dyer}}, \bibinfo {author}
  {\bibfnamefont {J.}~\bibnamefont {Sohl-Dickstein}}, \ and\ \bibinfo {author}
  {\bibfnamefont {G.}~\bibnamefont {Gur-Ari}},\ }\href@noop {} {\bibfield
  {journal} {\bibinfo  {journal} {arXiv preprint arXiv:2003.02218}\ } (\bibinfo
  {year} {2020})}\BibitemShut {NoStop}%
\bibitem [{\citenamefont {Shirai}\ \emph {et~al.}(2021)\citenamefont {Shirai},
  \citenamefont {Kubo}, \citenamefont {Mitarai},\ and\ \citenamefont
  {Fujii}}]{shirai2021quantum}%
  \BibitemOpen
  \bibfield  {author} {\bibinfo {author} {\bibfnamefont {N.}~\bibnamefont
  {Shirai}}, \bibinfo {author} {\bibfnamefont {K.}~\bibnamefont {Kubo}},
  \bibinfo {author} {\bibfnamefont {K.}~\bibnamefont {Mitarai}}, \ and\
  \bibinfo {author} {\bibfnamefont {K.}~\bibnamefont {Fujii}},\ }\href@noop {}
  {\enquote {\bibinfo {title} {Quantum tangent kernel},}\ } (\bibinfo {year}
  {2021}),\ \Eprint {http://arxiv.org/abs/2111.02951} {arXiv:2111.02951
  [quant-ph]} \BibitemShut {NoStop}%
\end{thebibliography}%

\pagebreak
\clearpage
\foreach \x in {1,...,\the\pdflastximagepages}
{
	\clearpage
	

\section*{Supplementary material}

# Supplemental Material: Representation Learning via Quantum Neural Tangent Kernels

Junyu Liu,[1, 2, 3, *] Francesco Tacchino,[4, †] Jennifer R. Glick,[5, ‡] Liang Jiang,[1, 2, §] and Antonio Mezzacapo[5, ¶]

[1]*Pritzker School of Molecular Engineering, The University of Chicago, Chicago, IL 60637, USA*
[2]*Chicago Quantum Exchange, Chicago, IL 60637, USA*
[3]*Kadanoff Center for Theoretical Physics, The University of Chicago, Chicago, IL 60637, USA*
[4]*IBM Quantum, IBM Research, Zurich, 8803 Rüschlikon, Switzerland*
[5]*IBM Quantum, IBM T. J. Watson Research Center, Yorktown Heights, New York 10598, USA*
(Dated: November 13, 2021)

## CONTENTS

## I. DETAILS ON THE LINEAR AND THE QUADRATIC MODELS

### A. Linear model

Before we discuss the quantum model setup, we start by reviewing the linear and quadratic models in the classical setup. The linear and quadratic models are originally solved in the framework presented by [1, 2], and in this section, we will give a brief review.

---

* junyuliu@uchicago.edu
† fta@zurich.ibm.com
‡ jennifer.r.glick@ibm.com
§ liang.jiang@uchicago.edu
¶ mezzacapo@ibm.com

We define the model as

$$z_{i;\delta}(\theta) = \sum_j W_{ij}\phi_j(\mathbf{x}_\delta) \ . \tag{1}$$

Here $\mathbf{x}_\delta$ is the data point $\delta$ in the space $\mathcal{D}$, and $W_{ij}$s are weights and biases. We use the slack notation such that $W_{0j} = b_j$ includes the bias. The model is called the linear model, which is linear in weights, but we wish to keep the feature map $\phi$ in general. Moreover, we write $\theta$ as a compact notation of the vectorized $W$.

We wish to optimize the following loss function

$$\mathcal{L}_\mathcal{A}(\theta) = \frac{1}{2}\sum_{i,\tilde{\alpha}}\left[y_{i;\tilde{\alpha}} - \sum_j W_{ij}\phi_j(\mathbf{x}_{\tilde{\alpha}})\right]^2 \ . \tag{2}$$

For the sample set $\mathcal{A}$. Again, we will use the notation $\tilde{\alpha}$ such that $\tilde{\alpha}$ is in the sample set $\mathcal{A}$. We define the kernel

$$k_{\delta_1\delta_2} \equiv k(\mathbf{x}_{\delta_1}, \mathbf{x}_{\delta_2}) \equiv \sum_i \phi_i(\mathbf{x}_{\delta_1})\phi_i(\mathbf{x}_{\delta_2}) = \sum_{a,b}\frac{dz_{i;\delta_1}}{dW_{ab}}\frac{dz_{i;\delta_2}}{dW_{ab}} \ . \tag{3}$$

Note that the right hand side does not contain a sum over $i$. It is equal for all $i$s. Moreover, We call $\delta \in \mathcal{D}$ as the whole data set, while $\mathcal{A} \subset \mathcal{D}$ is the sample set. We define $\widetilde{k}_{\tilde{\alpha}_1\tilde{\alpha}_2}$ with tilde to indicate the submatrix. We also define [3]

$$\sum_{\tilde{\alpha}_2 \in \mathcal{A}} \widetilde{k}^{\tilde{\alpha}_1\tilde{\alpha}_2}\widetilde{k}_{\tilde{\alpha}_2\tilde{\alpha}_3} = \delta^{\tilde{\alpha}_1}_{\tilde{\alpha}_3} \ . \tag{4}$$

namely, the upper index means the inverse. Similar to the main text, we consider the gradient descent algorithm,

$$d\theta_\mu = -\eta\frac{\partial \mathcal{L}_\mathcal{A}}{\partial \theta_\mu} \ . \tag{5}$$

The partial derivatives are computed as

$$\begin{aligned}\frac{\partial L_A(W)}{\partial W_{ab}} &= -\sum_{\tilde{\alpha},i,j}\delta_{ia}\delta_{jb}\phi_j(\mathbf{x}_{\tilde{\alpha}})\left[y_{i;\tilde{\alpha}} - \sum_j W_{ij}\phi_j(\mathbf{x}_{\tilde{\alpha}})\right] \\ &= \sum_{\tilde{\alpha}}\phi_b(\mathbf{x}_{\tilde{\alpha}})(z_{a;\tilde{\alpha}} - y_{a;\tilde{\alpha}}) = \sum_{\tilde{\alpha}}\varepsilon_{a;\tilde{\alpha}}\phi_b(\mathbf{x}_{\tilde{\alpha}})\end{aligned} \tag{6}$$

where $\varepsilon$ is the residual training error,

$$dW_{ij} = -\eta\sum_{\tilde{\alpha}}\phi_j(\mathbf{x}_{\tilde{\alpha}})\varepsilon_{i;\tilde{\alpha}} \ . \tag{7}$$

So we have

$$dz_{i;\delta} = \sum_{a,b}\frac{\partial z_{i;\delta}}{\partial W_{ab}}dW_{ab} = -\eta\sum_{\tilde{\alpha}}k_{\delta\tilde{\alpha}}\varepsilon_{i;\tilde{\alpha}} \ . \tag{8}$$

In the linear model, the kernel $k_{\delta\tilde{\alpha}}$ is static. The solution for the residual training error is

$$\varepsilon_{i;\tilde{\alpha}_1}(t) = \sum_{\tilde{\alpha}_2}U_{\tilde{\alpha}_1\tilde{\alpha}_2}(t)\varepsilon_{i;\tilde{\alpha}_2}(0) \ , \tag{9}$$

where

$$\begin{aligned}U_{\tilde{\alpha}_t\tilde{\alpha}_0}(t) &\equiv \left[(1-\eta k)^t\right]_{\tilde{\alpha}_t\tilde{\alpha}_0} \\ &= \sum_{\tilde{\alpha}_1,\ldots,\tilde{\alpha}_{t-1}}\left(\delta_{\tilde{\alpha}_t\tilde{\alpha}_{t-1}} - \eta k_{\tilde{\alpha}_t\tilde{\alpha}_{t-1}}\right)\cdots\left(\delta_{\tilde{\alpha}_1\tilde{\alpha}_0} - \eta k_{\tilde{\alpha}_1\tilde{\alpha}_0}\right) \ .\end{aligned} \tag{10}$$

Now we could predict the model output for an arbitrary $\delta$. We have

$$\begin{aligned}
z_{i;\delta}(\infty) &= z_{i;\delta}(0) - \sum_{\tilde{\alpha}} k_{\delta\tilde{\alpha}} \left\{ \eta \sum_{s=0}^{\infty} \varepsilon_{i;\tilde{\alpha}}(s) \right\} \\
&= z_{i;\delta}(0) - \sum_{\tilde{\alpha}} k_{\delta\tilde{\alpha}} \left\{ \eta \sum_{s=0}^{\infty} \left[ \sum_{\tilde{\alpha}_1} U_{\tilde{\alpha}\tilde{\alpha}_1}(s) \varepsilon_{i;\tilde{\alpha}_1}(0) \right] \right\} \\
&= z_{i;\delta}(0) - \sum_{\tilde{\alpha},\tilde{\alpha}_1} k_{\delta\tilde{\alpha}} \left\{ \eta \sum_{s=0}^{\infty} [(1-\eta k)^s]_{\tilde{\alpha}\tilde{\alpha}_1} \right\} \varepsilon_{i;\tilde{\alpha}_1}(0) \\
&= z_{i;\delta}(0) - \sum_{\tilde{\alpha},\tilde{\alpha}_1} k_{\delta\tilde{\alpha}} \left\{ \eta [1-(1-\eta k)]^{-1} \right\}^{\tilde{\alpha}\tilde{\alpha}_1} \varepsilon_{i;\tilde{\alpha}_1}(0) \\
&= z_{i;\delta}(0) - \sum_{\tilde{\alpha},\tilde{\alpha}_1} k_{\delta\tilde{\alpha}} \tilde{k}^{\tilde{\alpha}\tilde{\alpha}_1} \varepsilon_{i;\tilde{\alpha}_1}(0) ,
\end{aligned} \qquad (11)$$

where we have made use of geometric sums. It is easy to check that

$$z_{i;\tilde{\delta}}(\infty) = y_{i;\tilde{\delta}} , \qquad (12)$$

where we take $\tilde{\delta} \in \mathcal{A}$.

### B. Quadratic model

Now we start to study perturbative corrections about the linear model. We consider the model definition,

$$z_{i;\delta}(\theta) = \sum_j W_{ij} \phi_j(\mathbf{x}_\delta) + \frac{\sigma}{2} \sum_{j_1, j_2} W_{ij_1} W_{ij_2} \psi_{j_1 j_2}(\mathbf{x}_\delta) . \qquad (13)$$

Here $\sigma$ is a small number as a perturbative correction. Thus, the model prediction difference up to the quadratic order will be given by

$$d z_{i;\delta} = \sum_j \left[ \phi_j(\mathbf{x}_\delta) + \varepsilon \sum_{j_1=0} W_{ij_1} \psi_{j_1 j}(\mathbf{x}_\delta) \right] dW_{ij} + \frac{\sigma}{2} \sum_{j_1, j_2} \psi_{j_1 j_2}(\mathbf{x}_\delta) dW_{ij_1} dW_{ij_2} . \qquad (14)$$

The first term here is the *effective feature map*,

$$\phi_{ij}^E(\mathbf{x}_\delta) \equiv \frac{dz_{i;\delta}}{dW_{ij}} = \phi_j(\mathbf{x}_\delta) + \sigma \sum_k W_{ik} \psi_{kj}(\mathbf{x}_\delta) . \qquad (15)$$

Now, we could collect our dynamical equations for $z, \phi^E, W$ as

$$\begin{aligned}
dz_{i;\delta} &= \sum_j dW_{ij} \phi_{ij}^E(\mathbf{x}_\delta) + \frac{\sigma}{2} \sum_{j_1, j_2} dW_{ij_1} dW_{ij_2} \psi_{j_1 j_2}(\mathbf{x}_\delta) , \\
d\phi_{ij}^E(\mathbf{x}_\delta) &= \sigma \sum_k dW_{ik} \psi_{kj}(\mathbf{x}_\delta) , \\
dW_{ij} &= -\eta \sum_{\tilde{\alpha}} \phi_{ij}^E(\mathbf{x}_{\tilde{\alpha}}) \varepsilon_{i;\tilde{\alpha}} .
\end{aligned} \qquad (16)$$

Moreover, now we have an MSE loss as

$$\mathcal{L}_\mathcal{A} = \frac{1}{2} \sum_{i,\tilde{\alpha}} \left[ y_{i;\tilde{\alpha}} - \sum_j W_{ij} \phi_j(\mathbf{x}_{\tilde{\alpha}}) - \frac{\sigma}{2} \sum_{j_1, j_2} W_{ij_1} W_{ij_2} \psi_{j_1 j_2}(\mathbf{x}_{\tilde{\alpha}}) \right]^2 . \qquad (17)$$

So we could write the dynamics of $z$ more explicitly as

$$d z_{i;\delta} = -\eta \sum_{\tilde{\alpha}} \left[ \sum_j \phi_{ij}^E(\mathbf{x}_\delta) \phi_{ij}^E(\mathbf{x}_{\tilde{\alpha}}) \right] \varepsilon_{i;\tilde{\alpha}}$$
$$+ \frac{\eta^2}{2} \sum_{\tilde{\alpha}_1, \tilde{\alpha}_2} \left[ \sum_{j_1, j_2} \sigma \psi_{j_1 j_2}(\mathbf{x}_\delta) \phi_{ij_1}^E(\mathbf{x}_{\tilde{\alpha}_1}) \phi_{ij_2}^E(\mathbf{x}_{\tilde{\alpha}_2}) \right] \varepsilon_{i;\tilde{\alpha}_1} \varepsilon_{i;\tilde{\alpha}_2} . \tag{18}$$

We define the *effective kernel*:

$$k_{ii;\delta_1 \delta_2}^E \equiv \sum_j \phi_{ij}^E(\mathbf{x}_{\delta_1}) \phi_{ij}^E(\mathbf{x}_{\delta_2}) , \tag{19}$$

and the *meta kernel*,

$$\mu_{\delta_0 \delta_1 \delta_2} \equiv \sigma \sum_{j_1, j_2} \psi_{j_1 j_2}(\mathbf{x}_{\delta_0}) \phi_{j_1}(\mathbf{x}_{\delta_1}) \phi_{j_2}(\mathbf{x}_{\delta_2})$$
$$= \sigma \sum_{j_1, j_2} \varepsilon \psi_{j_1 j_2}(\mathbf{x}_{\delta_0}) \phi_{ij_1}^E(\mathbf{x}_{\delta_1}) \phi_{ij_2}^E(\mathbf{x}_{\delta_2}) + \mathcal{O}(\sigma^2) . \tag{20}$$

So we have

$$d z_{i;\delta} = -\eta \sum_{\tilde{\alpha}} k_{ii;\delta\tilde{\alpha}}^E \varepsilon_{i;\tilde{\alpha}} + \frac{\eta^2}{2} \sum_{\tilde{\alpha}_1, \tilde{\alpha}_2} \mu_{\delta \tilde{\alpha}_1 \tilde{\alpha}_2} \varepsilon_{i;\tilde{\alpha}_1} \varepsilon_{i;\tilde{\alpha}_2} + \mathcal{O}(\sigma^2) . \tag{21}$$

The meta kernel is fixed. The effective kernel will satisfy the following dynamics

$$d k_{ii;\delta_1 \delta_2}^E = -\eta \sum_{\tilde{\alpha}} (\mu_{\delta_1 \delta_2 \tilde{\alpha}} + \mu_{\delta_2 \delta_1 \tilde{\alpha}}) \varepsilon_{i;\tilde{\alpha}} + O(\sigma^2) . \tag{22}$$

Now we could try solving the output $z$. Based on perturbation theory, we could divide the whole output by the free term $z^F$ and the interacting term $z^I$,

$$z_{i;\delta}(t) \equiv z_{i;\delta}^F(t) + z_{i;\delta}^I(t) . \tag{23}$$

The free part follows the following exponential dynamics,

$$d z_{i;\delta}^F = -\eta \sum_{\tilde{\alpha}} k_{\delta\tilde{\alpha}} \varepsilon_{i;\tilde{\alpha}}^F \equiv z_{i;\delta}^F - \eta \sum_{\tilde{\alpha}} k_{\delta\tilde{\alpha}} \left[ z_{i;\tilde{\alpha}}^F - y_{i;\tilde{\alpha}} \right] . \tag{24}$$

Here, $k$ is the old definition of the kernel without quadratic terms, which is different from the effective kernel by $O(\sigma)$. Moreover, we have

$$k_{ii;\delta_1 \delta_2}^E(t)$$
$$= k_{ii;\delta_1 \delta_2}^E(0) - \sum_{\tilde{\alpha}} (\mu_{\delta_1 \delta_2 \tilde{\alpha}} + \mu_{\delta_2 \delta_1 \tilde{\alpha}}) a_{i;\tilde{\alpha}}(t) , \tag{25}$$

where

$$a_{i;\tilde{\alpha}}(t) = \sum_{\tilde{\alpha}_2} \tilde{k}^{\tilde{\alpha}\tilde{\alpha}_2} \left( \varepsilon_{i;\tilde{\alpha}}(t) - \varepsilon_{i;\tilde{\alpha}}^F(t) \right) . \tag{26}$$

where $\varepsilon^F$ is the free part of the residual training error $z^F - y$. Then, one could compute the interacting piece. We have

$$d z_{i;\delta}^I = -\sum_{j,\tilde{\alpha}} \eta k_{\delta\tilde{\alpha}} z_{j;\tilde{\alpha}}^I(t) + \eta \mathbb{F}_{i;\delta}(t) . \tag{27}$$

Here $\mathbb{F}$ is the damping force

$$\mathbb{F}_{i;\delta}(t) \equiv -\sum_{\tilde{\alpha}} \left[k^E_{\delta\tilde{\alpha}}(t) - k_{\delta\tilde{\alpha}}\right]\varepsilon^F_{i;\tilde{\alpha}}(t) + \frac{\eta}{2}\sum_{\tilde{\alpha}_1,\tilde{\alpha}_2} \mu_{\delta\tilde{\alpha}_1\tilde{\alpha}_2}\varepsilon^F_{i;\tilde{\alpha}_1}(t)\varepsilon^F_{i;\tilde{\alpha}_2}(t) \,, \tag{28}$$

and we have

$$z^I_{i;\delta}(t) = \eta \sum_{s=0}^{t-1} \left[\mathbb{F}_{i;\delta}(s) - \sum_{\tilde{\alpha}_1,\tilde{\alpha}_2} k_{\delta\tilde{\alpha}_1}\tilde{k}^{\tilde{\alpha}_1\tilde{\alpha}_2}\mathbb{F}_{i;\tilde{\alpha}_2}(s)\right] + \sum_{\tilde{\alpha}_1,\tilde{\alpha}_2} k_{\delta\tilde{\alpha}_1}\tilde{k}^{\tilde{\alpha}_1\tilde{\alpha}_2} z^I_{i;\tilde{\alpha}_2}(t) \,. \tag{29}$$

The whole system is a set of non-linear difference equations.

### 1. The residual training error

Now we could try to solve the residual training error. We note that, for $\tilde{\alpha} \in \mathcal{A}$, we have

$$\varepsilon_{i;\tilde{\alpha}}(t) - \varepsilon^F_{i;\tilde{\alpha}}(t) = z^I_{i;\tilde{\alpha}}(t) \,. \tag{30}$$

So we have

$$k^E_{ii;\delta_1\delta_2}(t) = k^E_{ii;\delta_1\delta_2}(0) - \sum_{\tilde{\alpha},\tilde{\alpha}_2} \left(\mu_{\delta_1\delta_2\tilde{\alpha}} + \mu_{\delta_2\delta_1\tilde{\alpha}}\right) \tilde{k}^{\tilde{\alpha}\tilde{\alpha}_2} z^I_{i;\tilde{\alpha}_2}(t) \,. \tag{31}$$

Moreover, we call

$$k^\Delta_{ii;\delta_1\delta_2} \equiv k^\Delta_{ii;\delta_1\delta_2} \equiv k^E_{ii;\delta_1\delta_2}(0) - k_{\delta_1\delta_2} \,. \tag{32}$$

We note that the effective kernel looks like,

$$\begin{aligned} k^E_{ii;\delta_1\delta_2} &\equiv \sum_j \phi^E_{ij}(\mathbf{x}_{\delta_1};0)\,\phi^E_{ij}(\mathbf{x}_{\delta_2};0) \\ &= \sum_j \left(\phi_j(\mathbf{x}_{\delta_1}) + \sigma\sum_k W_{ik}(0)\psi_{kj}(\mathbf{x}_{\delta_1})\right)\left(\phi_j(\mathbf{x}_{\delta_2}) + \sigma\sum_{k'} W_{ik}(0)\psi_{kj}(\mathbf{x}_{\delta_2})\right) \\ &= k_{\delta_1\delta_2} + \sigma\sum_{j,k} W_{ik}(0)\psi_{kj}(\mathbf{x}_{\delta_1})\,\phi_j(\mathbf{x}_{\delta_2}) + \sigma\sum_{j,k} W_{ik}(0)\psi_{kj}(\mathbf{x}_{\delta_2})\,\phi_j(\mathbf{x}_{\delta_1}) + \mathcal{O}(\sigma^2) \,. \end{aligned} \tag{33}$$

Keeping the leading order, we have

$$k^\Delta_{ii;\delta_1\delta_2} = \sigma\sum_{j,k} W_{ik}(0)\psi_{kj}(\mathbf{x}_{\delta_1})\,\phi_j(\mathbf{x}_{\delta_2}) + \sigma\sum_{j,k} W_{ik}(0)\psi_{kj}(\mathbf{x}_{\delta_2})\,\phi_j(\mathbf{x}_{\delta_1}) \,, \tag{34}$$

and

$$\begin{aligned} k^E_{ii;\delta_1\delta_2}(t) &= k^E_{ii;\delta_1\delta_2}(0) - \sum_{\tilde{\alpha},\tilde{\alpha}_2}(\mu_{\delta_1\delta_2\tilde{\alpha}} + \mu_{\delta_2\delta_1\tilde{\alpha}})\tilde{k}^{\tilde{\alpha}\tilde{\alpha}_2} z^I_{i;\tilde{\alpha}_2}(t) \\ &= k_{\delta_1\delta_2} + k^\Delta_{ii,\delta_1\delta_2} - \sum_{\tilde{\alpha},\tilde{\alpha}_2}(\mu_{\delta_1\delta_2\tilde{\alpha}} + \mu_{\delta_2\delta_1\tilde{\alpha}})\tilde{k}^{\tilde{\alpha}\tilde{\alpha}_2} z^I_{i;\tilde{\alpha}_2}(t) \,. \end{aligned} \tag{35}$$

Moreover, let us solve the damping force $\mathbb{F}$:

$$\mathbb{F}_{i;\delta}(t) = -\sum_{\tilde{\alpha}} \left[k^E_{\delta\tilde{\alpha}}(t) - k_{\delta\tilde{\alpha}}\right]\varepsilon^F_{i;\tilde{\alpha}}(t) + \frac{\eta}{2}\sum_{\tilde{\alpha}_1,\tilde{\alpha}_2}\mu_{\delta\tilde{\alpha}_1\tilde{\alpha}_2}\varepsilon^F_{i;\tilde{\alpha}_1}(t)\varepsilon^F_{i;\tilde{\alpha}_2}(t)$$

$$= -\sum_{\tilde{\alpha},\tilde{\alpha}_3}\left[k^E_{ii;\delta\tilde{\alpha}}(0) - \sum_{\tilde{\alpha}_1,\tilde{\alpha}_2}(\mu_{\delta\tilde{\alpha}\tilde{\alpha}_1} + \mu_{\tilde{\alpha}\delta\tilde{\alpha}_1})\tilde{k}^{\tilde{\alpha}_1\tilde{\alpha}_2}\varepsilon^I_{i;\tilde{\alpha}_2}(t) - k_{\delta\tilde{\alpha}}\right]U_{\tilde{\alpha}\tilde{\alpha}_3}(t)\varepsilon_{i;\tilde{\alpha}_3}(0)$$

$$+ \frac{\eta}{2}\sum_{\tilde{\alpha}_1,\tilde{\alpha}_2,\tilde{\alpha}_3,\tilde{\alpha}_4}\mu_{\delta\tilde{\alpha}_1\tilde{\alpha}_2}U_{\tilde{\alpha}_1\tilde{\alpha}_3}(t)U_{\tilde{\alpha}_2\tilde{\alpha}_4}(t)\varepsilon_{i;\tilde{\alpha}_3}(0)\varepsilon_{i;\tilde{\alpha}_4}(0)$$

$$= -\sum_{\tilde{\alpha},\tilde{\alpha}_3}k^\Delta_{ii;\delta\tilde{\alpha}}U_{\tilde{\alpha}\tilde{\alpha}_3}(t)\varepsilon_{i;\tilde{\alpha}_3}(0) + \sum_{\tilde{\alpha},\tilde{\alpha}_1,\tilde{\alpha}_2,\tilde{\alpha}_3}(\mu_{\delta\tilde{\alpha}\tilde{\alpha}_1} + \mu_{\tilde{\alpha}\delta\tilde{\alpha}_1})\tilde{k}^{\tilde{\alpha}_1\tilde{\alpha}_2}U_{\tilde{\alpha}\tilde{\alpha}_3}(t)\varepsilon_{i;\tilde{\alpha}_3}(0)z^I_{i;\tilde{\alpha}_2}(t)$$

$$+ \frac{\eta}{2}\sum_{\tilde{\alpha}_1,\tilde{\alpha}_2,\tilde{\alpha}_3,\tilde{\alpha}_4}\mu_{\delta\tilde{\alpha}_1\tilde{\alpha}_2}U_{\tilde{\alpha}_1\tilde{\alpha}_3}(t)U_{\tilde{\alpha}_2\tilde{\alpha}_4}(t)\varepsilon_{i;\tilde{\alpha}_3}(0)\varepsilon_{i;\tilde{\alpha}_4}(0) \ . \tag{36}$$

The second and the last term in the last equality contributes higher orders. Thus, if the initial $k^\Delta$ is not vanishing, we have

$$\mathbb{F}_{i;\delta}(t) = -\sum_{\tilde{\alpha},\tilde{\alpha}_3}k^\Delta_{ii;\delta\tilde{\alpha}}U_{\tilde{\alpha}\tilde{\alpha}_3}(t)\varepsilon_{i;\tilde{\alpha}_3}(0) + \mathcal{O}(\eta\sigma) \ . \tag{37}$$

The contribution in the first term will dominate at least in the early time when $k^\Delta$ is not vanishing. In the late time where $U$ decays significantly, we would have some non-perturbative effects.

Thus, we can plug the expression back to solve $z^I(t)$. We have

$$z^I_{i;\tilde{\alpha}}(t) = \eta\sum_{s=0}^{t-1}\sum_{\tilde{\alpha}_1}U_{\tilde{\alpha}\tilde{\alpha}_1}(t-1-s)\mathbb{F}_{i;\tilde{\alpha}_1}(s)$$

$$= -\eta\sum_{s=0}^{t-1}\sum_{\tilde{\alpha}_1,\tilde{\alpha}_2,\tilde{\alpha}_3}U_{\tilde{\alpha}\tilde{\alpha}_1}(t-1-s)U_{\tilde{\alpha}_2\tilde{\alpha}_3}(s)k^\Delta_{ii;\tilde{\alpha}_1\tilde{\alpha}_2}\varepsilon_{i;\tilde{\alpha}_3}(0)$$

$$= -\eta\sum_{s=0}^{t-1}(1-\eta k)^{t-1-s}k^\Delta_{ii}(\delta-\eta k)^s\varepsilon_i(0) \ , \tag{38}$$

The last formula is given in the following matrix form:

$$(1-\eta k)_{\tilde{\alpha}_1\tilde{\alpha}_2} = \delta_{\tilde{\alpha}_1\tilde{\alpha}_2} - \eta k_{\tilde{\alpha}_1\tilde{\alpha}_2} = \delta_{\tilde{\alpha}_1\tilde{\alpha}_2} - \eta\sum_j\phi_j(\mathbf{x}_{\tilde{\alpha}_1})\phi_j(\mathbf{x}_{\tilde{\alpha}_2}) \ ,$$

$$(k^\Delta_{ii})_{\tilde{\alpha}_1\tilde{\alpha}_2} = \sigma\sum_{j,k}W_{ik}(0)\psi_{kj}(\mathbf{x}_{\tilde{\alpha}_1})\phi_j(\mathbf{x}_{\tilde{\alpha}_2}) + \sigma\sum_{j,k}W_{ik}(0)\psi_{kj}(\mathbf{x}_{\tilde{\alpha}_2})\phi_j(\mathbf{x}_{\tilde{\alpha}_1}) \equiv \sigma(\mathbf{M}^i)_{\tilde{\alpha}_1,\tilde{\alpha}_2} \ ,$$

$$(\varepsilon_i(0))_{\tilde{\alpha}} = \varepsilon_{i;\tilde{\alpha}}(0) \ . \tag{39}$$

Moreover, we could indeed get more information by just making the bound. We have

$$\left\|z^I_{i;\tilde{\alpha}}(t)\right\| = \left\|\eta\sum_{s=0}^{t-1}(1-\eta k)^{t-1-s}k^\Delta_{ii}(1-\eta k)^s\varepsilon_i(0)\right\|$$

$$\leq \eta\sum_{s=0}^{t-1}\|1-\eta k\|^{t-1-s}\left\|k^\Delta_{ii}\right\|\|1-\eta k\|^s\|\varepsilon_i(0)\|$$

$$= \eta\left(\sum_{s=0}^{t-1}\|1-\eta k\|^{t-1}\right)\left\|k^\Delta_{ii}\right\|\|\varepsilon_i(0)\|$$

$$= \eta t\|1-\eta k\|^{t-1}\left\|k^\Delta_{ii}\right\|\|\varepsilon_i(0)\| = \sigma\eta t\|1-\eta k\|^{t-1}\left\|\mathbf{M}^i\right\|\|\varepsilon_i(0)\| \ . \tag{40}$$

Now we compare the convergence time noticing that

$$\varepsilon^F_{i;\tilde{\alpha}_1}(t) = \sum_{\tilde{\alpha}_2}U_{\tilde{\alpha}_1\tilde{\alpha}_2}(t)\varepsilon_{i;\tilde{\alpha}_2}(0) \ . \tag{41}$$

So
$$\left\|\varepsilon_i^F(t)\right\| \leq \|1 - \eta k\|^t \|\varepsilon_i(0)\| . \tag{42}$$

Schematically, we have
$$\frac{\|z_i^I(t)\|}{\|\varepsilon_i^F(t)\|} \sim \frac{\sigma \eta t \|\mathbf{M}^i\| \|\varepsilon_i(0)\|}{\|1 - \eta k\| \|\varepsilon_i(0)\|} \sim \frac{\sigma \eta t \|\mathbf{M}^i\|}{\|1 - \eta k\|} . \tag{43}$$

The relative perturbative error contribution will grow linearly in time.

### 2. The asymptotic regime

Now instead of only looking at the training set, we study the asymptotic regime for general inputs. We start from
$$z_{i;\delta}^I(t) = \eta \sum_{s=0}^{t-1} \left[ \mathbb{F}_{i;\delta}(s) - \sum_{\tilde{\alpha}_1,\tilde{\alpha}_2} k_{\delta\tilde{\alpha}_1} \tilde{k}^{\tilde{\alpha}_1\tilde{\alpha}_2} \mathbb{F}_{i;\tilde{\alpha}_2}(s) \right] + \sum_{\tilde{\alpha}_1,\tilde{\alpha}_2} k_{\delta\tilde{\alpha}_1} \tilde{k}^{\tilde{\alpha}_1\tilde{\alpha}_2} z_{i;\tilde{\alpha}_2}^I(t) . \tag{44}$$

At the asymptotic convergence, the interacting perturbative correction on the training set will converge to zero, so we have
$$z_{i;\delta}^I(\infty) = \left[ \eta \sum_{s=0}^{\infty} \mathbb{F}_{i;\delta}(s) \right] - \sum_{\tilde{\alpha}_1,\tilde{\alpha}_2} k_{\delta\tilde{\alpha}_1} \tilde{k}^{\tilde{\alpha}_1\tilde{\alpha}_2} \left[ \eta \sum_{s=0}^{\infty} \mathbb{F}_{i;\tilde{\alpha}_2}(s) \right] . \tag{45}$$

So we need to perform the sum
$$\eta \sum_{s=0}^{\infty} \mathbb{F}_{i;\delta}(s) = \sum_{\tilde{\alpha}_0,\tilde{\alpha}} (\mu_{\delta\tilde{\alpha}\tilde{\alpha}_0} + \mu_{\tilde{\alpha}\delta\tilde{\alpha}_0}) \left\{ \eta \sum_{s=0}^{\infty} a_{i;\tilde{\alpha}_0}(s) \varepsilon_{i;\tilde{\alpha}}^F(s) \right\}$$
$$+ \frac{\eta}{2} \sum_{\tilde{\alpha}_1,\tilde{\alpha}_2} \mu_{\delta\tilde{\alpha}_1\tilde{\alpha}_2} \left\{ \eta \sum_{s=0}^{\infty} \varepsilon_{i;\tilde{\alpha}_1}^F(s) \varepsilon_{i;\tilde{\alpha}_2}^F(s) \right\} . \tag{46}$$

We note that
$$\eta \sum_{s=0}^{\infty} \varepsilon_{i;\tilde{\alpha}_1}^F(t) \varepsilon_{i;\tilde{\alpha}_2}^F(t) = \eta \sum_{\tilde{\alpha}_3,\tilde{\alpha}_4} \sum_{s=0}^{\infty} [(1-\eta k)^s]_{\tilde{\alpha}_1\tilde{\alpha}_3} [(1-\eta k)^s]_{\tilde{\alpha}_2\tilde{\alpha}_4} \varepsilon_{i;\tilde{\alpha}_3}(0) \varepsilon_{i;\tilde{\alpha}_4}(0)$$
$$= \sum_{\tilde{\alpha}_3,\tilde{\alpha}_4} X_{\|}^{\tilde{\alpha}_1\tilde{\alpha}_2\tilde{\alpha}_3\tilde{\alpha}_4} \varepsilon_{i;\tilde{\alpha}_3}(0) \varepsilon_{i;\tilde{\alpha}_4}(0) . \tag{47}$$

Here, we define the following inverting tensor:
$$X_{\|}^{\tilde{\alpha}_1\tilde{\alpha}_2\tilde{\alpha}_3\tilde{\alpha}_4} = \sum_{s=0}^{\infty} [(1-\eta k)^s]_{\tilde{\alpha}_1\tilde{\alpha}_3} [(1-\eta k)^s]_{\tilde{\alpha}_2\tilde{\alpha}_4} , \tag{48}$$

which is implicitly defined as
$$\delta_{\tilde{\alpha}_5}^{\tilde{\alpha}_1} \delta_{\tilde{\alpha}_6}^{\tilde{\alpha}_2} = \sum_{\tilde{\alpha}_3,\tilde{\alpha}_4} X_{\|}^{\tilde{\alpha}_1\tilde{\alpha}_2\tilde{\alpha}_3\tilde{\alpha}_4} \frac{1}{\eta} \left[ \delta_{\tilde{\alpha}_3\tilde{\alpha}_5} \delta_{\tilde{\alpha}_4\tilde{\alpha}_6} - \left( \delta_{\tilde{\alpha}_3\tilde{\alpha}_5} - \eta \tilde{k}_{\tilde{\alpha}_3\tilde{\alpha}_5} \right) \left( \delta_{\tilde{\alpha}_4\tilde{\alpha}_6} - \eta \tilde{k}_{\tilde{\alpha}_4\tilde{\alpha}_6} \right) \right]$$
$$= \sum_{\tilde{\alpha}_3,\tilde{\alpha}_4} X_{\|}^{\tilde{\alpha}_1\tilde{\alpha}_2\tilde{\alpha}_3\tilde{\alpha}_4} \left( \tilde{k}_{\tilde{\alpha}_3\tilde{\alpha}_5} \delta_{\tilde{\alpha}_4\tilde{\alpha}_6} + \delta_{\tilde{\alpha}_3\tilde{\alpha}_5} \tilde{k}_{\tilde{\alpha}_4\tilde{\alpha}_6} - \eta \tilde{k}_{\tilde{\alpha}_3\tilde{\alpha}_5} \tilde{k}_{\tilde{\alpha}_4\tilde{\alpha}_6} \right) . \tag{49}$$

Using the inverting tensor, the final expression is given by
$$z_{i;\delta}(\infty) = z_{i;\delta}(0) - \sum_{\tilde{\alpha}_1,\tilde{\alpha}_2} k_{\delta\tilde{\alpha}_1} \tilde{k}^{\tilde{\alpha}_1\tilde{\alpha}_2} \varepsilon_{i;\tilde{\alpha}_2}(0)$$
$$+ \sum_{\tilde{\alpha}_1,\ldots,\tilde{\alpha}_4} \left[ \mu_{\tilde{\alpha}_1\delta\tilde{\alpha}_2} - \sum_{\tilde{\alpha}_5,\tilde{\alpha}_6 \in \mathcal{A}} k_{\delta\tilde{\alpha}_5} \tilde{k}^{\tilde{\alpha}_5\tilde{\alpha}_6} \mu_{\tilde{\alpha}_1\tilde{\alpha}_6\tilde{\alpha}_2} \right] Z_A^{\tilde{\alpha}_1\tilde{\alpha}_2\tilde{\alpha}_3\tilde{\alpha}_4} \varepsilon_{i;\tilde{\alpha}_3}(0) \varepsilon_{i;\tilde{\alpha}_3}(0)$$
$$+ \sum_{\tilde{\alpha}_1,\ldots,\tilde{\alpha}_4} \left[ \mu_{\delta\tilde{\alpha}_1\tilde{\alpha}_2} - \sum_{\tilde{\alpha}_5,\tilde{\alpha}_6 \in \mathcal{A}} k_{\delta\tilde{\alpha}_5} \tilde{k}^{\tilde{\alpha}_5\tilde{\alpha}_6} \mu_{\tilde{\alpha}_6\tilde{\alpha}_1\tilde{\alpha}_2} \right] Z_B^{\tilde{\alpha}_1\tilde{\alpha}_2\tilde{\alpha}_3\tilde{\alpha}_4} \varepsilon_{i;\tilde{\alpha}_3}(0) \varepsilon_{i;\tilde{\alpha}_3}(0) , \tag{50}$$

where $Z$s are the *algorithm projectors*,

$$Z_A^{\tilde{\alpha}_1\tilde{\alpha}_2\tilde{\alpha}_3\tilde{\alpha}_4} \equiv \tilde{k}^{\tilde{\alpha}_1\tilde{\alpha}_3}\tilde{k}^{\tilde{\alpha}_2\tilde{\alpha}_4} - \sum_{\tilde{\alpha}_5} \tilde{k}^{\tilde{\alpha}_2\tilde{\alpha}_5} X_\parallel^{\tilde{\alpha}_1\tilde{\alpha}_5\tilde{\alpha}_3\tilde{\alpha}_4} \,,$$

$$Z_B^{\tilde{\alpha}_1\tilde{\alpha}_2\tilde{\alpha}_3\tilde{\alpha}_4} \equiv \tilde{k}^{\tilde{\alpha}_1\tilde{\alpha}_3}\tilde{k}^{\tilde{\alpha}_2\tilde{\alpha}_4} - \sum_{\tilde{\alpha}_5} \tilde{k}^{\tilde{\alpha}_2\tilde{\alpha}_5} X_\parallel^{\tilde{\alpha}_1\tilde{\alpha}_5\tilde{\alpha}_3\tilde{\alpha}_4} + \frac{\eta}{2} X_\parallel^{\tilde{\alpha}_1\tilde{\alpha}_2\tilde{\alpha}_3\tilde{\alpha}_4} \,. \tag{51}$$

Finally, in the continuum limit, we could drop out the $\eta$ terms in the algorithm projectors,

$$Z_A^{\tilde{\alpha}_1\tilde{\alpha}_2\tilde{\alpha}_3\tilde{\alpha}_4} = Z_B^{\tilde{\alpha}_1\tilde{\alpha}_2\tilde{\alpha}_3\tilde{\alpha}_4} \equiv \tilde{k}^{\tilde{\alpha}_1\tilde{\alpha}_3}\tilde{k}^{\tilde{\alpha}_2\tilde{\alpha}_4} - \sum_{\tilde{\alpha}_5} \tilde{k}^{\tilde{\alpha}_2\tilde{\alpha}_5} X_\parallel^{\tilde{\alpha}_1\tilde{\alpha}_5\tilde{\alpha}_3\tilde{\alpha}_4} \,,$$

$$\sum_{\tilde{\alpha}_3,\tilde{\alpha}_4} X_\parallel^{\tilde{\alpha}_1\tilde{\alpha}_2\tilde{\alpha}_3\tilde{\alpha}_4} \left( \tilde{k}_{\tilde{\alpha}_3\tilde{\alpha}_5}\delta_{\tilde{\alpha}_4\tilde{\alpha}_6} + \delta_{\tilde{\alpha}_3\tilde{\alpha}_5}\tilde{k}_{\tilde{\alpha}_4\tilde{\alpha}_6} \right) = \delta_{\tilde{\alpha}_5}^{\tilde{\alpha}_1}\delta_{\tilde{\alpha}_6}^{\tilde{\alpha}_2} \,. \tag{52}$$

The existence of algorithm projectors shows algorithm dependence in those perturbative corrections. This will typically happen when the model has multiple (local) minimal.

## II. DETAILS ON THE QUANTUM OPTIMIZATION

### A. General setup

The quantum optimization problem we discuss here has a simpler formulation compared to quantum machine learning. Since the loss function does not contain the training data, we do not need to consider the difference between the training set and the whole space. It could be understood as a limiting case of the quantum machine learning problem.

We use the loss function,

$$\mathcal{L}(\theta) = \frac{1}{2}\left(\langle \Psi_0 | U^\dagger O U | \Psi_0 \rangle - O_0\right)^2 \equiv \frac{1}{2}\varepsilon^2 \,. \tag{53}$$

The trainable ansatz is,

$$U = \prod_{\ell=1}^L W_\ell U_\ell = \prod_{\ell=1}^L W_\ell \exp(i\theta_\ell X_\ell) \,. \tag{54}$$

Here, note that since the operator $O$ is Hermitian, the loss function is always real and non-negative. The gradient descent is

$$d\theta_\ell = -\eta \frac{d\mathcal{L}(\theta)}{d\theta_\ell} = -\eta \left(\langle \Psi_0 | U^\dagger O U | \Psi_0 \rangle - O_0\right) \frac{d\langle \Psi_0 | U^\dagger O U | \Psi_0 \rangle}{d\theta_\ell} = -\eta\varepsilon \frac{d\varepsilon}{d\theta_\ell} \,, \tag{55}$$

where

$$\varepsilon = \langle \Psi_0 | U^\dagger O U | \Psi_0 \rangle - O_0 \,. \tag{56}$$

Thus

$$d\varepsilon = \sum_\ell \frac{d\varepsilon}{d\theta_\ell} d\theta_\ell = -\eta \sum_\ell \frac{d\varepsilon}{d\theta_\ell}\frac{d\varepsilon}{d\theta_\ell} \varepsilon \,. \tag{57}$$

The object

$$K = \sum_\ell \frac{d\varepsilon}{d\theta_\ell}\frac{d\varepsilon}{d\theta_\ell} \,, \tag{58}$$

is the optimization analog of the QNTK. More precisely, we have

$$\frac{d\varepsilon}{d\theta_\ell} = \frac{d\langle \Psi_0 | U^\dagger O U | \Psi_0 \rangle}{d\theta_\ell} = -i \left\langle \Psi_0 \left| U_{+,\ell}^\dagger \left[ X_\ell, U_\ell^\dagger W_\ell^\dagger U_{-,\ell}^\dagger O U_{-,\ell} W_\ell U_\ell \right] U_{+,\ell} \right| \Psi_0 \right\rangle \,, \tag{59}$$

with the help of the definition,

$$U_{-,\ell} \equiv \prod_{\ell'=1}^{\ell-1} W_{\ell'} U_{\ell'} \,,$$

$$U_{+,\ell} \equiv \prod_{\ell'=\ell+1}^{L} W_{\ell'} U_{\ell'} \,. \tag{60}$$

So,

$$d\varepsilon = -\eta \sum_\ell \frac{d\varepsilon}{d\theta_\ell} \frac{d\varepsilon}{d\theta_\ell} \varepsilon = \eta\varepsilon \left\langle \Psi_0 \left| U_{+,\ell}^\dagger \left[ X_\ell, U_\ell^\dagger W_\ell^\dagger U_{-,\ell}^\dagger O U_{-,\ell} W_\ell U_\ell \right] U_{+,\ell} \right| \Psi_0 \right\rangle^2 \,. \tag{61}$$

Can we solve the difference equation? We have

$$d\theta_\ell = i\eta \left\langle \Psi_0 \left| U_{+,\ell}^\dagger \left[ X_\ell, U_\ell^\dagger W_\ell^\dagger U_{-,\ell}^\dagger O U_{-,\ell} W_\ell U_\ell \right] U_{+,\ell} \right| \Psi_0 \right\rangle \,. \tag{62}$$

In the continuum limit, we have

$$\frac{d\theta_\ell}{dt} = i\eta_c \left\langle \Psi_0 \left| U_{+,\ell}^\dagger \left[ X_\ell, U_\ell^\dagger W_\ell^\dagger U_{-,\ell}^\dagger O U_{-,\ell} W_\ell U_\ell \right] U_{+,\ell} \right| \Psi_0 \right\rangle \,, \tag{63}$$

where $\eta_c = \eta/dt$ is the continuous version of the learning rate. In principle, this is a coupled nonlinear ODE system, and one could use the ODE theory to study them. Moreover, we have

$$U_{+,\ell}^\dagger \left[ X_\ell, U_\ell^\dagger W_\ell^\dagger U_{-,\ell}^\dagger O U_{-,\ell} W_\ell U_\ell \right] U_{+,\ell} = U_{+,\ell}^\dagger U_\ell^\dagger \left[ X_\ell, W_\ell^\dagger U_{-,\ell}^\dagger O U_{-,\ell} W_\ell \right] U_\ell U_{+,\ell} \,. \tag{64}$$

Defining

$$A_\ell = \left[ X_\ell, W_\ell^\dagger U_{-,\ell}^\dagger O U_{-,\ell} W_\ell \right] \,, \tag{65}$$

and if we are expanding $\theta_\ell$ when it is slightly deviated from 0 (more generally, around a fixed angle, which is equivalent to a redefinition of constant gates), we have,

$$U_{+,\ell}^\dagger U_\ell^\dagger A_\ell U_\ell U_{+,\ell} = U_{+,\ell}^\dagger A_\ell U_{+,\ell} - i\theta_\ell U_{+,\ell}^\dagger [X_\ell, A_\ell] U_{+,\ell} + \mathcal{O}(\theta_\ell^2) \,. \tag{66}$$

Thus, we note that the second-order expansions of $\theta_\ell$ here means second-order commutators among $X_\ell$ and some unitary-addressed versions of the operator $O$, which is the operator we wish to optimize. Our experiences could be easily generalized to high orders.

### B. Frozen QNTK

Now, we rescale the original variational angles by a factor $\delta$ where the variational angles are around $\theta^* + \delta\varphi$. The constant term $\theta^*$ will produce a constant gate $\exp(i\theta_\ell^* X_\ell)$, such that $W_\ell U_\ell \to W_\ell \exp(i\theta_\ell^* X_\ell) \exp(i\delta\varphi_\ell X_\ell)$. So we could absorb the $\theta^*$ dependence to the definition of the constant gate by defining $W_\ell(\theta_\ell^*) \equiv W_\ell \exp(i\theta_\ell^* X_\ell)$, and we have $W_\ell U_\ell \to W_\ell(\theta_\ell^*) \exp(i\delta\varphi_\ell X_\ell)$. Thus, for simplicity, we could understand the trainable gate as $U_\ell \to \exp(i\delta\varphi_\ell X_\ell)$ with a redefined constant gate. In our calculation later, we will drop the $\theta^*$ notation and understand the variational angles as small parameters rescaled by $\delta$ for our notation convenience.

In the frozen QNTK limit, we have,

$$K = -\delta^2 \sum_\ell \left\langle \Psi_0 \left| W_{+,\ell}^\dagger \left[ X_\ell, W_\ell^\dagger W_{-,\ell}^\dagger O W_{-,\ell} W_\ell \right] W_{+,\ell} \right| \Psi_0 \right\rangle^2 \,. \tag{67}$$

And we define

$$W_{-,\ell} \equiv \prod_{\ell'=1}^{\ell-1} W_{\ell'} \,, \quad W_{+,\ell} \equiv \prod_{\ell'=\ell+1}^{L} W_{\ell'} \,. \tag{68}$$

Note that the frozen QNTK does not depend on the variational parameters. The gradient descent dynamics, in this case, is very easy to solve. We have

$$\varepsilon(t) = (1 - \eta K)^t \varepsilon(0) \equiv \left(1 + \eta \delta^2 \sum_\ell \left\langle \Psi_0 \left| W_{+,\ell}^\dagger \left[ X_\ell, W_\ell^\dagger W_{-,\ell}^\dagger O W_{-,\ell} W_\ell \right] W_{+,\ell} \right| \Psi_0 \right\rangle^2 \right)^t \varepsilon(0) . \qquad (69)$$

The convergence rate is given by,

$$\begin{aligned}
\tau_c &= -\log(1 - \eta K) \approx \eta K \\
&= \eta \delta^2 \sum_\ell \left\langle \Psi_0 \left| W_{+,\ell}^\dagger \left[ X_\ell, W_\ell^\dagger W_{-,\ell}^\dagger O W_{-,\ell} W_\ell \right] W_{+,\ell} \right| \Psi_0 \right\rangle^2 \\
&\leq 2\eta \delta^2 L \|O\|^2 \max_\ell \|X_\ell\|^2 .
\end{aligned} \qquad (70)$$

## C. dQNTK

Now let us focus on the second order to develop an analog of representation learning theory in the quantum optimization example. We have

$$\begin{aligned}
\varepsilon &= \langle \Psi_0 | U^\dagger O U | \Psi_0 \rangle - O_0 \\
&\approx \left\langle \Psi_0 \left| \left( \prod_{\ell'=L}^{1} W_{\ell'}^\dagger \right) O \left( \prod_{\ell=1}^{L} W_\ell \right) \right| \Psi_0 \right\rangle - O_0 \\
&- i\delta \sum_\ell \varphi_\ell \left\langle \Psi_0 \left| W_{+,\ell}^\dagger \left[ X_\ell, W_\ell^\dagger W_{-,\ell}^\dagger O W_{-,\ell} W_\ell \right] W_{+,\ell} \right| \Psi_0 \right\rangle \\
&- \frac{\delta^2}{2} \sum_{\ell_1, \ell_2} \varphi_{\ell_1} \varphi_{\ell_2} \times \begin{cases} \left\langle \Psi_0 \left| W_{+,\ell_1}^\dagger \left[ X_{\ell_1}, W_{\ell_1}^\dagger W_{\ell_1,\ell_2}^\dagger W_{+,\ell_2}^\dagger \left[ X_{\ell_2}, W_{\ell_2}^\dagger W_{-,\ell_2}^\dagger O W_{-,\ell_2} W_{\ell_2} \right] W_{\ell_2,\ell_1} W_{\ell_1} \right] W_{+,\ell_1} \right| \Psi_0 \right\rangle : \ell_1 \geq \ell_2 \\ \left\langle \Psi_0 \left| W_{+,\ell_2}^\dagger \left[ X_{\ell_2}, W_{\ell_2}^\dagger W_{\ell_2,\ell_1}^\dagger W_{+,\ell_1}^\dagger \left[ X_{\ell_1}, W_{\ell_1}^\dagger W_{-,\ell_1}^\dagger O W_{-,\ell_1} W_{\ell_1} \right] W_{\ell_1,\ell_2} W_{\ell_2} \right] W_{+,\ell_2} \right| \Psi_0 \right\rangle : \ell_1 < \ell_2 \end{cases} ,
\end{aligned} \qquad (71)$$

where

$$W_{\ell_1, \ell_2} = \prod_{\ell = \ell_1 + 1}^{\ell_2 - 1} W_\ell . \qquad (72)$$

So we could take the derivative as,

$$\begin{aligned}
\frac{d\varepsilon}{d\varphi_\ell} &= -i\delta \left\langle \Psi_0 \left| W_{+,\ell}^\dagger \left[ X_\ell, W_\ell^\dagger W_{-,\ell}^\dagger O W_{-,\ell} W_\ell \right] W_{+,\ell} \right| \Psi_0 \right\rangle - \delta^2 \times \sum_{\ell'} \varphi_{\ell'} \times \\
&\begin{cases} \left\langle \Psi_0 \left| W_{+,\ell'}^\dagger \left[ X_{\ell'}, W_{\ell'}^\dagger W_{\ell',\ell}^\dagger W_{+,\ell}^\dagger \left[ X_\ell, W_\ell^\dagger W_{-,\ell}^\dagger O W_{-,\ell} W_\ell \right] W_{\ell,\ell'} W_{\ell'} \right] W_{+,\ell'} \right| \Psi_0 \right\rangle : \ell' \geq \ell \\ \left\langle \Psi_0 \left| W_{+,\ell}^\dagger \left[ X_\ell, W_\ell^\dagger W_{\ell,\ell'}^\dagger W_{+,\ell'}^\dagger \left[ X_{\ell'}, W_{\ell'}^\dagger W_{-,\ell'}^\dagger O W_{-,\ell'} W_{\ell'} \right] W_{\ell',\ell} W_\ell \right] W_{+,\ell} \right| \Psi_0 \right\rangle : \ell' < \ell \end{cases} .
\end{aligned} \qquad (73)$$

Thus

$$\begin{aligned}
d\varepsilon = \sum_\ell &\left( \begin{array}{l} -i\delta \left\langle \Psi_0 \left| W_{+,\ell}^\dagger \left[ X_\ell, W_\ell^\dagger W_{-,\ell}^\dagger O W_{-,\ell} W_\ell \right] W_{+,\ell} \right| \Psi_0 \right\rangle - \delta^2 \sum_{\ell'} \varphi_{\ell'} \times \\ \begin{cases} \left\langle \Psi_0 \left| W_{+,\ell'}^\dagger \left[ X_{\ell'}, W_{\ell'}^\dagger W_{\ell',\ell}^\dagger W_{+,\ell}^\dagger \left[ X_\ell, W_\ell^\dagger W_{-,\ell}^\dagger O W_{-,\ell} W_\ell \right] W_{\ell,\ell'} W_{\ell'} \right] W_{+,\ell'} \right| \Psi_0 \right\rangle : \ell' \geq \ell \\ \left\langle \Psi_0 \left| W_{+,\ell}^\dagger \left[ X_\ell, W_\ell^\dagger W_{\ell,\ell'}^\dagger W_{+,\ell'}^\dagger \left[ X_{\ell'}, W_{\ell'}^\dagger W_{-,\ell'}^\dagger O W_{-,\ell'} W_{\ell'} \right] W_{\ell',\ell} W_\ell \right] W_{+,\ell} \right| \Psi_0 \right\rangle : \ell' < \ell \end{cases} \end{array} \right) d\varphi_\ell \\
&- \frac{\delta^2}{2} \sum_{\ell_1, \ell_2} d\varphi_{\ell_1} d\varphi_{\ell_2} \times \\
&\left( \begin{cases} \left\langle \Psi_0 \left| W_{+,\ell_1}^\dagger \left[ X_{\ell_1}, W_{\ell_1}^\dagger W_{\ell_1,\ell_2}^\dagger W_{+,\ell_2}^\dagger \left[ X_{\ell_2}, W_{\ell_2}^\dagger W_{-,\ell_2}^\dagger O W_{-,\ell_2} W_{\ell_2} \right] W_{\ell_2,\ell_1} W_{\ell_1} \right] W_{+,\ell_1} \right| \Psi_0 \right\rangle : \ell_1 \geq \ell_2 \\ \left\langle \Psi_0 \left| W_{+,\ell_2}^\dagger \left[ X_{\ell_2}, W_{\ell_2}^\dagger W_{\ell_2,\ell_1}^\dagger W_{+,\ell_1}^\dagger \left[ X_{\ell_1}, W_{\ell_1}^\dagger W_{-,\ell_1}^\dagger O W_{-,\ell_1} W_{\ell_1} \right] W_{\ell_1,\ell_2} W_{\ell_2} \right] W_{+,\ell_2} \right| \Psi_0 \right\rangle : \ell_1 < \ell_2 \end{cases} \right) .
\end{aligned} \qquad (74)$$

Moreover, we have

$$d\varphi_\ell = -\eta\varepsilon\frac{d\varepsilon}{d\varphi_\ell} . \tag{75}$$

Thus

$$d\varepsilon = -\eta\sum_\ell \frac{d\varepsilon}{d\varphi_\ell}\frac{d\varepsilon}{d\varphi_\ell}\varepsilon + \frac{1}{2}\eta^2\varepsilon^2 \sum_{\ell_1,\ell_2} \frac{d^2\varepsilon}{d\varphi_{\ell_1}d\varphi_{\ell_2}}\frac{d\varepsilon}{d\varphi_{\ell_1}}\frac{d\varepsilon}{d\varphi_{\ell_2}} . \tag{76}$$

The structure of the gradient descent equation for the residual optimization error is very similar to the case in the quadratic model and representation learning context. We define the free part and the interacting part of $\varepsilon$ as

$$\varepsilon = \varepsilon^F + \varepsilon^I . \tag{77}$$

The free part is given by,

$$\varepsilon^F = (1-\eta K)^t\varepsilon(0) ,$$
$$K = -\delta^2 \sum_\ell \left\langle \Psi_0 \left| W_{+,\ell}^\dagger \left[ X_\ell, W_\ell^\dagger W_{-,\ell}^\dagger O W_{-,\ell} W_\ell \right] W_{+,\ell} \right| \Psi_0 \right\rangle^2 , \tag{78}$$

and the interacting part is given by

$$\varepsilon^I(t) = -\eta t(1-\eta K)^{t-1} K^\Delta \varepsilon(0) , \tag{79}$$

where we define the effective kernel up to the dQNTK order,

$$K^E = \sum_\ell \frac{d\varepsilon}{d\varphi_\ell}\frac{d\varepsilon}{d\varphi_\ell} . \tag{80}$$

And we have

$$K^\Delta \equiv K^E(0) - K = \left(\sum_\ell \frac{d\varepsilon}{d\varphi_\ell}\frac{d\varepsilon}{d\varphi_\ell}\right)(0) - \sum_\ell \frac{d\varepsilon^F}{d\varphi_\ell}\frac{d\varepsilon^F}{d\varphi_\ell}$$
$$= 2i\delta^3 \sum_\ell \left\langle \Psi_0 \left| W_{+,\ell}^\dagger \left[ X_\ell, W_\ell^\dagger W_{-,\ell}^\dagger O W_{-,\ell} W_\ell \right] W_{+,\ell} \right| \Psi_0 \right\rangle$$
$$\sum_{\ell'} \begin{cases} \left\langle \Psi_0 \left| W_{+,\ell'}^\dagger \left[ X_{\ell'}, W_{\ell'}^\dagger W_{\ell',\ell}^\dagger W_{+,\ell}^\dagger \left[ X_\ell, W_\ell^\dagger W_{-,\ell}^\dagger O W_{-,\ell} W_\ell \right] W_{\ell,\ell'} W_{\ell'} \right] W_{+,\ell'} \right| \Psi_0 \right\rangle : \ell' \geq \ell \\ \left\langle \Psi_0 \left| W_{+,\ell}^\dagger \left[ X_\ell, W_\ell^\dagger W_{\ell,\ell'}^\dagger W_{+,\ell'}^\dagger \left[ X_{\ell'}, W_{\ell'}^\dagger W_{-,\ell'}^\dagger O W_{-,\ell'} W_{\ell'} \right] W_{\ell',\ell} W_\ell \right] W_{+,\ell} \right| \Psi_0 \right\rangle : \ell' < \ell \end{cases} \varphi_{\ell'}(0) . \tag{81}$$

Similarly, one define the quantum meta-kernel (dQNTK) as

$$\mu = \sum_{\ell_1,\ell_2} \frac{d^2\varepsilon}{d\varphi_{\ell_1}d\varphi_{\ell_2}}\frac{d\varepsilon}{d\varphi_{\ell_1}}\frac{d\varepsilon}{d\varphi_{\ell_2}}$$
$$= \delta^4 \sum_{\ell_1,\ell_2} \begin{matrix} \left\langle \Psi_0 \left| W_{+,\ell_1}^\dagger \left[ X_{\ell_1}, W_{\ell_1}^\dagger W_{-,\ell_1}^\dagger O W_{-,\ell_1} W_{\ell_1} \right] W_{+,\ell_1} \right| \Psi_0 \right\rangle \\ \left\langle \Psi_0 \left| W_{+,\ell_2}^\dagger \left[ X_{\ell_2}, W_{\ell_2}^\dagger W_{-,\ell_2}^\dagger O W_{-,\ell_2} W_{\ell_2} \right] W_{+,\ell_2} \right| \Psi_0 \right\rangle \end{matrix} \times$$
$$\left( \begin{cases} \left\langle \Psi_0 \left| W_{+,\ell_1}^\dagger \left[ X_{\ell_1}, W_{\ell_1}^\dagger W_{\ell_1,\ell_2}^\dagger W_{+,\ell_2}^\dagger \left[ X_{\ell_2}, W_{\ell_2}^\dagger W_{-,\ell_2}^\dagger O W_{-,\ell_2} W_{\ell_2} \right] W_{\ell_2,\ell_1} W_{\ell_1} \right] W_{+,\ell_1} \right| \Psi_0 \right\rangle : \ell_1 \geq \ell_2 \\ \left\langle \Psi_0 \left| W_{+,\ell_2}^\dagger \left[ X_{\ell_2}, W_{\ell_2}^\dagger W_{\ell_2,\ell_1}^\dagger W_{+,\ell_1}^\dagger \left[ X_{\ell_1}, W_{\ell_1}^\dagger W_{-,\ell_1}^\dagger O W_{-,\ell_1} W_{\ell_1} \right] W_{\ell_1,\ell_2} W_{\ell_2} \right] W_{+,\ell_2} \right| \Psi_0 \right\rangle : \ell_1 < \ell_2 \end{cases} \right) . \tag{82}$$

Finally, we wish to mention that in [1], for the classical neural networks they study, the leading order perturbative contribution $\mathcal{O}(1/\text{width})$ is both given by dNTK and ddNTK in dynamics. In our frozen QNTK limit (in the context of lazy training), this does not happen because of power counting in $\delta$.

## III. DETAILS ON THE QUANTUM MACHINE LEARNING: HERMITIAN OPERATOR EXPECTATION VALUE EVALUATION

### A. General setup

We define our model as

$$z_{i;\delta} \equiv z_i(\theta, \mathbf{x}_\delta) = \langle \phi(\mathbf{x}_\delta) | U^\dagger O_i U | \phi(\mathbf{x}_\delta) \rangle \ . \tag{83}$$

where $O_i$ is the $i$-th Hermitian observable. We assume that $O_i$ is taken from a subset of Hermitian operators of the Hilbert space $\mathcal{H}$, denoted by $\mathcal{O}(\mathcal{H})$. The dimension of $\mathcal{O}(\mathcal{H})$ is upper bounded by polynomials of the dimension of the Hilbert space, $\dim \mathcal{H}$. The trainable ansatz is,

$$U = \prod_{\ell=1}^{L} W_\ell U_\ell = \prod_{\ell=1}^{L} W_\ell \exp(i\theta_\ell X_\ell) \ . \tag{84}$$

One could take the derivative,

$$dz_{i;\delta} = \sum_\ell \frac{dz_{i;\delta}}{d\theta_\ell} d\theta_\ell \ , \tag{85}$$

The loss is

$$L_{\mathcal{A}}(\theta) = \frac{1}{2} \sum_{\tilde{\alpha},i} (y_{i;\tilde{\alpha}} - z_{i;\tilde{\alpha}})^2 = \frac{1}{2} \sum_{\tilde{\alpha},i} \varepsilon_{i;\tilde{\alpha}}^2 \ . \tag{86}$$

So

$$\frac{dL_{\mathcal{A}}(\theta)}{d\theta_\ell} = \sum_{\tilde{\alpha},i} \varepsilon_{i;\tilde{\alpha}} \frac{dz_{i;\tilde{\alpha}}}{d\theta_\ell} \ . \tag{87}$$

The gradient descent rule is

$$d\theta_\ell = -\eta \frac{dL_{\mathcal{A}}}{d\theta_\ell} = -\eta \sum_{\tilde{\alpha},i} \varepsilon_{i;\tilde{\alpha}} \frac{dz_{i;\tilde{\alpha}}}{d\theta_\ell} \ , \tag{88}$$

so we have

$$dz_{i;\delta} = -\eta \sum_{\ell,i',\tilde{\alpha}} \varepsilon_{i';\tilde{\alpha}} \frac{dz_{i;\delta}}{d\theta_\ell} \frac{dz_{i';\tilde{\alpha}}}{d\theta_\ell} \ . \tag{89}$$

Since we measure the expectation values of operators, our $\varepsilon$s are always real. Defining the kernel,

$$K_{\delta,\tilde{\alpha}}^{ii'} = \sum_\ell \frac{dz_{i;\delta}}{d\theta_\ell} \frac{dz_{i';\tilde{\alpha}}}{d\theta_\ell} \ , \tag{90}$$

we have

$$dz_{i;\delta} = -\eta \sum_{\tilde{\alpha},i'} K_{\delta,\tilde{\alpha}}^{ii'} \varepsilon_{i';\tilde{\alpha}} \ . \tag{91}$$

We could also make the joint indices

$$(\delta, i) = \bar{a} \ , \quad (\tilde{\alpha}, i') = \hat{b} \ , \tag{92}$$

which are running in the space $\mathcal{D} \times \mathcal{O}(\mathcal{H})$ and $\mathcal{A} \times \mathcal{O}(\mathcal{H})$ respectively. The notation $\hat{a}$ is indicating that the data point component belongs to the training set $\mathcal{A}$, while the notation $\bar{a}$ means that the data point component is general in $\mathcal{D}$. And we have

$$dz_{\bar{a}} = -\eta \sum_{\hat{b}} K_{\bar{a}\hat{b}} \varepsilon_{\hat{b}} \ . \tag{93}$$

In general, one could prove the following statement.

**Theorem 1.** *The matrix $K_{\bar{a}\bar{b}}$ is non-negative and symmetric.*

*Proof.* It is symmetric by definition. Moreover, we consider an arbitrary vector $f_{\bar{a}}$. We have

$$\sum_{\bar{a},\bar{b}} f_{\bar{a}} K_{\bar{a}\bar{b}} f_{\bar{b}} = \sum_{\delta,\delta',i,i'} f_{\delta,i} K_{\delta,\delta'}^{ii'} f_{\delta',i'} =$$

$$= \sum_{\delta,\delta',i,i',\ell} f_{\delta,i} f_{\delta',i'} \frac{dz_{i';\delta'}}{d\theta_\ell} \frac{dz_{i;\delta}}{d\theta_\ell} = \sum_\ell \left( \sum_{\delta,i} f_{\delta,i} \frac{dz_{i;\delta}}{d\theta_\ell} \right)^2 \geq 0 . \quad (94)$$

□

Thus, the matrix $K$ is a proper version of the positive semi-definite symmetric (PDS) kernel [4] in the sense of the classical learning theory.

Now, putting the variational ansatz inside the kernel, we get

$$\frac{dz_{i;\delta}}{d\theta_\ell} = \left\langle \phi(\mathbf{x}_\delta) \left| U_{+,\ell}^\dagger \left[ X_\ell, U_\ell^\dagger W_\ell^\dagger U_{-,\ell}^\dagger O_i U_{-,\ell} W_\ell U_\ell \right] U_{+,\ell} \right| \phi(\mathbf{x}_\delta) \right\rangle ,$$

$$\frac{dz_{i';\tilde{\alpha}}}{d\theta_\ell} = \left\langle \phi(\mathbf{x}_{\tilde{\alpha}}) \left| U_{+,\ell}^\dagger \left[ X_\ell, U_\ell^\dagger W_\ell^\dagger U_{-,\ell}^\dagger O_{i'} U_{-,\ell} W_\ell U_\ell \right] U_{+,\ell} \right| \phi(\mathbf{x}_{\tilde{\alpha}}) \right\rangle , \quad (95)$$

So we have

$$K_{\delta,\tilde{\alpha}}^{ii'} = \sum_\ell \frac{dz_{i;\delta}}{d\theta_\ell} \frac{dz_{i';\tilde{\alpha}}}{d\theta_\ell} = -\sum_\ell \left( \begin{array}{l} \left\langle \phi(\mathbf{x}_\delta) \left| U_{+,\ell}^\dagger \left[ X_\ell, U_\ell^\dagger W_\ell^\dagger U_{-,\ell}^\dagger O_i U_{-,\ell} W_\ell U_\ell \right] U_{+,\ell} \right| \phi(\mathbf{x}_\delta) \right\rangle \times \\ \left\langle \phi(\mathbf{x}_{\tilde{\alpha}}) \left| U_{+,\ell}^\dagger \left[ X_\ell, U_\ell^\dagger W_\ell^\dagger U_{-,\ell}^\dagger O_{i'} U_{-,\ell} W_\ell U_\ell \right] U_{+,\ell} \right| \phi(\mathbf{x}_{\tilde{\alpha}}) \right\rangle \end{array} \right) . \quad (96)$$

### B. No representation learning

The statement of *no representation learning* corresponds to the limit where all the change of variational angles are sufficiently close to zero. In this case, the QNTK becomes frozen (static), similar to the optimization problem. With the variational angle redefined, and the frozen QNTK limit, we have,

$$K_{\delta,\tilde{\alpha}}^{ii'} = -\delta^2 \sum_\ell \left( \begin{array}{l} \left\langle \phi(\mathbf{x}_\delta) \left| W_{+,\ell}^\dagger \left[ X_\ell, W_\ell^\dagger W_{-,\ell}^\dagger O_i W_{-,\ell} W_\ell \right] W_{+,\ell} \right| \phi(\mathbf{x}_\delta) \right\rangle \times \\ \left\langle \phi(\mathbf{x}_{\tilde{\alpha}}) \left| W_{+,\ell}^\dagger \left[ X_\ell, W_\ell^\dagger W_{-,\ell}^\dagger O_{i'} W_{-,\ell} W_\ell \right] W_{+,\ell} \right| \phi(\mathbf{x}_{\tilde{\alpha}}) \right\rangle \end{array} \right) . \quad (97)$$

Here, $\delta$ is the factor we are used to redefine the variational angles $\theta$ by $\theta^* + \delta \times \varphi$. Moreover, one could exactly solve the gradient descent dynamics. We start from the equation

$$d\varepsilon_{\hat{a}} = -\eta \sum_{\hat{b}} K_{\hat{a}\hat{b}} \varepsilon_{\hat{b}} . \quad (98)$$

Thus we have

$$\varepsilon_{\hat{a}_1}(t) = \sum_{\hat{a}_2} U_{\hat{a}_1 \hat{a}_2}(t) \varepsilon_{\hat{a}_2}(0) , \quad (99)$$

where

$$U_{\hat{a}_1 \hat{a}_2}(t) = \left[ (1 - \eta K)^t \right]_{\hat{a}_1 \hat{a}_2} . \quad (100)$$

One can compute the convergence time as

$$\tau_c = \| -\log(1 - \eta K) \| \approx \eta \left\| K_{\delta,\tilde{\alpha}}^{ii'} \right\| . \quad (101)$$

And we could compute the prediction similarly by

$$
\begin{aligned}
z_{\bar{a}}(\infty) &= z_{\bar{a}}(0) - \eta \sum_{\hat{b}} K_{\bar{a}\hat{b}} \sum_{t=0}^{+\infty} \varepsilon_{\hat{b}}(t) \\
&= z_{\bar{a}}(0) - \eta \sum_{\hat{b}} K_{\bar{a}\hat{b}} \left( \sum_{t=0,\hat{b}_0}^{+\infty} \left[(1-\eta K)^t\right]_{\hat{b}\hat{b}_0} \right) \varepsilon_{\hat{b}_0}(0) \\
&= z_{\bar{a}}(0) - \eta \sum_{\hat{b}} K_{\bar{a}\hat{b}} \sum_{\hat{b}_0} \left[(1-(1-\eta K))^{-1}\right]_{\hat{b}\hat{b}_0} \varepsilon_{\hat{b}_0}(0) \\
&= z_{\bar{a}}(0) - \sum_{\hat{b}} K_{\bar{a}\hat{b}} \sum_{\hat{b}_0} \left[K^{i,-1}\right]_{\hat{b}\hat{b}_0} \varepsilon_{\hat{b}_0}(0) .
\end{aligned}
\tag{102}
$$

Moreover, we could compute the prediction by firstly defining the kernel inverse. We define

$$
\sum_{\hat{a} \in \mathcal{A} \times \mathcal{O}(\mathcal{H})} \tilde{K}^{\hat{a}_1 \hat{a}_2} \tilde{K}_{\hat{a}_2 \hat{a}_3} = \delta^{\hat{a}_1}_{\hat{a}_3} ,
\tag{103}
$$

so we get

$$
z_{\bar{a}}(\infty) = z_{\bar{a}}(0) - \sum_{\hat{a}_1, \hat{a}_2} \tilde{K}^{\hat{a}_1 \hat{a}_2} K_{\bar{a}\hat{a}_1} \varepsilon_{\hat{a}_2}(0) .
\tag{104}
$$

### C. Representation learning

Now we start to develop our quantum representation learning theory at the dQNTK order. We make a quadratic expansion:

$$
\begin{aligned}
\varepsilon_{i;\tilde{\alpha}} &= \left\langle \phi(\mathbf{x}_{\tilde{\alpha}}) \left| U^\dagger O_i U \right| \phi(\mathbf{x}_{\tilde{\alpha}}) \right\rangle - y_{i;\tilde{\alpha}} \\
&\approx \left\langle \phi(\mathbf{x}_{\tilde{\alpha}}) \left| \left( \prod_{\ell'=L}^{1} W_{\ell'}^\dagger \right) O_i \left( \prod_{\ell=1}^{L} W_\ell \right) \right| \phi(\mathbf{x}_{\tilde{\alpha}}) \right\rangle - y_{i;\tilde{\alpha}} \\
&- i\delta \sum_\ell \varphi_\ell \left\langle \phi(\mathbf{x}_{\tilde{\alpha}}) \left| W_{+,\ell}^\dagger \left[ X_\ell, W_\ell^\dagger W_{-,\ell}^\dagger O_i W_{-,\ell} W_\ell \right] W_{+,\ell} \right| \phi(\mathbf{x}_{\tilde{\alpha}}) \right\rangle - \frac{\delta^2}{2} \sum_{\ell_1, \ell_2} \varphi_{\ell_1} \varphi_{\ell_2} \times \\
&\begin{cases} \left\langle \psi(\mathbf{x}_{\tilde{\alpha}}) \left| W_{+,\ell_1}^\dagger \left[ X_{\ell_1}, W_{\ell_1}^\dagger W_{\ell_1,\ell_2}^\dagger W_{+,\ell_2}^\dagger \left[ X_{\ell_2}, W_{\ell_2}^\dagger W_{-,\ell_2}^\dagger O_i W_{-,\ell_2} W_{\ell_2} \right] W_{\ell_2,\ell_1} W_{\ell_1} \right] W_{+,\ell_1} \right| \psi(\mathbf{x}_{\tilde{\alpha}}) \right\rangle : \ell_1 \geq \ell_2 \\ \left\langle \psi(\mathbf{x}_{\tilde{\alpha}}) \left| W_{+,\ell_2}^\dagger \left[ X_{\ell_2}, W_{\ell_2}^\dagger W_{\ell_2,\ell_1}^\dagger W_{+,\ell_1}^\dagger \left[ X_{\ell_1}, W_{\ell_1}^\dagger W_{-,\ell_1}^\dagger O_i W_{-,\ell_1} W_{\ell_1} \right] W_{\ell_1,\ell_2} W_{\ell_2} \right] W_{+,\ell_2} \right| \psi(\mathbf{x}_{\tilde{\alpha}}) \right\rangle : \ell_1 < \ell_2 \end{cases} .
\end{aligned}
\tag{105}
$$

So we could write down the derivatives

$$
\begin{aligned}
\frac{d\varepsilon_{i;\tilde{\alpha}}}{d\varphi_\ell} &= -i\delta \left\langle \phi(\mathbf{x}_{\tilde{\alpha}}) \left| W_{+,\ell}^\dagger \left[ X_\ell, W_\ell^\dagger W_{-,\ell}^\dagger O_i W_{-,\ell} W_\ell \right] W_{+,\ell} \right| \phi(\mathbf{x}_{\tilde{\alpha}}) \right\rangle - \delta^2 \times \sum_{\ell'} \varphi_{\ell'} \times \\
&\begin{cases} \left\langle \phi(\mathbf{x}_{\tilde{\alpha}}) \left| W_{+,\ell'}^\dagger \left[ X_{\ell'}, W_{\ell'}^\dagger W_{\ell',\ell}^\dagger W_{+,\ell}^\dagger \left[ X_\ell, W_\ell^\dagger W_{-,\ell}^\dagger O_i W_{-,\ell} W_\ell \right] W_{\ell,\ell'} W_{\ell'} \right] W_{+,\ell'} \right| \phi(\mathbf{x}_{\tilde{\alpha}}) \right\rangle : \ell' \geq \ell \\ \left\langle \phi(\mathbf{x}_{\tilde{\alpha}}) \left| W_{+,\ell}^\dagger \left[ X_\ell, W_\ell^\dagger W_{\ell,\ell'}^\dagger W_{+,\ell'}^\dagger \left[ X_{\ell'}, W_{\ell'}^\dagger W_{-,\ell'}^\dagger O_i W_{-,\ell'} W_{\ell'} \right] W_{\ell',\ell} W_\ell \right] W_{+,\ell} \right| \phi(\mathbf{x}_{\tilde{\alpha}}) \right\rangle : \ell' < \ell \end{cases} .
\end{aligned}
\tag{106}
$$

Similar derivation works for $z_{i;\delta}$ in general. So we have

$$đz_{i;\delta} = \sum_{\ell} \begin{pmatrix} -i\delta \left\langle \phi(\mathbf{x}_\delta) \left| W^\dagger_{+,\ell} \left[ X_\ell, W^\dagger_\ell W^\dagger_{-,\ell} O_i W_{-,\ell} W_\ell \right] W_{+,\ell} \right| \phi(\mathbf{x}_\delta) \right\rangle - \delta^2 \sum_{\ell'} \varphi_{\ell'} \times \\ \begin{cases} \left\langle \phi(\mathbf{x}_\delta) \left| W^\dagger_{+,\ell'} \left[ X_{\ell'}, W^\dagger_{\ell'} W^\dagger_{\ell',\ell} W^\dagger_{+,\ell} \left[ X_\ell, W^\dagger_\ell W^\dagger_{-,\ell} O_i W_{-,\ell} W_\ell \right] W_{\ell,\ell'} W_{\ell'} \right] W_{+,\ell'} \right| \phi(\mathbf{x}_\delta) \right\rangle : \ell' \geq \ell \\ \left\langle \phi(\mathbf{x}_\delta) \left| W^\dagger_{+,\ell} \left[ X_\ell, W^\dagger_\ell W^\dagger_{\ell,\ell'} W^\dagger_{+,\ell'} \left[ X_{\ell'}, W^\dagger_{\ell'} W^\dagger_{-,\ell'} O_i W_{-,\ell'} W_{\ell'} \right] W_{\ell',\ell} W_\ell \right] W_{+,\ell} \right| \phi(\mathbf{x}_\delta) \right\rangle : \ell' < \ell \end{cases} \end{pmatrix} đ\varphi_\ell$$

$$- \frac{\delta^2}{2} \sum_{\ell_1, \ell_2} đ\varphi_{\ell_1} đ\varphi_{\ell_2} \times$$

$$\left( \begin{cases} \left\langle \phi(\mathbf{x}_\delta) \left| W^\dagger_{+,\ell_1} \left[ X_{\ell_1}, W^\dagger_{\ell_1} W^\dagger_{\ell_1,\ell_2} W^\dagger_{+,\ell_2} \left[ X_{\ell_2}, W^\dagger_{\ell_2} W^\dagger_{-,\ell_2} O_i W_{-,\ell_2} W_{\ell_2} \right] W_{\ell_2,\ell_1} W_{\ell_1} \right] W_{+,\ell_1} \right| \phi(\mathbf{x}_\delta) \right\rangle : \ell_1 \geq \ell_2 \\ \left\langle \phi(\mathbf{x}_\delta) \left| W^\dagger_{+,\ell_2} \left[ X_{\ell_2}, W^\dagger_{\ell_2} W^\dagger_{\ell_2,\ell_1} W^\dagger_{+,\ell_1} \left[ X_{\ell_1}, W^\dagger_{\ell_1} W^\dagger_{-,\ell_1} O_i W_{-,\ell_1} W_{\ell_1} \right] W_{\ell_1,\ell_2} W_{\ell_2} \right] W_{+,\ell_2} \right| \phi(\mathbf{x}_\delta) \right\rangle : \ell_1 < \ell_2 \end{cases} \right). \tag{107}$$

The leading order piece is exactly the effective feature map. A more compact version is,

$$đz_{i;\delta} = \sum_\ell \frac{dz_{i;\delta}}{d\varphi_\ell} đ\varphi_\ell - \frac{1}{2}\delta^2 \sum_{\ell_1,\ell_2} đ\varphi_{\ell_1} đ\varphi_{\ell_2} G^{\delta,i}_{\ell_1,\ell_2}, \tag{108}$$

where we define

$$G^{\delta,i}_{\ell_1,\ell_2} \equiv G_{\ell_1,\ell_2}(\phi(\mathbf{x}_\delta), O_i) =$$
$$\left( \begin{cases} \left\langle \phi(\mathbf{x}_\delta) \left| W^\dagger_{+,\ell_1} \left[ X_{\ell_1}, W^\dagger_{\ell_1} W^\dagger_{\ell_1,\ell_2} W^\dagger_{+,\ell_2} \left[ X_{\ell_2}, W^\dagger_{\ell_2} W^\dagger_{-,\ell_2} O_i W_{-,\ell_2} W_{\ell_2} \right] W_{\ell_2,\ell_1} W_{\ell_1} \right] W_{+,\ell_1} \right| \phi(\mathbf{x}_\delta) \right\rangle : \ell_1 \geq \ell_2 \\ \left\langle \phi(\mathbf{x}_\delta) \left| W^\dagger_{+,\ell_2} \left[ X_{\ell_2}, W^\dagger_{\ell_2} W^\dagger_{\ell_2,\ell_1} W^\dagger_{+,\ell_1} \left[ X_{\ell_1}, W^\dagger_{\ell_1} W^\dagger_{-,\ell_1} O_i W_{-,\ell_1} W_{\ell_1} \right] W_{\ell_1,\ell_2} W_{\ell_2} \right] W_{+,\ell_2} \right| \phi(\mathbf{x}_\delta) \right\rangle : \ell_1 < \ell_2 \end{cases} \right). \tag{109}$$

Moreover, we want to define

$$\Theta^{\delta,i}_\ell \equiv \Theta_\ell(\phi(\mathbf{x}_\delta), O_i) = \left\langle \phi(\mathbf{x}_\delta) \left| W^\dagger_{+,\ell} \left[ X_\ell, W^\dagger_\ell W^\dagger_{-,\ell} O_i W_{-,\ell} W_\ell \right] W_{+,\ell} \right| \phi(\mathbf{x}_\delta) \right\rangle. \tag{110}$$

We will now compute the effective kernel. We have,

$$\frac{dz_{i;\delta}}{d\varphi_\ell} = -i\delta \Theta^{\delta,i}_\ell - \delta^2 \sum_{\ell'} \varphi_{\ell'} G^{\delta,i}_{\ell',\ell},$$
$$\frac{dz_{i';\tilde{\alpha}}}{d\varphi_\ell} = -i\delta \Theta^{\tilde{\alpha},i'}_\ell - \delta^2 \sum_{\ell'} \varphi_{\ell'} G^{\tilde{\alpha},i'}_{\ell',\ell}. \tag{111}$$

Thus, we define

$$K^{E,ii'}_{\delta,\tilde{\alpha}} = \sum_\ell \frac{dz_{i;\delta}}{d\varphi_\ell} \frac{dz_{i';\tilde{\alpha}}}{d\varphi_\ell}, \tag{112}$$

and

$$K^{E,ii'}_{\delta,\tilde{\alpha}}(0) = K^{ii'}_{\delta,\tilde{\alpha}} + K^{\Delta,ii'}_{\delta,\tilde{\alpha}}. \tag{113}$$

We find

$$K^{\Delta,ii'}_{\delta,\tilde{\alpha}} = i\delta^3 \sum_{\ell,\ell'} \varphi_{\ell'}(0) G^{\delta,i}_{\ell',\ell} \Theta^{\tilde{\alpha},i'}_\ell + i\delta^3 \sum_{\ell,\ell'} \varphi_{\ell'}(0) G^{\tilde{\alpha},i'}_{\ell',\ell} \Theta^{\delta,i}_\ell. \tag{114}$$

Now we could write down the prediction on the training set. We have

$$\varepsilon_{\hat{a}}(t) = \varepsilon^F_{\hat{a}}(t) + \varepsilon^I_{\hat{a}}(t), \tag{115}$$

where

$$\varepsilon_{\hat{a}}^{F}(t) = \sum_{\hat{a}_1} U_{\hat{a}\hat{a}_1}(t)\varepsilon_{\hat{a}_1}(0) ,$$

$$U_{\hat{a}_1\hat{a}_2}(t) = \left[(1-\eta K)^t\right]_{\hat{a}_1\hat{a}_2} , \tag{116}$$

and

$$\varepsilon_{\hat{a}}^{I}(t) = \left(-\eta \sum_{s=0}^{t-1}(1-\eta K)^{t-1-s} K^{\Delta}(1-\eta K)^s \varepsilon(0)\right)_{\hat{a}} . \tag{117}$$

It is the compact matrix product form in the space $\mathcal{A} \times \mathcal{O}(\mathcal{H})$. Moreover, we have

$$\left\|\varepsilon^I(t)\right\| \leq \eta t \|1-\eta K\|^{t-1} \left\|K^\Delta\right\| \|\varepsilon(0)\| . \tag{118}$$

Finally, we discuss the asymptotic convergence regime. We could compute the quantum meta-kernel. We notice that

$$\check{d}z_{i;\delta} = -\eta \sum_{\ell,i',\tilde{\alpha}} \frac{dz_{i;\delta}}{d\varphi_\ell} \frac{dz_{i';\tilde{\alpha}}}{d\varphi_\ell} \varepsilon_{i;\tilde{\alpha}} + \eta^2 \sum_{\ell_1,\ell_2,\tilde{\alpha}_1,i_1,\tilde{\alpha}_2,i_2} \frac{d^2 z_{i;\delta}}{d\varphi_{\ell_1} d\varphi_{\ell_2}} \frac{dz_{i_1;\tilde{\alpha}_1}}{d\varphi_{\ell_1}} \frac{dz_{i_2;\tilde{\alpha}_2}}{d\varphi_{\ell_2}} \varepsilon_{i_1;\tilde{\alpha}_1} \varepsilon_{i_2;\tilde{\alpha}_2} . \tag{119}$$

So we define

$$\mu_{\delta\tilde{\alpha}_1\tilde{\alpha}_2}^{ii_1i_2} = \mu_{\bar{a}\hat{a}_1\hat{a}_2} = \sum_{\ell_1,\ell_2} \frac{d^2 z_{i;\delta}}{d\varphi_{\ell_1} d\varphi_{\ell_2}} \left(\frac{dz_{i_1;\tilde{\alpha}_1}}{d\varphi_{\ell_1}} \frac{dz_{i_2;\tilde{\alpha}_2}}{d\varphi_{\ell_2}}\right)\bigg|_{\varphi=0}, \tag{120}$$

in the leading order. Note that it is natural to extend the definition to more general inputs

$$\mu_{\delta_0\delta_1\delta_2}^{i_0i_1i_2} = \mu_{\bar{a}_0\bar{a}_1\bar{a}_2} = \sum_{\ell_1,\ell_2} \frac{d^2 z_{i_0;\delta_0}}{d\varphi_{\ell_1} d\varphi_{\ell_2}} \left(\frac{dz_{i_1;\delta_1}}{d\varphi_{\ell_1}} \frac{dz_{i_2;\delta_2}}{d\varphi_{\ell_2}}\right)\bigg|_{\varphi=0} . \tag{121}$$

So the asymptotic convergence is given by

$$z_{\bar{a}}(\infty) = z_{\bar{a}}(0) - \sum_{\hat{a}_1,\hat{a}_2} K_{\bar{a}\hat{a}_1} \tilde{K}^{\hat{a}_1\hat{a}_2} \varepsilon_{\hat{a}_2}(0)$$

$$+ \sum_{\hat{a}_1,\hat{a}_2,\hat{a}_3,\hat{a}_4} \left[\mu_{\hat{a}_1\bar{a}\hat{a}_2} - \sum_{\hat{a}_5,\hat{a}_6} K_{\bar{a}\hat{a}_5} \tilde{K}^{\hat{a}_5\hat{a}_6} \mu_{\hat{a}_1\hat{a}_6\hat{a}_2}\right] Z_A^{\hat{a}_1\hat{a}_2\hat{a}_3\hat{a}_4} \varepsilon_{\hat{a}_3}(0)\varepsilon_{\hat{a}_4}(0)$$

$$+ \sum_{\hat{a}_1,\hat{a}_2,\hat{a}_3,\hat{a}_4} \left[\mu_{\bar{a}\hat{a}_1\hat{a}_2} - \sum_{\hat{a}_5,\hat{a}_6} K_{\bar{a}\hat{a}_5} \tilde{K}^{\hat{a}_5\hat{a}_6} \mu_{\hat{a}_6\hat{a}_1\hat{a}_2}\right] Z_B^{\hat{a}_1\hat{a}_2\hat{a}_3\hat{a}_4} \varepsilon_{\hat{a}_3}(0)\varepsilon_{\hat{a}_4}(0) , \tag{122}$$

where the algorithm projector $Z_{A,B}$s are

$$Z_A^{\hat{a}_1\hat{a}_2\hat{a}_3\hat{a}_4} \equiv \tilde{K}^{\hat{a}_1\hat{a}_3} \tilde{K}^{\hat{a}_2\hat{a}_4} - \sum_{\hat{a}_5} \tilde{K}^{\hat{a}_2\hat{a}_5} X_\parallel^{\hat{a}_1\hat{a}_5\hat{a}_3\hat{a}_4} ,$$

$$Z_B^{\hat{a}_1\hat{a}_2\hat{a}_3\hat{a}_4} \equiv \tilde{K}^{\hat{a}_1\hat{a}_3} \tilde{K}^{\hat{a}_2\hat{a}_4} - \sum_{\hat{a}_5} \tilde{K}^{\hat{a}_2\hat{a}_5} X_\parallel^{\hat{a}_1\hat{a}_5\hat{a}_3\hat{a}_4} + \frac{\eta}{2} X_\parallel^{\hat{a}_1\hat{a}_2\hat{a}_3\hat{a}_4} , \tag{123}$$

and

$$X_\parallel^{\hat{a}_1\hat{a}_2\hat{a}_3\hat{a}_4} = \sum_{s=0}^{\infty} [(1-\eta K)^s]_{\hat{a}_1\hat{a}_3} [(1-\eta K)^s]_{\hat{a}_2\hat{a}_4} , \tag{124}$$

which is defined implicitly as

$$\delta_{\hat{a}_5}^{\hat{a}_1} \delta_{\hat{a}_6}^{\hat{a}_2} = \sum_{\hat{a}_3,\hat{a}_4} X_\parallel^{\hat{a}_1\hat{a}_2\hat{a}_3\hat{a}_4} \left(\tilde{K}_{\hat{a}_3\hat{a}_5} \delta_{\hat{a}_4\hat{a}_6} + \delta_{\hat{a}_3\hat{a}_5} \tilde{K}_{\hat{a}_4\hat{a}_6} - \eta \tilde{K}_{\hat{a}_3\hat{a}_5} \tilde{K}_{\hat{a}_4\hat{a}_6}\right) . \tag{125}$$

Now we write down all components of $\mu$. We have

$$\frac{dz_{i_1;\delta_1}}{d\varphi_{\ell_1}}(\varphi=0) = -i\delta\Theta^{\delta_1,i_1}_{\ell_1} ,$$

$$\frac{dz_{i_2;\delta_2}}{d\varphi_{\ell_2}}(\varphi=0) = -i\delta\Theta^{\delta_2,i_2}_{\ell_2} ,$$

$$\frac{d^2 z_{i_0;\delta_0}}{d\varphi_{\ell_1}d\varphi_{\ell_2}} = -\delta^2 G^{\delta_0,i_0}_{\ell_1,\ell_2} . \tag{126}$$

So

$$\mu^{i_0 i_1 i_2}_{\delta_0 \delta_1 \delta_2} = \sum_{\ell_1,\ell_2} \frac{d^2 z_{i_0;\delta_0}}{d\varphi_{\ell_1} d\varphi_{\ell_2}} \left( \frac{dz_{i_1;\delta_1}}{d\varphi_{\ell_1}} \frac{dz_{i_2;\delta_2}}{d\varphi_{\ell_2}} \right) \bigg|_{\varphi=0} = \delta^4 \sum_{\ell_1,\ell_2} \Theta^{\delta_1,i_1}_{\ell_1} \Theta^{\delta_2,i_2}_{\ell_2} G^{\delta_0,i_0}_{\ell_1,\ell_2} . \tag{127}$$

## IV. DETAILS ON THE QUANTUM MACHINE LEARNING: AMPLITUDE ENCODING

### A. General setup

Here, we consider alternative quantum machine learning models with the amplitude encoding. In this case, we naturally extend the NTK formalism to the complex version. We consider the variational setup where

$$z_{i;\delta} \equiv z_i(\theta, \mathbf{x}_\delta) = \left\langle i \left| \prod_{\ell=1}^{L} W_\ell U_\ell \right| \phi(\mathbf{x}_\delta) \right\rangle . \tag{128}$$

One could take the derivative,

$$dz_{i;\delta} = \sum_\ell \frac{dz_{i;\delta}}{d\theta_\ell} d\theta_\ell , \tag{129}$$

The loss is

$$L_\mathcal{A}(\theta) = \frac{1}{2} \sum_{\tilde{\alpha},i} |y_{i;\tilde{\alpha}} - z_{i;\tilde{\alpha}}|^2 = \frac{1}{2} \sum_{\tilde{\alpha},i} (y_{i;\tilde{\alpha}} - z_{i;\tilde{\alpha}})(y_{i;\tilde{\alpha}} - z_{i;\tilde{\alpha}})^* . \tag{130}$$

So

$$\frac{dL_\mathcal{A}(\theta)}{d\theta_\ell} = \frac{1}{2} \sum_{\tilde{\alpha},i} \varepsilon_{i;\tilde{\alpha}} \frac{dz^*_{i;\tilde{\alpha}}}{d\theta_\ell} + \frac{1}{2} \sum_{\tilde{\alpha},i} \varepsilon^*_{i;\tilde{\alpha}} \frac{dz_{i;\tilde{\alpha}}}{d\theta_\ell} = \text{Re} \sum_{\tilde{\alpha},i} \varepsilon_{i;\tilde{\alpha}} \frac{dz^*_{i;\tilde{\alpha}}}{d\theta_\ell} . \tag{131}$$

Note that if we count the number of times that $U$ appears in the loss function, the amplitude encoding model is a *squareroot* of the operator expectation value model. The gradient descent rule is

$$d\theta_\ell = -\eta \frac{d\mathcal{L}_\mathcal{A}}{d\theta_\ell} = -\frac{\eta}{2} \sum_{\tilde{\alpha},i} \varepsilon_{i;\tilde{\alpha}} \frac{dz^*_{i;\tilde{\alpha}}}{d\theta_\ell} - \frac{\eta}{2} \sum_{\tilde{\alpha},i} \varepsilon^*_{i;\tilde{\alpha}} \frac{dz_{i;\tilde{\alpha}}}{d\theta_\ell} , \tag{132}$$

so we have

$$dz_{i;\delta} = -\eta \sum_{\ell,i',\tilde{\alpha}} \frac{dz_{i;\delta}}{d\theta_\ell} \text{Re}\left( \varepsilon_{i;\tilde{\alpha}} \frac{dz^*_{i';\tilde{\alpha}}}{d\theta_\ell} \right)$$

$$= -\frac{\eta}{2} \sum_{\ell,i',\tilde{\alpha}} \varepsilon_{i';\tilde{\alpha}} \frac{dz_{i;\delta}}{d\theta_\ell} \frac{dz^*_{i';\tilde{\alpha}}}{d\theta_\ell} - \frac{\eta}{2} \sum_{\ell,i',\tilde{\alpha}} \varepsilon^*_{i';\tilde{\alpha}} \frac{dz_{i;\delta}}{d\theta_\ell} \frac{dz_{i';\tilde{\alpha}}}{d\theta_\ell} . \tag{133}$$

We notice that the variable $z$ is complex in general. So we could write,

$$dz_{i;\delta} = -\frac{\eta}{2} \sum_{\ell,i',\tilde{\alpha}} \varepsilon_{i';\tilde{\alpha}} \frac{dz_{i;\delta}}{d\theta_\ell} \frac{dz^*_{i';\tilde{\alpha}}}{d\theta_\ell} - \frac{\eta}{2} \sum_{\ell,i',\tilde{\alpha}} \varepsilon^*_{i';\tilde{\alpha}} \frac{dz_{i;\delta}}{d\theta_\ell} \frac{dz_{i';\tilde{\alpha}}}{d\theta_\ell} ,$$

$$dz^*_{i;\delta} = -\frac{\eta}{2} \sum_{\ell,i',\tilde{\alpha}} \varepsilon^*_{i';\tilde{\alpha}} \frac{dz^*_{i;\delta}}{d\theta_\ell} \frac{dz_{i';\tilde{\alpha}}}{d\theta_\ell} - \frac{\eta}{2} \sum_{\ell,i',\tilde{\alpha}} \varepsilon_{i';\tilde{\alpha}} \frac{dz^*_{i;\delta}}{d\theta_\ell} \frac{dz^*_{i';\tilde{\alpha}}}{d\theta_\ell} . \tag{134}$$

Defining the kernel

$$\begin{pmatrix} K^{+,ii'}_{\delta,\tilde{\alpha}} & K^{-,ii'}_{\delta,\tilde{\alpha}} \\ K^{*,-,ii'}_{\delta,\tilde{\alpha}} & K^{*,+,ii'}_{\delta,\tilde{\alpha}} \end{pmatrix} = \begin{pmatrix} \sum_\ell \frac{dz_{i;\delta}}{d\theta_\ell} \frac{dz^*_{i';\tilde{\alpha}}}{d\theta_\ell} & \sum_\ell \frac{dz_{i;\delta}}{d\theta_\ell} \frac{dz_{i';\tilde{\alpha}}}{d\theta_\ell} \\ \sum_\ell \frac{dz^*_{i;\delta}}{d\theta_\ell} \frac{dz^*_{i';\tilde{\alpha}}}{d\theta_\ell} & \sum_\ell \frac{dz^*_{i;\delta}}{d\theta_\ell} \frac{dz_{i';\tilde{\alpha}}}{d\theta_\ell} \end{pmatrix}, \tag{135}$$

we have

$$\begin{aligned} \check{d}z_{i;\delta} &= -\frac{\eta}{2} \sum_{\tilde{\alpha},i'} K^{+,ii'}_{\delta,\tilde{\alpha}} \varepsilon_{i';\tilde{\alpha}} - \frac{\eta}{2} \sum_{\tilde{\alpha},i'} K^{-,ii'}_{\delta,\tilde{\alpha}} \varepsilon^*_{i';\tilde{\alpha}}, \\ \check{d}z^*_{i;\delta} &= -\frac{\eta}{2} \sum_{\tilde{\alpha},i'} K^{*,-,ii'}_{\delta,\tilde{\alpha}} \varepsilon_{i';\tilde{\alpha}} - \frac{\eta}{2} \sum_{\tilde{\alpha},i'} K^{*,+,ii'}_{\delta,\tilde{\alpha}} \varepsilon^*_{i';\tilde{\alpha}}, \end{aligned} \tag{136}$$

Here we also use $K^*$ to denote the complex conjugate. We introduce the worldsheet index $\dot{\alpha}, \dot{\beta}$ to denote the two-component of the complex variable $(\check{d}z, \check{d}z^*)$. So we have [5]

$$\check{d}z^{\dot{\alpha}}_{i;\delta} = -\frac{\eta}{2} \sum_{\tilde{\alpha},\dot{\beta},i'} K^{\dot{\alpha}\dot{\beta},ii'}_{\delta,\tilde{\alpha}} \varepsilon^{\dot{\beta}}_{i';\tilde{\alpha}}, \tag{137}$$

where

$$\varepsilon^{\dot{\alpha}}_{i;\delta} = \begin{pmatrix} \varepsilon_{i;\delta} \\ \varepsilon^*_{i;\delta} \end{pmatrix}, \quad K^{\dot{\alpha}\dot{\beta},ii'}_{\delta,\tilde{\alpha}} = \begin{pmatrix} K^{+,ii'}_{\delta,\tilde{\alpha}} & K^{-,ii'}_{\delta,\tilde{\alpha}} \\ K^{*,-,ii'}_{\delta,\tilde{\alpha}} & K^{*,+,ii'}_{\delta,\tilde{\alpha}} \end{pmatrix}. \tag{138}$$

We could also make the joint index

$$(\delta, \dot{\alpha}, i) = \bar{\mu}, \quad (\tilde{\alpha}, \dot{\beta}, i') = \hat{\nu}, \tag{139}$$

which is running in the space $\mathcal{D} \times \mathbb{Z}_2 \times \mathcal{H}$ and $\mathcal{A} \times \mathbb{Z}_2 \times \mathcal{H}$ respectively. The notation $\hat{\mu}$ is indicating that the data point component belongs to the training set $\mathcal{A}$, while the notation $\bar{\mu}$ means that the data point component is general in $\mathcal{D}$. And we have

$$\check{d}z_{\bar{\mu}} = -\frac{\eta}{2} \sum_{\hat{\nu}} K_{\bar{\mu}\hat{\nu}} \varepsilon_{\hat{\nu}}. \tag{140}$$

In general, one could prove the following statement.

**Theorem 2.** *The matrix $K_{\bar{\mu}\bar{\nu}}$ is non-negative and Hermitian.*

*Proof.* One could easily check that the matrix form

$$\begin{aligned} K^\dagger_{\bar{\mu}\bar{\nu}} &= \begin{pmatrix} K^{+,ii'}_{\delta,\tilde{\alpha}} & K^{-,ii'}_{\delta,\tilde{\alpha}} \\ K^{*,-,ii'}_{\delta,\tilde{\alpha}} & K^{*,+,ii'}_{\delta,\tilde{\alpha}} \end{pmatrix}^{*,T} = \begin{pmatrix} K^{*,+,ii'}_{\delta,\tilde{\alpha}} & K^{*,-,ii'}_{\delta,\tilde{\alpha}} \\ K^{-,ii'}_{\delta,\tilde{\alpha}} & K^{+,ii'}_{\delta,\tilde{\alpha}} \end{pmatrix}^T \\ &= \begin{pmatrix} (K^{*,+,ii'}_{\delta,\tilde{\alpha}})^T & (K^{-,ii'}_{\delta,\tilde{\alpha}})^T \\ (K^{*,-,ii'}_{\delta,\tilde{\alpha}})^T & (K^{+,ii'}_{\delta,\tilde{\alpha}})^T \end{pmatrix} = \begin{pmatrix} K^{+,ii'}_{\delta,\tilde{\alpha}} & K^{-,i'i}_{\delta,\tilde{\alpha}} \\ K^{*,-,ii'}_{\delta,\tilde{\alpha}} & K^{*,+,ii'}_{\delta,\tilde{\alpha}} \end{pmatrix} = K_{\bar{\mu}\bar{\nu}}, \end{aligned} \tag{141}$$

is Hermitian. Moreover, consider an arbitrary vector $f_{\bar{\mu}}$, we have

$$\begin{aligned} \sum_{\bar{\mu},\bar{\nu}} f^*_{\bar{\mu}} K_{\bar{\mu}\bar{\nu}} f_{\bar{\nu}} &= \sum_{\delta,\delta',i,i'} \begin{pmatrix} f^*_{\delta,i} & f_{\delta,i} \end{pmatrix} \begin{pmatrix} K^{+,ii'}_{\delta,\delta'} & K^{-,ii'}_{\delta,\delta'} \\ K^{*,-,ii'}_{\delta,\delta'} & K^{*,+,ii'}_{\delta,\delta'} \end{pmatrix} \begin{pmatrix} f_{\delta',i'} \\ f^*_{\delta',i'} \end{pmatrix} \\ &= \sum_{\delta,\delta',i,i'} \left( f^*_{\delta,i} f_{\delta',i'} K^{+,ii'}_{\delta,\delta'} + f_{\delta,i} f_{\delta',i'} K^{*,-,ii'}_{\delta,\delta'} + f^*_{\delta,i} f^*_{\delta',i'} K^{-,ii'}_{\delta,\delta'} + f_{\delta,i} f^*_{\delta',i'} K^{*,+,ii'}_{\delta,\delta'} \right) \\ &= \sum_{\delta,\delta',i,i',\ell} \begin{pmatrix} f^*_{\delta,i} f_{\delta',i'} \frac{dz_{i;\delta}}{d\theta_\ell} \frac{dz^*_{i';\delta'}}{d\theta_\ell} + f_{\delta,i} f_{\delta',i'} \frac{dz^*_{i;\delta}}{d\theta_\ell} \frac{dz^*_{i';\delta'}}{d\theta_\ell} \\ + f^*_{\delta,i} f^*_{\delta',i'} \frac{dz_{i;\delta}}{d\theta_\ell} \frac{dz_{i';\delta'}}{d\theta_\ell} + f_{\delta,i} f^*_{\delta',i'} \frac{dz^*_{i;\delta}}{d\theta_\ell} \frac{dz_{i';\delta'}}{d\theta_\ell} \end{pmatrix} \\ &= \sum_{\delta,\delta',i,i',\ell} \left( f_{\delta,i} \frac{dz^*_{i;\delta}}{d\theta_\ell} + f^*_{\delta,i} \frac{dz_{i;\delta}}{d\theta_\ell} \right) \left( f_{\delta',i'} \frac{dz^*_{i';\delta'}}{d\theta_\ell} + f^*_{\delta',i'} \frac{dz_{i';\delta'}}{d\theta_\ell} \right) \\ &= \sum_\ell \left( \sum_{\delta,i} f_{\delta,i} \frac{dz^*_{i;\delta}}{d\theta_\ell} + f^*_{\delta,i} \frac{dz_{i;\delta}}{d\theta_\ell} \right)^2 = 4 \sum_\ell \left( \text{Re} \sum_{\delta,i} f_{\delta,i} \frac{dz^*_{i;\delta}}{d\theta_\ell} \right)^2 \geq 0. \end{aligned} \tag{142}$$

□

Thus, the matrix $K$ is a complexified version of the positive semi-definite symmetric (PDS) kernel in the sense of the kernel method of statistical learning theory (see, for instance, [4]), but running during the training dynamics. Moreover, our work provides a complexified version of the NTK theory that could be useful for machine learning itself.

Now, putting the variational ansatz inside the kernel, we get

$$\frac{dz_{i;\alpha}}{d\theta_\ell} = \frac{d}{d\theta_\ell} \left\langle i \left| \prod_{\ell'=1}^{L} W_{\ell'} U_{\ell'} \right| \phi(\mathbf{x}_\alpha) \right\rangle = i \left\langle i \left| \prod_{\ell'=1}^{\ell-1} W_{\ell'} U_{\ell'} X_\ell \prod_{\ell''=\ell}^{L} W_{\ell''} U_{\ell''} \right| \phi(\mathbf{x}_\alpha) \right\rangle , \tag{143}$$

$$\frac{dz^*_{i;\alpha}}{d\theta_\ell} = \frac{d}{d\theta_\ell} \left\langle \phi(\mathbf{x}_\alpha) \left| \prod_{\ell'=L}^{1} U^\dagger_{\ell'} W^\dagger_{\ell'} \right| i \right\rangle = -i \left\langle \phi(\mathbf{x}_\alpha) \left| \prod_{\ell'=L}^{\ell+1} U^\dagger_{\ell'} W^\dagger_{\ell'} X_\ell \prod_{\ell''=1}^{\ell} U^\dagger_{\ell''} W^\dagger_{\ell''} \right| i \right\rangle . \tag{144}$$

And we have

$$\begin{aligned} \frac{dz_{i;\alpha}}{d\theta_\ell} &= i \left\langle i \left| U_{-,\ell} X_\ell W_\ell U_\ell U_{+,\ell} \right| \phi(\mathbf{x}_\alpha) \right\rangle , \\ \frac{dz^*_{i;\alpha}}{d\theta_\ell} &= -i \left\langle \phi(\mathbf{x}_\alpha) \left| U^\dagger_{+,\ell} U^\dagger_\ell W^\dagger_\ell X_\ell U^\dagger_{-,\ell} \right| i \right\rangle . \end{aligned} \tag{145}$$

So we have

$$\begin{aligned} \sum_\ell \frac{dz_{i;\delta}}{d\theta_\ell} \frac{dz^*_{i';\tilde{\alpha}}}{d\theta_\ell} &= \sum_\ell \left\langle i \left| U_{-,\ell} X_\ell W_\ell U_\ell U_{+,\ell} \right| \phi(\mathbf{x}_\delta) \right\rangle \left\langle \phi(\mathbf{x}_{\tilde{\alpha}}) \left| U^\dagger_{+,\ell} U^\dagger_\ell W^\dagger_\ell X_\ell U^\dagger_{-,\ell} \right| i' \right\rangle , \\ \sum_\ell \frac{dz_{i;\delta}}{d\theta_\ell} \frac{dz_{i';\tilde{\alpha}}}{d\theta_\ell} &= -\sum_\ell \left\langle i \left| U_{-,\ell} X_\ell W_\ell U_\ell U_{+,\ell} \right| \phi(\mathbf{x}_\delta) \right\rangle \left\langle i' \left| U_{-,\ell} X_\ell W_\ell U_\ell U_{+,\ell} \right| \phi(\mathbf{x}_{\tilde{\alpha}}) \right\rangle , \end{aligned} \tag{146}$$

and their conjugates. Now, we define some notations. We define the *feature density matrix*

$$\rho_{\delta\tilde{\alpha}} = |\phi(\mathbf{x}_\delta)\rangle \langle \phi(\mathbf{x}_{\tilde{\alpha}})| , \tag{147}$$

and the *feature projector*

$$\Lambda_{i';\delta} = |\phi(\mathbf{x}_\delta)\rangle \langle i'| . \tag{148}$$

We have

$$\begin{aligned} K^{+,ii'}_{\delta,\tilde{\alpha}} &= \sum_\ell \left\langle i \left| U_{-,\ell} X_\ell W_\ell U_\ell U_{+,\ell} \rho_{\delta\tilde{\alpha}} U^\dagger_{+,\ell} U^\dagger_\ell W^\dagger_\ell X_\ell U^\dagger_{-,\ell} \right| i' \right\rangle , \\ K^{-,ii'}_{\delta,\tilde{\alpha}} &= -\sum_\ell \left\langle i' \left| U_{-,\ell} X_\ell W_\ell U_\ell U_{+,\ell} \Lambda_{i;\tilde{\alpha}} U_{-,\ell} X_\ell W_\ell U_\ell U_{+,\ell} \right| \phi(\mathbf{x}_\delta) \right\rangle . \end{aligned} \tag{149}$$

### B. No representation learning

In the frozen QNTK limit with the variational angle redefined, one could compute the expressions of the kernel as

$$\begin{aligned} K^{+,ii'}_{\delta,\tilde{\alpha}} &= \delta^2 \sum_\ell \left\langle i \left| W_{-,\ell} X_\ell W_\ell W_{+,\ell} \rho_{\delta\tilde{\alpha}} W^\dagger_{+,\ell} W^\dagger_\ell X_\ell W^\dagger_{-,\ell} \right| i' \right\rangle , \\ K^{-,ii'}_{\delta,\tilde{\alpha}} &= -\delta^2 \sum_\ell \left\langle i' \left| W_{-,\ell} X_\ell W_\ell U_{+,\ell} \Lambda_{i;\tilde{\alpha}} W_{-,\ell} X_\ell W_\ell W_{+,\ell} \right| \phi(\mathbf{x}_\delta) \right\rangle . \end{aligned} \tag{150}$$

Here, $\delta$ is the factor we are used to redefine the variational angles $\theta$ by $\theta^* + \delta \times \varphi$, and we define

$$W_{-,\ell} \equiv \prod_{\ell'=1}^{\ell-1} W_{\ell'} , \quad W_{+,\ell} \equiv \prod_{\ell'=\ell+1}^{L} W_{\ell'} . \tag{151}$$

Moreover, one could exactly solve the gradient descent dynamics. We start from the equation

$$\check{d}\varepsilon_{\hat{\mu}} = -\frac{\eta}{2} \sum_{\hat{\nu}} K_{\hat{\mu}\hat{\nu}} \varepsilon_{\hat{\nu}} . \tag{152}$$

Thus we have

$$\varepsilon_{\hat{\mu}_1}(t) = \sum_{\hat{\mu}_2} U_{\hat{\mu}_1\hat{\mu}_2}(t)\varepsilon_{\hat{\mu}_2}(0) \ , \tag{153}$$

where

$$U_{\hat{\mu}_1\hat{\mu}_2}(t) = \left[\left(1 - \frac{\eta}{2}K\right)^t\right]_{\hat{\mu}_1\hat{\mu}_2} \ . \tag{154}$$

One can compute the convergence time as

$$\tau_c = \left\|-\log\left(1 - \frac{\eta}{2}K\right)\right\| \approx \frac{\eta}{2}\left\|\begin{pmatrix} K_{\delta,\tilde{\alpha}}^{+,ii'} & K_{\delta,\tilde{\alpha}}^{-,ii'} \\ K_{\delta,\tilde{\alpha}}^{*,-,ii'} & K_{\delta,\tilde{\alpha}}^{*,+,ii'} \end{pmatrix}\right\| \ . \tag{155}$$

And we could compute the prediction similarly by

$$\begin{aligned} z_{\bar{\mu}}(\infty) &= z_{\bar{\mu}}(0) - \frac{\eta}{2}\sum_{\hat{\nu}} K_{\bar{\mu}\hat{\nu}} \sum_{t=0}^{+\infty} \varepsilon_{\hat{\nu}}(t) \\ &= z_{\bar{\mu}}(0) - \frac{\eta}{2}\sum_{\hat{\nu}} K_{\bar{\mu}\hat{\nu}} \left(\sum_{t=0,\hat{\nu}_0}^{+\infty}\left[\left(1 - \frac{\eta}{2}K\right)^t\right]_{\hat{\nu}\hat{\nu}_0}\right) \varepsilon_{\hat{\nu}_0}(0) \\ &= z_{\bar{\mu}}(0) - \frac{\eta}{2}\sum_{\hat{\nu}} K_{\bar{\mu}\hat{\nu}} \sum_{\hat{\nu}_0}\left[\left(1 - \left(1 - \frac{\eta}{2}K\right)\right)^{-1}\right]_{\hat{\nu}\hat{\nu}_0} \varepsilon_{\hat{\nu}_0}(0) \\ &= z_{\bar{\mu}}(0) - \sum_{\hat{\nu}} K_{\bar{\mu}\hat{\nu}} \sum_{\hat{\nu}_0}\left[K^{i,-1}\right]_{\hat{\nu}\hat{\nu}_0} \varepsilon_{\hat{\nu}_0}(0) \ . \end{aligned} \tag{156}$$

Moreover, we could compute the prediction by firstly defining the kernel inverse. We define

$$\sum_{\hat{\mu}\in\mathcal{A}\times\mathbb{Z}_2\times\mathcal{H}} \tilde{K}^{\hat{\mu}_1\hat{\mu}_2}\tilde{K}_{\hat{\mu}_2\hat{\mu}_3} = \delta_{\hat{\mu}_3}^{\hat{\mu}_1} \ , \tag{157}$$

so we get

$$z_{\bar{\mu}}(\infty) = z_{\bar{\mu}}(0) - \sum_{\hat{\mu}_1,\hat{\mu}_2} \tilde{K}^{\hat{\mu}_1\hat{\mu}_2} K_{\bar{\mu}\hat{\mu}_1}\varepsilon_{\hat{\mu}_2}(0) \ . \tag{158}$$

### C. Representation learning

Now we start to develop our quantum representation learning theory at the dQNTK order. We make a quadratic expansion:

$$\begin{aligned} U(\theta) \to \prod_\ell W_\ell \exp\left(i\delta\varphi_\ell X_\ell\right) &= \prod_\ell W_\ell + i\delta\sum_\ell \varphi_\ell W_{\ell,-} W_\ell X_\ell W_{\ell,+} \\ &- \frac{1}{2}\delta^2 \sum_{\ell_1,\ell_2} \varphi_{\ell_1}\varphi_{\ell_2} \begin{cases} W_{\ell_1,-}W_{\ell_1}X_{\ell_1}W_{\ell_1,\ell_2}W_{\ell_2}X_{\ell_2}W_{\ell_2,+} & : \ell_1 \le \ell_2 \\ W_{\ell_2,-}W_{\ell_2}X_{\ell_2}W_{\ell_2,\ell_1}W_{\ell_1}X_{\ell_1}W_{\ell_1,+} & : \ell_1 > \ell_2 \end{cases} \ . \end{aligned} \tag{159}$$

Now, we could call

$$\begin{aligned} G_\ell^{(1)} &\equiv W_\ell X_\ell \ , \\ G_{\ell_1\ell_2}^{(2)} &\equiv \begin{cases} W_{\ell_1,-}W_{\ell_1}X_{\ell_1}W_{\ell_1,\ell_2}W_{\ell_2}X_{\ell_2}W_{\ell_2,+} & : \ell_1 \le \ell_2 \\ W_{\ell_2,-}W_{\ell_2}X_{\ell_2}W_{\ell_2,\ell_1}W_{\ell_1}X_{\ell_1}W_{\ell_1,+} & : \ell_1 > \ell_2 \end{cases} \ . \end{aligned} \tag{160}$$

So the model will look like

$$z_{j;\delta} = \sum_i (\prod_\ell W_\ell)_{ji}\phi_i(\mathbf{x}_\delta) + i\delta\sum_{i,\ell}\varphi_\ell G_{\ell,ji}^{(1)}\phi_i(\mathbf{x}_\delta) - \frac{1}{2}\delta^2\sum_{i,\ell_1,\ell_2}\varphi_{\ell_1}\varphi_{\ell_2}G_{\ell_1\ell_2,ji}^{(2)}\phi_i(\mathbf{x}_\delta) \ . \tag{161}$$

So we could write down the derivatives

$$\frac{dz_{j;\delta}}{d\varphi_\ell} = i\delta \sum_i G^{(1)}_{\ell,ji}\phi_i(\mathbf{x}_\delta) - \frac{1}{2}\delta^2 \sum_{i,\ell'} \varphi_{\ell'}(G^{(2)}_{\ell\ell',ji} + G^{(2)}_{\ell'\ell,ji})\phi_i(\mathbf{x}_\delta) \,. \tag{162}$$

Moreover, we have

$$\begin{aligned}
đz_{j;\delta} &= i\delta \sum_i \sum_\ell đ\varphi_\ell G^{(1)}_{\ell,ji}\phi_i(\mathbf{x}_\delta) - \frac{1}{2}\delta^2 \sum_i \sum_{\ell_1,\ell_2} đ\varphi_{\ell_1}\varphi_{\ell_2} G^{(2)}_{\ell_1\ell_2,ji}\phi_i(\mathbf{x}_\delta) \\
&\quad - \frac{1}{2}\delta^2 \sum_i \sum_{\ell_1,\ell_2} \varphi_{\ell_1} đ\varphi_{\ell_2} G^{(2)}_{\ell_1\ell_2,ji}\phi_i(\mathbf{x}_\delta) - \frac{1}{2}\delta^2 \sum_i \sum_{\ell_1,\ell_2} đ\varphi_{\ell_1} đ\varphi_{\ell_2} G^{(2)}_{\ell_1\ell_2,ji}\phi_i(\mathbf{x}_\delta) \,.
\end{aligned} \tag{163}$$

The leading order piece is exactly the effective feature map. A more compact version is,

$$đz_{j;\delta} = \sum_\ell \frac{dz_{j;\delta}}{d\varphi_\ell} đ\varphi_\ell - \frac{1}{2}\delta^2 \sum_i \sum_{\ell_1,\ell_2} đ\varphi_{\ell_1} đ\varphi_{\ell_2} G^{(2)}_{\ell_1\ell_2,ji}\phi_i(\mathbf{x}_\delta) \,. \tag{164}$$

Firstly we need to compute the effective kernel, we have

$$\begin{aligned}
\frac{dz_{j;\delta}}{d\varphi_\ell} &= i\delta \sum_i G^{(1)}_{\ell,ji}\phi_i(\mathbf{x}_\delta) - \frac{1}{2}\delta^2 \sum_{i,\ell'} \varphi_{\ell'}\left(G^{(2)}_{\ell\ell',ji} + G^{(2)}_{\ell'\ell,ji}\right)\phi_i(\mathbf{x}_\delta) \,, \\
\frac{dz_{i';\tilde\alpha}}{d\varphi_\ell} &= i\delta \sum_i G^{(1)}_{\ell,i'i}\phi_i(\mathbf{x}_{\tilde\alpha}) - \frac{1}{2}\delta^2 \sum_{i,\ell'} \varphi_{\ell'}\left(G^{(2)}_{\ell\ell',i'i} + G^{(2)}_{\ell'\ell,i'i}\right)\phi_i(\mathbf{x}_{\tilde\alpha}) \,, \\
\frac{dz^*_{i';\tilde\alpha}}{d\varphi_\ell} &= -i\delta \sum_i \phi^*_i(\mathbf{x}_{\tilde\alpha}) G^{\dagger,(1)}_{\ell,ii'} - \frac{1}{2}\delta^2 \sum_{i,\ell'} \varphi_{\ell'} \phi^*_i(\mathbf{x}_{\tilde\alpha})\left(G^{\dagger,(2)}_{\ell\ell',ii'} + G^{\dagger,(2)}_{\ell'\ell,ii'}\right) \,.
\end{aligned} \tag{165}$$

Thus, we define

$$K^{E,\dot\alpha\dot\beta,ii'}_{\delta,\tilde\alpha} = \begin{pmatrix} K^{E,+,ii'}_{\delta,\tilde\alpha} & K^{E,-,ii'}_{\delta,\tilde\alpha} \\ K^{*,E,-,ii'}_{\delta,\tilde\alpha} & K^{*,E,+,ii'}_{\delta,\tilde\alpha} \end{pmatrix}(\varphi) = \begin{pmatrix} \sum_\ell \frac{dz_{i;\delta}}{d\varphi_\ell} \frac{dz^*_{i';\tilde\alpha}}{d\varphi_\ell} & \sum_\ell \frac{dz_{i;\delta}}{d\varphi_\ell} \frac{dz_{i';\tilde\alpha}}{d\varphi_\ell} \\ \sum_\ell \frac{dz^*_{i;\delta}}{d\varphi_\ell} \frac{dz^*_{i';\tilde\alpha}}{d\varphi_\ell} & \sum_\ell \frac{dz^*_{i;\delta}}{d\varphi_\ell} \frac{dz_{i';\tilde\alpha}}{d\varphi_\ell} \end{pmatrix}, \tag{166}$$

and

$$K^{E,\dot\alpha\dot\beta,ii'}_{\delta,\tilde\alpha}(0) = K^{\dot\alpha\dot\beta,ii'}_{\delta,\tilde\alpha} + K^{\Delta,\dot\alpha\dot\beta,ii'}_{\delta,\tilde\alpha} = \begin{pmatrix} K^{+,ii'}_{\delta,\tilde\alpha} & K^{-,ii'}_{\delta,\tilde\alpha} \\ K^{*,-,ii'}_{\delta,\tilde\alpha} & K^{*,+,j}_{\delta,\tilde\alpha} \end{pmatrix} + \begin{pmatrix} K^{\Delta,+,ii'}_{\delta,\tilde\alpha} & K^{\Delta,-,ii'}_{\delta,\tilde\alpha} \\ K^{*,\Delta,-,ii'}_{\delta,\tilde\alpha} & K^{*,\Delta,+,ii'}_{\delta,\tilde\alpha} \end{pmatrix} \,. \tag{167}$$

We find

$$\begin{aligned}
K^{\Delta,+,ii'}_{\delta,\tilde\alpha} &= \frac{i}{2}\delta^3 \sum_{j,\ell',i''} \varphi_{\ell'}(0)\left(G^{(2)}_{\ell\ell',ij} + G^{(2)}_{\ell'\ell,ij}\right) G^{\dagger,(1)}_{\ell,i''i'} \phi_j(\mathbf{x}_\delta) \phi^*_{i''}(\mathbf{x}_{\tilde\alpha}) \\
&\quad - \frac{i}{2}\delta^3 \sum_{j,\ell',i''} \varphi_{\ell'}(0)\left(G^{\dagger,(2)}_{\ell\ell',i''i'} + G^{\dagger,(2)}_{\ell'\ell,i''i'}\right) G^{(1)}_{\ell,ij} \phi_j(\mathbf{x}_\delta) \phi^*_{i''}(\mathbf{x}_{\tilde\alpha}) \,, \\
K^{\Delta,-,ii'}_{\delta,\tilde\alpha} &= -\frac{i}{2}\delta^3 \sum_{j,\ell',i''} \varphi_{\ell'}(0)\left(G^{(2)}_{\ell\ell',ij} + G^{(2)}_{\ell'\ell,ij}\right) G^{(1)}_{\ell,i'i''} \phi_j(\mathbf{x}_\delta) \phi_{i''}(\mathbf{x}_{\tilde\alpha}) \\
&\quad - \frac{i}{2}\delta^3 \sum_{j,\ell',i''} \varphi_{\ell'}(0)\left(G^{(2)}_{\ell\ell',i'i''} + G^{(2)}_{\ell'\ell,i'i''}\right) G^{(1)}_{\ell,ij} \phi_j(\mathbf{x}_\delta) \phi_{i''}(\mathbf{x}_{\tilde\alpha}) \,.
\end{aligned} \tag{168}$$

We could also write it in the bracket fashion

$$\begin{aligned}
K^{\Delta,+,ii'}_{\delta,\tilde\alpha} &= \frac{i}{2}\delta^3 \sum_{\ell'} \varphi_{\ell'}(0)\begin{pmatrix} \langle i|\left(G^{(2)}_{\ell\ell'} + G^{(2)}_{\ell'\ell}\right)|\phi(\mathbf{x}_\delta)\rangle \langle \phi(\mathbf{x}_{\tilde\alpha})| G^{\dagger,(1)}_\ell |i'\rangle \\ -\langle \phi(\mathbf{x}_{\tilde\alpha})|\left(G^{\dagger,(2)}_{\ell\ell'} + G^{\dagger,(2)}_{\ell'\ell}\right)|i'\rangle \langle i| G^{(1)}_\ell |\phi(\mathbf{x}_\delta)\rangle \end{pmatrix}, \\
K^{\Delta,-,ii'}_{\delta,\tilde\alpha} &= -\frac{i}{2}\delta^3 \sum_{\ell'} \varphi_{\ell'}(0)\begin{pmatrix} \langle i|\left(G^{(2)}_{\ell\ell'} + G^{(2)}_{\ell'\ell}\right)|\phi(\mathbf{x}_\delta)\rangle \langle i'| G^{(1)}_\ell |\phi(\mathbf{x}_{\tilde\alpha})\rangle \\ +\langle i'|\left(G^{(2)}_{\ell\ell'} + G^{(2)}_{\ell'\ell}\right)|\phi_{i''}(\mathbf{x}_{\tilde\alpha})\rangle \langle i| G^{(1)}_\ell |\phi(\mathbf{x}_\delta)\rangle \end{pmatrix}.
\end{aligned} \tag{169}$$

Now we could write down the prediction on the training set. We have

$$\varepsilon_{\hat{\mu}}(t) = \varepsilon_{\hat{\mu}}^F(t) + \varepsilon_{\hat{\mu}}^I(t) , \tag{170}$$

where

$$\varepsilon_{\hat{\mu}}^F(t) = \sum_{\hat{\mu}_1} U_{\hat{\mu}\hat{\mu}_1}(t)\varepsilon_{\hat{\mu}_1}(0) ,$$

$$U_{\hat{\mu}_1\hat{\mu}_2}(t) = \left[\left(1 - \frac{\eta}{2}K\right)^t\right]_{\hat{\mu}_1\hat{\mu}_2} , \tag{171}$$

and

$$\varepsilon_{\hat{\mu}}^I(t) = \left(-\frac{\eta}{2}\sum_{s=0}^{t-1}\left(1 - \frac{\eta}{2}K\right)^{t-1-s} K^{\Delta} \left(1 - \frac{\eta}{2}K\right)^s \varepsilon(0)\right)_{\hat{\mu}} . \tag{172}$$

It is the compact matrix product form in the space $\mathcal{A} \times \mathbb{Z}_2 \times \mathcal{H}$. Moreover, we have

$$\left\|\varepsilon^I(t)\right\| \leq \frac{\eta}{2}t\left\|1 - \frac{\eta}{2}K\right\|^{t-1} \left\|K^{\Delta}\right\| \|\varepsilon(0)\| . \tag{173}$$

Finally, we could discuss the asymptotic convergence. We could start by computing the meta-kernel. We notice that

$$\begin{aligned}
\check{d}z_{i;\delta} = & -\frac{\eta}{2}\sum_{\ell,i',\tilde{\alpha}} \varepsilon_{i;\tilde{\alpha}} \frac{dz_{i;\delta}}{d\varphi_\ell} \frac{dz^*_{i';\tilde{\alpha}}}{d\varphi_\ell} - \frac{\eta}{2}\sum_{\ell,i',\tilde{\alpha}} \varepsilon^*_{i;\tilde{\alpha}} \frac{dz_{i;\delta}}{d\varphi_\ell} \frac{dz_{i';\tilde{\alpha}}}{d\varphi_\ell} \\
& + \left(\frac{\eta}{2}\right)^2 \sum_{\ell_1,\ell_2} \frac{d^2 z_{i;\delta}}{d\varphi_{\ell_1} d\varphi_{\ell_2}} \left(\sum_{\tilde{\alpha}_1,i_1} \varepsilon_{i_1;\tilde{\alpha}_1} \frac{dz^*_{i_1;\tilde{\alpha}_1}}{d\varphi_{\ell_1}} + \sum_{\tilde{\alpha}_1,i_1} \varepsilon^*_{i_1;\tilde{\alpha}_1} \frac{dz_{i_1;\tilde{\alpha}_1}}{d\varphi_{\ell_2}}\right) \left(\sum_{\tilde{\alpha}_2,i_2} \varepsilon_{i_2;\tilde{\alpha}_2} \frac{dz^*_{i_2;\tilde{\alpha}_2}}{d\varphi_{\ell_2}} + \sum_{\tilde{\alpha}_2,i_2} \varepsilon^*_{i_2;\tilde{\alpha}_2} \frac{dz_{i_2;\tilde{\alpha}_2}}{d\varphi_{\ell_2}}\right) ,
\end{aligned} \tag{174}$$

We have

$$\begin{aligned}
& \sum_{\ell_1,\ell_2} \frac{d^2 z_{i;\delta}}{d\varphi_{\ell_1} d\varphi_{\ell_2}} \left(\sum_{\tilde{\alpha}_1,i_1} \varepsilon_{i_1;\tilde{\alpha}_1} \frac{dz^*_{i_1;\tilde{\alpha}_1}}{d\varphi_{\ell_1}} + \sum_{\tilde{\alpha}_1,i_1} \varepsilon^*_{i_1;\tilde{\alpha}_1} \frac{dz_{i_1;\tilde{\alpha}_1}}{d\varphi_{\ell_1}}\right) \left(\sum_{\tilde{\alpha}_2,i_2} \varepsilon_{i_2;\tilde{\alpha}_2} \frac{dz^*_{i_2;\tilde{\alpha}_2}}{d\varphi_{\ell_2}} + \sum_{\tilde{\alpha}_2,i_2} \varepsilon^*_{i_2;\tilde{\alpha}_2} \frac{dz_{i_2;\tilde{\alpha}_2}}{d\varphi_{\ell_2}}\right) \\
& = \sum_{\ell_1,\ell_2,\tilde{\alpha}_1,i_1,\tilde{\alpha}_2,i_2,\dot{\alpha}_1,\dot{\alpha}_2} \frac{d^2 z_{i;\delta}}{d\varphi_{\ell_1} d\varphi_{\ell_2}} \frac{dz^{1-\dot{\alpha}_1}_{i_1;\tilde{\alpha}_1}}{d\varphi_{\ell_1}} \frac{dz^{1-\dot{\alpha}_2}_{i_2;\tilde{\alpha}_2}}{d\varphi_{\ell_2}} \varepsilon^{\dot{\alpha}_1}_{i_1;\tilde{\alpha}_1} \varepsilon^{\dot{\alpha}_2}_{i_2;\tilde{\alpha}_2} .
\end{aligned} \tag{175}$$

So we define

$$\mu^{ii_1 i_2;\dot{\beta}\dot{\alpha}_1\dot{\alpha}_2}_{\delta\tilde{\alpha}_1\tilde{\alpha}_2} = \mu_{\bar{\mu}\hat{\mu}_1\hat{\mu}_2} = \sum_{\ell_1,\ell_2} \frac{d^2 z^{\dot{\beta}}_{i;\delta}}{d\varphi_{\ell_1} d\varphi_{\ell_2}} \left(\frac{dz^{1-\dot{\alpha}_1}_{i_1;\tilde{\alpha}_1}}{d\varphi_{\ell_1}} \frac{dz^{1-\dot{\alpha}_2}_{i_2;\tilde{\alpha}_2}}{d\varphi_{\ell_2}}\right)\bigg|_{\varphi=0} , \tag{176}$$

in the leading order. Moreover, in general we have,

$$\mu^{i_0 i_1 i_2;\dot{\alpha}_0\dot{\alpha}_1\dot{\alpha}_2}_{\delta_0\delta_1\delta_2} = \mu_{\bar{\mu}_0\bar{\mu}_1\bar{\mu}_2} = \sum_{\ell_1,\ell_2} \frac{d^2 z^{\dot{\alpha}_0}_{i_0;\delta_0}}{d\varphi_{\ell_1} d\varphi_{\ell_2}} \left(\frac{dz^{1-\dot{\alpha}_1}_{i_1;\delta_1}}{d\varphi_{\ell_1}} \frac{dz^{1-\dot{\alpha}_2}_{i_2;\delta_2}}{d\varphi_{\ell_2}}\right)\bigg|_{\varphi=0} . \tag{177}$$

So the asymptotic convergence is given by

$$\begin{aligned}
z_{\bar{\mu}}(\infty) = & z_{\bar{\mu}}(0) - \sum_{\hat{\mu}_1,\hat{\mu}_2} K_{\bar{\mu}\hat{\mu}_1} \tilde{K}^{\hat{\mu}_1\hat{\mu}_2} \varepsilon_{\hat{\mu}_2}(0) \\
& + \sum_{\hat{\mu}_1,\hat{\mu}_2,\hat{\mu}_3,\hat{\mu}_4} \left[\mu_{\hat{\mu}_1\bar{\mu}\hat{\mu}_2} - \sum_{\hat{\mu}_5,\hat{\mu}_6} K_{\bar{\mu}\hat{\mu}_5} \tilde{K}^{\hat{\mu}_5\hat{\mu}_6} \mu_{\hat{\mu}_1\hat{\mu}_6\hat{\mu}_2}\right] Z_A^{\hat{\mu}_1\hat{\mu}_2\hat{\mu}_3\hat{\mu}_4} \varepsilon_{\hat{\mu}_3}(0)\varepsilon_{\hat{\mu}_4}(0) \\
& + \sum_{\hat{\mu}_1,\hat{\mu}_2,\hat{\mu}_3,\hat{\mu}_4} \left[\mu_{\bar{\mu}\hat{\mu}_1\hat{\mu}_2} - \sum_{\hat{\mu}_5,\hat{\mu}_6} K_{\bar{\mu}\hat{\mu}_5} \tilde{K}^{\hat{\mu}_5\hat{\mu}_6} \mu_{\hat{\mu}_6\hat{\mu}_1\hat{\mu}_2}\right] Z_B^{\hat{\mu}_1\hat{\mu}_2\hat{\mu}_3\hat{\mu}_4} \varepsilon_{\hat{\mu}_3}(0)\varepsilon_{\hat{\mu}_4}(0) ,
\end{aligned} \tag{178}$$

where the algorithm projector $Z_{A,B}$s are

$$Z_A^{\hat{\mu}_1\hat{\mu}_2\hat{\mu}_3\hat{\mu}_4} \equiv \tilde{K}^{\hat{\mu}_1\hat{\mu}_3}\tilde{K}^{\hat{\mu}_2\hat{\mu}_4} - \sum_{\hat{\mu}_5}\tilde{K}^{\hat{\mu}_2\hat{\mu}_5}X_{\parallel}^{\hat{\mu}_1\hat{\mu}_5\hat{\mu}_3\hat{\mu}_4},$$
$$Z_B^{\hat{\mu}_1\hat{\mu}_2\hat{\mu}_3\hat{\mu}_4} \equiv \tilde{K}^{\hat{\mu}_1\hat{\mu}_3}\tilde{K}^{\hat{\mu}_2\hat{\mu}_4} - \sum_{\hat{\mu}_5}\tilde{K}^{\hat{\mu}_2\hat{\mu}_5}X_{\parallel}^{\hat{\mu}_1\hat{\mu}_5\hat{\mu}_3\hat{\mu}_4} + \frac{\eta}{4}X_{\parallel}^{\hat{\mu}_1\hat{\mu}_2\hat{\mu}_3\hat{\mu}_4}, \quad (179)$$

and

$$X_{\parallel}^{\hat{\mu}_1\hat{\mu}_2\hat{\mu}_3\hat{\mu}_4} = \sum_{s=0}^{\infty}\left[(1-\frac{\eta}{2}K)^s\right]_{\hat{\mu}_1\hat{\mu}_3}\left[(1-\frac{\eta}{2}K)^s\right]_{\hat{\mu}_2\hat{\mu}_4}, \quad (180)$$

which is defined implicitly as

$$\delta_{\hat{\mu}_5}^{\hat{\mu}_1}\delta_{\hat{\mu}_6}^{\hat{\mu}_2} = \sum_{\hat{\mu}_3,\hat{\mu}_4}X_{\parallel}^{\hat{\mu}_1\hat{\mu}_2\hat{\mu}_3\hat{\mu}_4}\left(\tilde{K}_{\hat{\mu}_3\hat{\mu}_5}\delta_{\hat{\mu}_4\hat{\mu}_6} + \delta_{\hat{\mu}_3\hat{\mu}_5}\tilde{K}_{\hat{\mu}_4\hat{\mu}_6} - \frac{\eta}{2}\tilde{K}_{\hat{\mu}_3\hat{\mu}_5}\tilde{K}_{\hat{\mu}_4\hat{\mu}_6}\right). \quad (181)$$

Now we write down all components of $\mu$. We have

$$\frac{d^2 z_{i_0;\delta_0}}{d\varphi_{\ell_1}d\varphi_{\ell_2}} = -\frac{1}{2}\delta^2\sum_i\left(G^{(2)}_{\ell_1\ell_2,i_0i} + G^{(2)}_{\ell_2\ell_1,i_0i}\right)\phi_i(\mathbf{x}_{\delta_0}),$$
$$\frac{d^2 z^*_{i_0;\delta_0}}{d\varphi_{\ell_1}d\varphi_{\ell_2}} = -\frac{1}{2}\delta^2\sum_i\phi_i^*(\mathbf{x}_{\delta_0})\left(G^{\dagger,(2)}_{\ell_1\ell_2,ii_0} + G^{\dagger,(2)}_{\ell_2\ell_1,ii_0}\right),$$
$$\frac{dz^*_{i_1;\delta_1}}{d\varphi_{\ell_1}} = -i\delta\sum_{i'}\phi_{i'}^*(\mathbf{x}_{\delta_1})G^{\dagger,(1)}_{\ell_1,i'i_1}, \qquad \frac{dz_{i_1;\delta_1}}{d\varphi_{\ell_1}} = i\delta\sum_{i'}G^{(1)}_{\ell_1,i_1i'}\phi_{i'}(\mathbf{x}_{\delta_1}),$$
$$\frac{dz^*_{i_2;\delta_2}}{d\varphi_{\ell_2}} = -i\delta\sum_{i''}\phi_{i''}^*(\mathbf{x}_{\delta_2})G^{\dagger,(1)}_{\ell_2,i''i_2}, \qquad \frac{dz_{i_2;\delta_2}}{d\varphi_{\ell_2}} = i\delta\sum_{i''}G^{(1)}_{\ell_2,i_2i''}\phi_{i''}(\mathbf{x}_{\delta_2}). \quad (182)$$

So

$$\mu^{i_0i_1i_2;000}_{\delta_0\delta_1\delta_2} = \frac{1}{2}\delta^4\sum_{i,i',i'',\ell_1,\ell_2}\left(G^{(2)}_{\ell_1\ell_2,i_0i} + G^{(2)}_{\ell_2\ell_1,i_0i}\right)G^{\dagger,(1)}_{\ell_1,i'i_1}G^{\dagger,(2)}_{\ell_2,i''i_2}\phi_i(\mathbf{x}_{\delta_0})\phi_{i'}^*(\mathbf{x}_{\delta_1})\phi_{i''}^*(\mathbf{x}_{\delta_2}),$$
$$\mu^{i_0i_1i_2;001}_{\delta_0\delta_1\delta_2} = -\frac{1}{2}\delta^4\sum_{i,i',i'',\ell_1,\ell_2}\left(G^{(2)}_{\ell_1\ell_2,i_0i} + G^{(2)}_{\ell_2\ell_1,i_0i}\right)G^{\dagger,(1)}_{\ell_1,i'i_1}G^{(2)}_{\ell_2,i_2i''}\phi_i(\mathbf{x}_{\delta_0})\phi_{i'}^*(\mathbf{x}_{\delta_1})\phi_{i''}(\mathbf{x}_{\delta_2}),$$
$$\mu^{i_0i_1i_2;010}_{\delta_0\delta_1\delta_2} = -\frac{1}{2}\delta^4\sum_{i,i',i'',\ell_1,\ell_2}\left(G^{(2)}_{\ell_1\ell_2,i_0i} + G^{(2)}_{\ell_2\ell_1,i_0i}\right)G^{(1)}_{\ell_1,i_1i'}G^{\dagger,(2)}_{\ell_2,i''i_2}\phi_i(\mathbf{x}_{\delta_0})\phi_{i'}(\mathbf{x}_{\delta_1})\phi_{i''}^*(\mathbf{x}_{\delta_2}),$$
$$\mu^{i_0i_1i_2;011}_{\delta_0\delta_1\delta_2} = \frac{1}{2}\delta^4\sum_{i,i',i'',\ell_1,\ell_2}\left(G^{(2)}_{\ell_1\ell_2,i_0i} + G^{(2)}_{\ell_2\ell_1,i_0i}\right)G^{(1)}_{\ell_1,i_1i'}G^{(2)}_{\ell_2,i_2i''}\phi_i(\mathbf{x}_{\delta_0})\phi_{i'}(\mathbf{x}_{\delta_1})\phi_{i''}(\mathbf{x}_{\delta_2}),$$
$$\mu^{i_0i_1i_2;100}_{\delta_0\delta_1\delta_2} = \frac{1}{2}\delta^4\sum_{i,i',i'',\ell_1,\ell_2}\left(G^{\dagger,(2)}_{\ell_1\ell_2,ii_0} + G^{\dagger,(2)}_{\ell_2\ell_1,ii_0}\right)G^{\dagger,(1)}_{\ell_1,i'i_1}G^{\dagger,(2)}_{\ell_2,i''i_2}\phi_i^*(\mathbf{x}_{\delta_0})\phi_{i'}^*(\mathbf{x}_{\delta_1})\phi_{i''}^*(\mathbf{x}_{\delta_2}),$$
$$\mu^{i_0i_1i_2;101}_{\delta_0\delta_1\delta_2} = -\frac{1}{2}\delta^4\sum_{i,i',i'',\ell_1,\ell_2}\left(G^{\dagger,(2)}_{\ell_1\ell_2,ii_0} + G^{\dagger,(2)}_{\ell_2\ell_1,ii_0}\right)G^{\dagger,(1)}_{\ell_1,i'i_1}G^{(2)}_{\ell_2,i_2i''}\phi_i^*(\mathbf{x}_{\delta_0})\phi_{i'}^*(\mathbf{x}_{\delta_1})\phi_{i''}(\mathbf{x}_{\delta_2}),$$
$$\mu^{i_0i_1i_2;110}_{\delta_0\delta_1\delta_2} = -\frac{1}{2}\delta^4\sum_{i,i',i'',\ell_1,\ell_2}\left(G^{\dagger,(2)}_{\ell_1\ell_2,ii_0} + G^{\dagger,(2)}_{\ell_2\ell_1,ii_0}\right)G^{(1)}_{\ell_1,i_1i'}G^{\dagger,(2)}_{\ell_2,i''i_2}\phi_i^*(\mathbf{x}_{\delta_0})\phi_{i'}(\mathbf{x}_{\delta_1})\phi_{i''}^*(\mathbf{x}_{\delta_2}),$$
$$\mu^{i_0i_1i_2;111}_{\delta_0\delta_1\delta_2} = \frac{1}{2}\delta^4\sum_{i,i',i'',\ell_1,\ell_2}\left(G^{\dagger,(2)}_{\ell_1\ell_2,ii_0} + G^{\dagger,(2)}_{\ell_2\ell_1,ii_0}\right)G^{(1)}_{\ell_1,i_1i'}G^{(2)}_{\ell_2,i_2i''}\phi_i^*(\mathbf{x}_{\delta_0})\phi_{i'}(\mathbf{x}_{\delta_1})\phi_{i''}(\mathbf{x}_{\delta_2}). \quad (183)$$

One could also write them in the bracket notations.

$$
\begin{aligned}
\mu^{i_0 i_1 i_2;000}_{\delta_0\delta_1\delta_2} &= \frac{1}{2}\delta^4 \sum_{\ell_1,\ell_2} \langle i_0| \left(G^{(2)}_{\ell_1\ell_2} + G^{(2)}_{\ell_2\ell_1}\right) |\phi(\mathbf{x}_{\delta_0})\rangle \langle \phi(\mathbf{x}_{\delta_1})| G^{\dagger,(1)}_{\ell_1} |i_1\rangle \langle \phi(\mathbf{x}_{\delta_2})| G^{\dagger,(1)}_{\ell_2} |i_2\rangle, \\
\mu^{i_0 i_1 i_2;001}_{\delta_0\delta_1\delta_2} &= -\frac{1}{2}\delta^4 \sum_{\ell_1,\ell_2} \langle i_0| \left(G^{(2)}_{\ell_1\ell_2} + G^{(2)}_{\ell_2\ell_1}\right) |\phi(\mathbf{x}_{\delta_0})\rangle \langle \phi(\mathbf{x}_{\delta_1})| G^{\dagger,(1)}_{\ell_1} |i_1\rangle \langle i_2| G^{(1)}_{\ell_2} |\phi(\mathbf{x}_{\delta_2})\rangle, \\
\mu^{i_0 i_1 i_2;010}_{\delta_0\delta_1\delta_2} &= -\frac{1}{2}\delta^4 \sum_{\ell_1,\ell_2} \langle i_0| \left(G^{(2)}_{\ell_1\ell_2} + G^{(2)}_{\ell_2\ell_1}\right) |\phi(\mathbf{x}_{\delta_0})\rangle \langle i_1| G^{(1)}_{\ell_1} |\phi(\mathbf{x}_{\delta_1})\rangle \langle \phi(\mathbf{x}_{\delta_2})| G^{\dagger,(1)}_{\ell_2} |i_2\rangle, \\
\mu^{i_0 i_1 i_2;011}_{\delta_0\delta_1\delta_2} &= \frac{1}{2}\delta^4 \sum_{\ell_1,\ell_2} \langle i_0| \left(G^{(2)}_{\ell_1\ell_2} + G^{(2)}_{\ell_2\ell_1}\right) |\phi(\mathbf{x}_{\delta_0})\rangle \langle i_1| G^{(1)}_{\ell_1} |\phi(\mathbf{x}_{\delta_1})\rangle \langle i_2| G^{(1)}_{\ell_2} |\phi(\mathbf{x}_{\delta_2})\rangle, \\
\mu^{i_0 i_1 i_2;100}_{\delta_0\delta_1\delta_2} &= \frac{1}{2}\delta^4 \sum_{\ell_1,\ell_2} \langle \phi(\mathbf{x}_{\delta_0})| \left(G^{\dagger,(2)}_{\ell_1\ell_2} + G^{\dagger,(2)}_{\ell_2\ell_1}\right) |i_0\rangle \langle \phi(\mathbf{x}_{\delta_1})| G^{\dagger,(1)}_{\ell_1} |i_1\rangle \langle \phi(\mathbf{x}_{\delta_2})| G^{\dagger,(1)}_{\ell_2} |i_2\rangle, \\
\mu^{i_0 i_1 i_2;101}_{\delta_0\delta_1\delta_2} &= -\frac{1}{2}\delta^4 \sum_{\ell_1,\ell_2} \langle \phi(\mathbf{x}_{\delta_0})| \left(G^{\dagger,(2)}_{\ell_1\ell_2} + G^{\dagger,(2)}_{\ell_2\ell_1}\right) |i_0\rangle \langle \phi(\mathbf{x}_{\delta_1})| G^{\dagger,(1)}_{\ell_1} |i_1\rangle \langle i_2| G^{(1)}_{\ell_2} |\phi(\mathbf{x}_{\delta_2})\rangle, \\
\mu^{i_0 i_1 i_2;110}_{\delta_0\delta_1\delta_2} &= -\frac{1}{2}\delta^4 \sum_{\ell_1,\ell_2} \langle \phi(\mathbf{x}_{\delta_0})| \left(G^{\dagger,(2)}_{\ell_1\ell_2} + G^{\dagger,(2)}_{\ell_2\ell_1}\right) |i_0\rangle \langle i_1| G^{(1)}_{\ell_1} |\phi(\mathbf{x}_{\delta_1})\rangle \langle \phi(\mathbf{x}_{\delta_2})| G^{\dagger,(1)}_{\ell_2} |i_2\rangle, \\
\mu^{i_0 i_1 i_2;111}_{\delta_0\delta_1\delta_2} &= \frac{1}{2}\delta^4 \sum_{\ell_1,\ell_2} \langle \phi(\mathbf{x}_{\delta_0})| \left(G^{\dagger,(2)}_{\ell_1\ell_2} + G^{\dagger,(2)}_{\ell_2\ell_1}\right) |i_0\rangle \langle i_1| G^{(1)}_{\ell_1} |\phi(\mathbf{x}_{\delta_1})\rangle \langle i_2| G^{(1)}_{\ell_2} |\phi(\mathbf{x}_{\delta_2})\rangle.
\end{aligned}
\tag{184}
$$

### D. Reading the amplitude

Here, we review the amplitude protocol for realizing the evaluation of the inner product of two states: $\langle x|y\rangle$ [6]. The protocol is very similar to the celebrated Hadamard test. For a given pair of quantum states $|x\rangle$ and $|y\rangle$, we need to get access to the state

$$|\varphi\rangle = \frac{1}{\sqrt{2}}(|0\rangle|x\rangle + |1\rangle|y\rangle). \tag{185}$$

The first qubit is serving as an ancillary qubit. Applying the Hadamard gate to the first qubit, we get

$$|\varphi\rangle \to \frac{1}{2}(|0\rangle(|x\rangle + |y\rangle) + |1\rangle(|x\rangle - |y\rangle)). \tag{186}$$

Now we could measure the probability to obtain $|0\rangle$ in the ancillary system. We have

$$p = \frac{1}{2}(1 + \text{Re}(\langle x|y\rangle)). \tag{187}$$

Thus, we could use the probability to estimate the real part of the inner product. Moreover, we could add a phase rotation to $|\varphi\rangle$ to get

$$|\varphi\rangle = \frac{1}{\sqrt{2}}(|0\rangle|x\rangle - i|1\rangle|y\rangle). \tag{188}$$

Then, we apply the same Hadamard gate and measure the first qubit in the Pauli-$Z$ basis. We get the probability,

$$p = \frac{1}{2}(1 + \text{Im}(\langle x|y\rangle)), \tag{189}$$

to obtain $|0\rangle$. Similar to the Hadamard test where we need to get access to a controlled unitary acting on an arbitrary state, here we need to get access to the state $|\varphi\rangle$. The inner product evaluation operation has a statistical error coming from the measurement. If we measure $N$ times, the error scales as $1/\sqrt{N}$.

## V. SUPPRESSION OF NON-GAUSSIANITY IN THE LARGE-WIDTH LIMIT

In this section, we visit the statistics of hybrid quantum-classical neural networks. The model is defined as the following. First, we initialize the neural network by a quantum model,

$$z^Q_{1;\alpha;j_1} = \langle \phi_1(\mathbf{x}_\alpha) | U^{\dagger,1}(\theta^1) O^1_{j_1} U^1(\theta^1) | \phi_1(\mathbf{x}_\alpha) \rangle . \tag{190}$$

Here we use the notation $j_\omega$ to denote the index of the operator space $\mathcal{O}^\omega(\mathcal{H}^\omega)$, where $\omega \in [1, 2, \cdots, \Omega]$ is denoting the layer of the hybrid quantum-classical neural network. Here we are starting our first layer, so $\omega = 1$. We use $z^Q_{\omega;\alpha;j_\omega}$ to denote our quantum model output. We also use

$$U^\omega(\theta^\omega) = \prod_{\ell_\omega=1}^{L_\omega} W^\omega_{\ell_\omega} \exp\left(i\theta^\omega_{\ell_\omega} X^\omega_{\ell_\omega}\right) , \tag{191}$$

to denote our $\omega$th quantum ansatz, and $\phi_\omega$ is used to denote the $\omega$th feature map. In each layer, after the quantum network, we connect it with a classical neural network by

$$w^C_{\omega;\alpha;j^C_\omega} = \sigma^\omega_{j^C_\omega}\left(\sum_{j_\omega=1}^{\dim \mathcal{O}^\omega(\mathcal{H}^\omega)} W^\omega_{j^C_\omega, j_\omega} z^Q_{\omega;\alpha;j_\omega} + b^\omega_{j^C_\omega}\right) \equiv \sigma^\omega_{j^C_\omega}\left(z^C_{\omega;\alpha;j^C_\omega}\right) . \tag{192}$$

Here, we call

$$z^C_{\omega;\alpha;j^C_\omega} = \sum_{j_\omega=1}^{\dim \mathcal{O}^\omega(\mathcal{H}^\omega)} W^\omega_{j^C_\omega, j_\omega} z^Q_{\omega;\alpha;j_\omega} + b^\omega_{j^C_\omega} , \tag{193}$$

as classical preactivation in each layer, and we use the non-linear activation $\sigma$. At initialization, we will set all $W$s and $b$s distributed randomly from the following Gaussian statistics:

$$\mathbb{E}\left(W^\omega_{j^C_{1,\omega}, j_{1,\omega}} W^\omega_{j^C_{2,\omega}, j_{2,\omega}}\right) = \delta_{j^C_{1,\omega}, j^C_{2,\omega}} \delta_{j_{1,\omega}, j_{2,\omega}} \frac{C^\omega_W}{\dim \mathcal{O}^\omega(\mathcal{H}^\omega)} ,$$
$$\mathbb{E}\left(b^\omega_{j^C_{1,\omega}} b^\omega_{j^C_{2,\omega}}\right) = \delta_{j^C_{1,\omega}, j^C_{2,\omega}} C^\omega_b . \tag{194}$$

Moreover, after the classical network in each layer, we can move to the next quantum layer by doing the following encoding,

$$z^Q_{\omega;\alpha;j_\omega} = \langle \phi_\omega(\mathbf{w}_{\omega-1;\alpha}) | U^{\dagger,\omega}(\theta^\omega) O^\omega_{j_\omega} U^\omega(\theta^\omega) | \phi_\omega(\mathbf{w}_{\omega-1;\alpha}) \rangle , \tag{195}$$

where we use the vector notation $\mathbf{w}_{\omega;\alpha} = (w_{\omega;\alpha})_{j^C_\omega}$. Moreover, we will assume that the initial distribution of quantum variational angles is given by a statistical ensemble. We denote all those ensemble averages as $\mathbb{E}$. One could, for instance, compute the two-point function as

$$\mathbb{E}\left(z^C_{\omega;\alpha;j^C_{1,\omega}} z^C_{\omega;\beta;j^C_{2,\omega}}\right) = \mathbb{E}\left(\left(\sum_{j_{1,\omega}=1}^{\dim \mathcal{O}^\omega(\mathcal{H}^\omega)} W^\omega_{j^C_{1,\omega} j_{1,\omega}} z^Q_{\omega;\alpha;j_{1,\omega}} + b^\omega_{j^C_{1,\omega}}\right)\left(\sum_{j_{2,\omega}=1}^{\dim \mathcal{O}^\omega(\mathcal{H}^\omega)} W^\omega_{j^C_{2,\omega} j_{2,\omega}} z^Q_{\omega;\beta;j_{2,\omega}} + b^\omega_{j^C_{2,\omega}}\right)\right)$$

$$= \sum_{j_{1,\omega}, j_{2,\omega}=1}^{\dim \mathcal{O}^\omega(\mathcal{H}^\omega)} \mathbb{E}\left(W^\omega_{j^C_{1,\omega} j_{1,\omega}} W^\omega_{j^C_{2,\omega} j_{2,\omega}}\right) \mathbb{E}\left(z^Q_{\omega;\alpha;j_{1,\omega}} z^Q_{\omega;\beta;j_{2,\omega}}\right) + \mathbb{E}\left(b^\omega_{j^C_{1,\omega}} b^\omega_{j^C_{2,\omega}}\right)$$

$$= \frac{C^\omega_W}{\dim \mathcal{O}^\omega(\mathcal{H}^\omega)} \sum_{j_{1,\omega}, j_{2,\omega}=1}^{\dim \mathcal{O}^\omega(\mathcal{H}^\omega)} \delta_{j^C_{1,\omega}, j^C_{2,\omega}} \delta_{j_{1,\omega}, j_{2,\omega}} \mathbb{E}(z^Q_{\omega;\alpha;j_{1,\omega}} z^Q_{\omega;\beta;j_{2,\omega}}) + \delta_{j^C_{1,\omega}, j^C_{2,\omega}} C^\omega_b$$

$$= \delta_{j^C_{1,\omega}, j^C_{2,\omega}} \left(\frac{C^\omega_W}{\dim \mathcal{O}^\omega(\mathcal{H}^\omega)} \sum_{j_\omega=1}^{\dim \mathcal{O}^\omega(\mathcal{H}^\omega)} \mathbb{E}(z^Q_{\omega;\alpha;j_\omega} z^Q_{\omega;\beta;j_\omega}) + C^\omega_b\right) . \tag{196}$$

In order to compute non-Gaussianities of preactivations, we start from the connected part of the two-point function, which is given by

$$\begin{aligned}
&\mathbb{E}_{\text{conn}}\left(z^C_{\omega;\alpha_1;j^C_{1,\omega}} z^C_{\omega;\alpha_2;j^C_{2,\omega}} z^C_{\omega;\alpha_3;j^C_{3,\omega}} z^C_{\omega;\alpha_4;j^C_{4,\omega}}\right) \\
&\equiv \mathbb{E}\left(z^C_{\omega;\alpha_1;j^C_{1,\omega}} z^C_{\omega;\alpha_2;j^C_{2,\omega}} z^C_{\omega;\alpha_3;j^C_{3,\omega}} z^C_{\omega;\alpha_4;j^C_{4,\omega}}\right) - \mathbb{E}\left(z^C_{\omega;\alpha_1;j^C_{1,\omega}} z^C_{\omega;\alpha_2;j^C_{2,\omega}}\right) \mathbb{E}\left(z^C_{\omega;\alpha_3;j^C_{3,\omega}} z^C_{\omega;\alpha_4;j^C_{4,\omega}}\right) \\
&- \mathbb{E}\left(z^C_{\omega;\alpha_1;j^C_{1,\omega}} z^C_{\omega;\alpha_3;j^C_{3,\omega}}\right) \mathbb{E}\left(z^C_{\omega;\alpha_2;j^C_{2,\omega}} z^C_{\omega;\alpha_4;j^C_{4,\omega}}\right) - \mathbb{E}\left(z^C_{\omega;\alpha_1;j^C_{1,\omega}} z^C_{\omega;\alpha_4;j^C_{4,\omega}}\right) \mathbb{E}\left(z^C_{\omega;\alpha_2;j^C_{2,\omega}} z^C_{\omega;\alpha_3;j^C_{3,\omega}}\right).
\end{aligned} \quad (197)$$

We proceed by direct computation. We have

$$\begin{aligned}
&\mathbb{E}\left(z^C_{\omega;\alpha_1;j^C_{1,\omega}} z^C_{\omega;\alpha_2;j^C_{2,\omega}} z^C_{\omega;\alpha_3;j^C_{3,\omega}} z^C_{\omega;\alpha_4;j^C_{4,\omega}}\right) \\
&= \mathbb{E}\left(\begin{pmatrix}\left(\sum_{j_{1,\omega}=1}^{\dim \mathcal{O}^\omega(\mathcal{H}^\omega)} W^\omega_{j^C_{1,\omega},j_{1,\omega}} z^Q_{\omega;\alpha_1;j_{1,\omega}} + b^\omega_{j^C_{1,\omega}}\right)\left(\sum_{j_{2,\omega}=1}^{\dim \mathcal{O}^\omega(\mathcal{H}^\omega)} W^\omega_{j^C_{2,\omega},j_{2,\omega}} z^Q_{\omega;\alpha_2;j_{2,\omega}} + b^\omega_{j^C_{2,\omega}}\right) \\ \left(\sum_{j_{3,\omega}=1}^{\dim \mathcal{O}^\omega(\mathcal{H}^\omega)} W^\omega_{j^C_{3,\omega},j_{3,\omega}} z^Q_{\omega;\alpha_3;j_{3,\omega}} + b^\omega_{j^C_{3,\omega}}\right)\left(\sum_{j_{4,\omega}=1}^{\dim \mathcal{O}^\omega(\mathcal{H}^\omega)} W^\omega_{j^C_{4,\omega},j_{4,\omega}} z^Q_{\omega;\alpha_4;j_{4,\omega}} + b^\omega_{j^C_{4,\omega}}\right)\end{pmatrix}\right) \\
&= \mathbb{E}\left(\sum_{j_{1,\omega},j_{2,\omega},j_{3,\omega},j_{4,\omega}=1}^{\dim \mathcal{O}^\omega(\mathcal{H}^\omega)} W^\omega_{j^C_{1,\omega},j_{1,\omega}} W^\omega_{j^C_{2,\omega},j_{2,\omega}} W^\omega_{j^C_{3,\omega},j_{3,\omega}} W^\omega_{j^C_{4,\omega},j_{4,\omega}} z^Q_{\omega;\alpha_1;j_{1,\omega}} z^Q_{\omega;\alpha_2;j_{2,\omega}} z^Q_{\omega;\alpha_3;j_{3,\omega}} z^Q_{\omega;\alpha_4;j_{4,\omega}}\right) \\
&+ \mathbb{E}\left(b^\omega_{j^C_{3,\omega}} b^\omega_{j^C_{4,\omega}} \sum_{j_{1,\omega},j_{2,\omega}=1}^{\dim \mathcal{O}^\omega(\mathcal{H}^\omega)} W^\omega_{j^C_{1,\omega},j_{1,\omega}} W^\omega_{j^C_{2,\omega},j_{2,\omega}} z^Q_{\omega;\alpha_1;j_{1,\omega}} z^Q_{\omega;\alpha_2;j_{2,\omega}}\right) + (5 \text{ perms.}) \\
&+ \mathbb{E}\left(b^\omega_{j^C_{1,\omega}} b^\omega_{j^C_{2,\omega}} b^\omega_{j^C_{3,\omega}} b^\omega_{j^C_{4,\omega}}\right).
\end{aligned} \quad (198)$$

The notation 5 perms means that $\binom{4}{2} - 1 = 5$ permutations of indices 1,2,3,4. Now, combining with the disconnected four-point function result,

$$\begin{aligned}
&\mathbb{E}\left(z^C_{\omega;\alpha_1;j^C_{1,\omega}} z^C_{\omega;\alpha_2;j^C_{2,\omega}}\right) \mathbb{E}\left(z^C_{\omega;\alpha_3;j^C_{3,\omega}} z^C_{\omega;\alpha_4;j^C_{4,\omega}}\right) \\
&= \delta_{j^C_{1,\omega},j^C_{2,\omega}} \delta_{j^C_{3,\omega},j^C_{4,\omega}} \left(\begin{pmatrix}\frac{C^\omega_W}{\dim \mathcal{O}^\omega(\mathcal{H}^\omega)} \sum_{j_{1,\omega}=1}^{\dim \mathcal{O}^\omega(\mathcal{H}^\omega)} \mathbb{E}\left(z^Q_{\omega;\alpha_1;j_{1,\omega}} z^Q_{\omega;\alpha_2;j_{1,\omega}}\right) + C^\omega_b\end{pmatrix}\begin{pmatrix}\frac{C^\omega_W}{\dim \mathcal{O}^\omega(\mathcal{H}^\omega)} \sum_{j_{2,\omega}=1}^{\dim \mathcal{O}^\omega(\mathcal{H}^\omega)} \mathbb{E}\left(z^Q_{\omega;\alpha_3;j_{2,\omega}} z^Q_{\omega;\alpha_4;j_{2,\omega}}\right) + C^\omega_b\end{pmatrix}\right) \\
&= \delta_{j^C_{1,\omega},j^C_{2,\omega}} \delta_{j^C_{3,\omega},j^C_{4,\omega}} \begin{pmatrix}\left(\frac{C^\omega_W}{\dim \mathcal{O}^\omega(\mathcal{H}^\omega)}\right)^2 \left(\sum_{j_{1,\omega}=1}^{\dim \mathcal{O}^\omega(\mathcal{H}^\omega)} \mathbb{E}\left(z^Q_{\omega;\alpha_1;j_{1,\omega}} z^Q_{\omega;\alpha_2;j_{1,\omega}}\right)\right)\left(\sum_{j_{2,\omega}=1}^{\dim \mathcal{O}^\omega(\mathcal{H}^\omega)} \mathbb{E}\left(z^Q_{\omega;\alpha_3;j_{2,\omega}} z^Q_{\omega;\alpha_4;j_{2,\omega}}\right)\right) \\ + \frac{C^\omega_W C^\omega_b}{\dim \mathcal{O}^\omega(\mathcal{H}^\omega)}\left(\sum_{j_{1,\omega}=1}^{\dim \mathcal{O}^\omega(\mathcal{H}^\omega)} \mathbb{E}\left(z^Q_{\omega;\alpha_1;j_{1,\omega}} z^Q_{\omega;\alpha_2;j_{1,\omega}}\right) + \sum_{j_{2,\omega}=1}^{\dim \mathcal{O}^\omega(\mathcal{H}^\omega)} \mathbb{E}\left(z^Q_{\omega;\alpha_3;j_{2,\omega}} z^Q_{\omega;\alpha_4;j_{2,\omega}}\right)\right) \\ +(C^\omega_b)^2\end{pmatrix},
\end{aligned} \quad (199)$$

and its two other $t$ (14-23) and $u$-channel (13-24) permutations, we have

$$\begin{aligned}
&\mathbb{E}_{\text{conn}}\left(z^C_{\omega;\alpha_1;j^C_{1,\omega}} z^C_{\omega;\alpha_2;j^C_{2,\omega}} z^C_{\omega;\alpha_3;j^C_{3,\omega}} z^C_{\omega;\alpha_4;j^C_{4,\omega}}\right) \\
&= \left(\frac{C^\omega_W}{\dim \mathcal{O}^\omega(\mathcal{H}^\omega)}\right)^2 \begin{pmatrix}\delta_{j^C_{1,\omega},j^C_{2,\omega}} \delta_{j^C_{3,\omega},j^C_{4,\omega}} \sum_{j_{1,\omega},j_{2,\omega}=1}^{\dim \mathcal{O}^\omega(\mathcal{H}^\omega)} \mathbb{E}_{\text{conn}}\left(z^Q_{\omega;\alpha_1;j_{1,\omega}} z^Q_{\omega;\alpha_2;j_{1,\omega}} z^Q_{\omega;\alpha_3;j_{2,\omega}} z^Q_{\omega;\alpha_4;j_{2,\omega}}\right) \\ +\delta_{j^C_{1,\omega},j^C_{3,\omega}} \delta_{j^C_{2,\omega},j^C_{4,\omega}} \sum_{j_{1,\omega},j_{2,\omega}=1}^{\dim \mathcal{O}^\omega(\mathcal{H}^\omega)} \mathbb{E}_{\text{conn}}\left(z^Q_{\omega;\alpha_1;j_{1,\omega}} z^Q_{\omega;\alpha_3;j_{1,\omega}} z^Q_{\omega;\alpha_2;j_{2,\omega}} z^Q_{\omega;\alpha_4;j_{2,\omega}}\right) \\ +\delta_{j^C_{1,\omega},j^C_{4,\omega}} \delta_{j^C_{2,\omega},j^C_{3,\omega}} \sum_{j_{1,\omega},j_{2,\omega}=1}^{\dim \mathcal{O}^\omega(\mathcal{H}^\omega)} \mathbb{E}_{\text{conn}}\left(z^Q_{\omega;\alpha_1;j_{1,\omega}} z^Q_{\omega;\alpha_4;j_{1,\omega}} z^Q_{\omega;\alpha_2;j_{2,\omega}} z^Q_{\omega;\alpha_3;j_{2,\omega}}\right)\end{pmatrix}.
\end{aligned} \quad (200)$$

Here, the connected piece made by quantum circuits is given by

$$\mathbb{E}_{\mathrm{conn}}\left(z^Q_{\omega;\alpha_1;j_{1,\omega}} z^Q_{\omega;\alpha_2;j_{1,\omega}} z^Q_{\omega;\alpha_3;j_{2,\omega}} z^Q_{\omega;\alpha_4;j_{2,\omega}}\right)$$
$$\equiv \mathbb{E}\left(z^Q_{\omega;\alpha_1;j_{1,\omega}} z^Q_{\omega;\alpha_2;j_{1,\omega}} z^Q_{\omega;\alpha_3;j_{2,\omega}} z^Q_{\omega;\alpha_4;j_{2,\omega}}\right)$$
$$- \mathbb{E}\left(z^Q_{\omega;\alpha_1;j_{1,\omega}} z^Q_{\omega;\alpha_2;j_{1,\omega}}\right) \mathbb{E}\left(z^Q_{\omega;\alpha_3;j_{2,\omega}} z^Q_{\omega;\alpha_4;j_{2,\omega}}\right) . \tag{201}$$

Now, we note that since we have $\mathcal{O}\left((\dim \mathcal{O}^\omega(\mathcal{H}^\omega)))^2\right)$ terms in the sum, the connected part will at most scale as $\mathcal{O}(1)$ since the quantum outputs are made by normalized states whose norms are bounded by 1. However, if we assume

$$\mathbb{E}_{\mathrm{conn}}\left(z^Q_{\omega;\alpha_1;j_{1,\omega}} z^Q_{\omega;\alpha_2;j_{1,\omega}} z^Q_{\omega;\alpha_3;j_{2,\omega}} z^Q_{\omega;\alpha_4;j_{2,\omega}}\right) = \mathcal{O}(1) \times \delta_{j_{1,\omega},j_{2,\omega}} , \tag{202}$$

and their permutations for all $\omega$s, we have

$$\mathbb{E}_{\mathrm{conn}}\left(z^Q_{\omega;\alpha_1;j_{1,\omega}} z^Q_{\omega;\alpha_2;j_{1,\omega}} z^Q_{\omega;\alpha_3;j_{2,\omega}} z^Q_{\omega;\alpha_4;j_{2,\omega}}\right)$$
$$= \left(\frac{C_W^\omega}{\dim \mathcal{O}^\omega(\mathcal{H}^\omega)}\right)^2 \left(\delta_{j^C_{1,\omega},j^C_{2,\omega}}\delta_{j^C_{3,\omega},j^C_{4,\omega}} + \delta_{j^C_{1,\omega},j^C_{3,\omega}}\delta_{j^C_{2,\omega},j^C_{4,\omega}} + \delta_{j^C_{1,\omega},j^C_{4,\omega}}\delta_{j^C_{2,\omega},j^C_{3,\omega}}\right) \times \sum_{j_{1,\omega},j_{2,\omega}=1}^{\dim \mathcal{O}^\omega(\mathcal{H}^\omega)} \delta_{j_{1,\omega},j_{2,\omega}} \times \mathcal{O}(1)$$
$$= \left(\frac{C_W^\omega}{\dim \mathcal{O}^\omega(\mathcal{H}^\omega)}\right)^2 \times \mathcal{O}(\dim \mathcal{O}^\omega(\mathcal{H}^\omega)) \times \left(\delta_{j^C_{1,\omega},j^C_{2,\omega}}\delta_{j^C_{3,\omega},j^C_{4,\omega}} + \delta_{j^C_{1,\omega},j^C_{3,\omega}}\delta_{j^C_{2,\omega},j^C_{4,\omega}} + \delta_{j^C_{1,\omega},j^C_{4,\omega}}\delta_{j^C_{2,\omega},j^C_{3,\omega}}\right)$$
$$= \frac{(C_W^\omega)^2}{\dim \mathcal{O}^\omega(\mathcal{H}^\omega)} \times \mathcal{O}(1) \times \left(\delta_{j^C_{1,\omega},j^C_{2,\omega}}\delta_{j^C_{3,\omega},j^C_{4,\omega}} + \delta_{j^C_{1,\omega},j^C_{3,\omega}}\delta_{j^C_{2,\omega},j^C_{4,\omega}} + \delta_{j^C_{1,\omega},j^C_{4,\omega}}\delta_{j^C_{2,\omega},j^C_{3,\omega}}\right) . \tag{203}$$

Thus, the connected part is suppressed by the large width $1/\dim \mathcal{O}^\omega(\mathcal{H}^\omega)$.

The orthogonal condition we impose for quantum neural networks might be satisfied by random assumptions of choices of operators $O$ or random variational ansätze (say, averaging over the Pauli group or the 1-design). Similar arguments could be made for the dynamical NTK. It is beyond our scope in this work to discuss how to connect the suppressed non-Gaussian correlations randomized over initialization, and the suppression of dNTK (dQNTK) in dynamics in the large width. Thus, we hereby make more general arguments and leave the detailed work in the future. Consider the schematic formula of dQNTK in the hybrid network

$$\mu^{i_0 i_1 i_2}_{\delta_0 \delta_1 \delta_2} = \sum_{\ell_1,\ell_2} \frac{d^2 z_{i_0;\delta_0}}{d\theta_{\ell_1} d\theta_{\ell_2}} \left(\frac{dz_{i_1;\delta_1}}{d\theta_{\ell_1}} \frac{dz_{i_2;\delta_2}}{d\theta_{\ell_2}}\right) , \tag{204}$$

We schematically use notation $\theta_\ell$ to denote both classical and quantum training variables. The sum will lead to a combination of the following three cases: classical contribution, quantum contribution, and the classical-quantum mixed contribution. For pure classical contribution, the formula from Gaussian correlations will lead to 1/width suppression. For quantum contribution and its mixture with classical ones, the meta-kernel will be naturally suppressed by $1/\dim \mathcal{O}(\mathcal{H})$ for its corresponding operator space dimension $\dim \mathcal{O}(\mathcal{H})$. Otherwise, there are non-negligible terms that give $\mathcal{O}(1)$ modification to NTK during the gradient descent, leading to a non-linear regime of training dynamics and representation learning. Otherwise, the training dynamics will linearize, and we get an exponential convergence of the residual training error. The observations indicate a possible connection between barren plateau, random unitary, and large-width limit in classical neural networks, where we will leave details to our future works.

## VI. ON THE LECUN PARAMETRIZATION AND THE NTK PARAMETRIZATION

In this section, we will clarify the issue of the initialization (parametrization) convention we use comparing to different literature. In several original papers about classical NTK [7], there is a convention which is called the NTK parametrization, which defines the Gaussian correlation of weights as

$$\mathbb{E}\left(W_{ij} W_{kl}\right) = \delta_{ik} \delta_{jl} C_W . \tag{205}$$

The initialization will lead to rescaling of parameters as the following table,

In our work, we use the standard (LeCun) scaling, where we will still naturally obtain the natural NTK, and the dNTK suppression results [1]. However, there are also studies pointing out alternative parametrizations [8].

| Type | Weight | Single-layer | Learning rate |
|------|--------|--------------|---------------|
| LeCun | $\mathbb{E}(W_{ij}W_{kl}) = \delta_{ik}\delta_{jl}C_W/\text{width}$ | $Wx+b$ | $\eta = \mathcal{O}(1)/\text{width}$ |
| NTK | $\mathbb{E}(W_{ij}W_{kl}) = \delta_{ik}\delta_{jl}C_W$ | $(Wx+b)/\sqrt{\text{width}}$ | $\eta = \mathcal{O}(1)$ |

TABLE I. Comparison between the LeCun (standard) parametrization and the NTK parametrization.

## VII. SIMULATION DETAILS AND ADDITIONAL DATA

We simulate the NTK dynamics in a concrete example using the Qiskit python library. In the main text, we consider a binary classification task carried out with a 3-qubit quantum neural network (QNN) in a supervised learning setting, using an ad-hoc data set provided in Qiskit Machine Learning. The input data points have 3-dimensional features, and are partitioned in a training set (20 elements) and a test set (5 elements). Each input $\mathbf{x} = (\text{x[0],x[1],x[2]})$ is encoded through the `ZZFeatureMap`

$$\text{ZZFeatureMap} = \left( \text{[circuit diagram]} \right)^2, \quad (206)$$

where $\phi(\theta)$ is the single-qubit phase gate and $\hat{x}_i = \pi - x_i$. The trainable part of the quantum neural network, which appended to the quantum circuit after the feature map, is represented by a 3-layer `RealAmplitudes` variational ansatz, combining parametrized $R_y$ rotations and entangling CNOT operations:

`RealAmplitudes = `

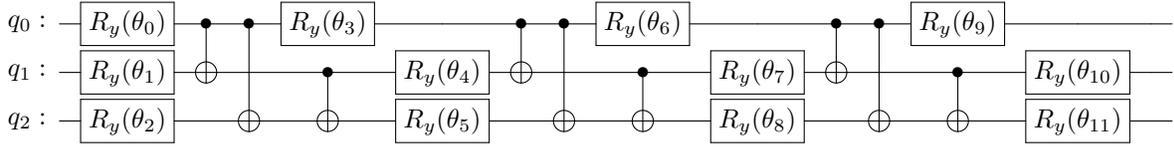

$$. \quad (207)$$

The QNN output is obtained as the expectation value of the 3-qubit observable $Z_0 Z_1 Z_2$, where $Z_i$ is the single qubit Pauli $Z$ operator. Our model is

$$z_\delta = \langle \phi(\mathbf{x}_\delta)| U^\dagger O U |\phi(\mathbf{x}_\delta)\rangle , \quad (208)$$

where $O$ is given by

$$O = \text{diag}(1, -1, -1, 1, -1, 1, 1, -1) . \quad (209)$$

Note that we only have one observable $O$, so our quantum neural network only has a scalar as the output. In this setup, the QNTK is given by

$$K_{\delta,\tilde{\alpha}} = \sum_\ell \frac{dz_\delta}{d\theta_\ell}\frac{dz_{\tilde{\alpha}}}{d\theta_\ell} = -\sum_\ell \begin{pmatrix} \langle \phi(\mathbf{x}_\delta)|U^\dagger_{+,\ell}\left[X_\ell, U^\dagger_\ell W^\dagger_\ell U^\dagger_{-,\ell}OU_{-,\ell}W_\ell U_\ell\right]U_{+,\ell}|\phi(\mathbf{x}_\delta)\rangle \times \\ \langle \phi(\mathbf{x}_{\tilde{\alpha}})|U^\dagger_{+,\ell}\left[X_\ell, U^\dagger_\ell W^\dagger_\ell U^\dagger_{-,\ell}OU_{-,\ell}W_\ell U_\ell\right]U_{+,\ell}|\phi(\mathbf{x}_{\tilde{\alpha}})\rangle \end{pmatrix}, \quad (210)$$

and the frozen QNTK is given by

$$K_{\delta,\tilde{\alpha}} = -\delta^2 \sum_\ell \begin{pmatrix} \langle \phi(\mathbf{x}_\delta)|W^\dagger_{+,\ell}\left[X_\ell, W^\dagger_\ell W^\dagger_{-,\ell}OW_{-,\ell}W_\ell\right]W_{+,\ell}|\phi(\mathbf{x}_\delta)\rangle \times \\ \langle \phi(\mathbf{x}_{\tilde{\alpha}})|W^\dagger_{+,\ell}\left[X_\ell, W^\dagger_\ell W^\dagger_{-,\ell}OW_{-,\ell}W_\ell\right]W_{+,\ell}|\phi(\mathbf{x}_{\tilde{\alpha}})\rangle \end{pmatrix}, \quad (211)$$

with the help of the initial angle redefinition. All the theories we have discussed could be applied in a straightforward way in our simulation case. For the theoretical prediction, we take the variational angle $\theta^*$ at the last step of our training.

29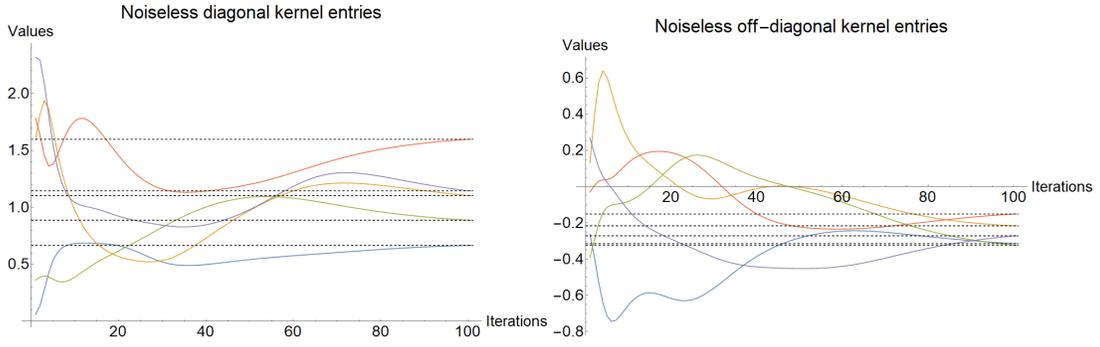

FIG. 1. Noiseless diagonal and off-diagonal entries of QNTK during the noiseless gradient descent dynamics. We compute the gradient descent evolution of five random diagonal and off-diagonal elements in the QNTK. The solid line is the actual value of the QNTK entries during the experiment, and the dashed line is the theoretical prediction of the frozen QNTK.

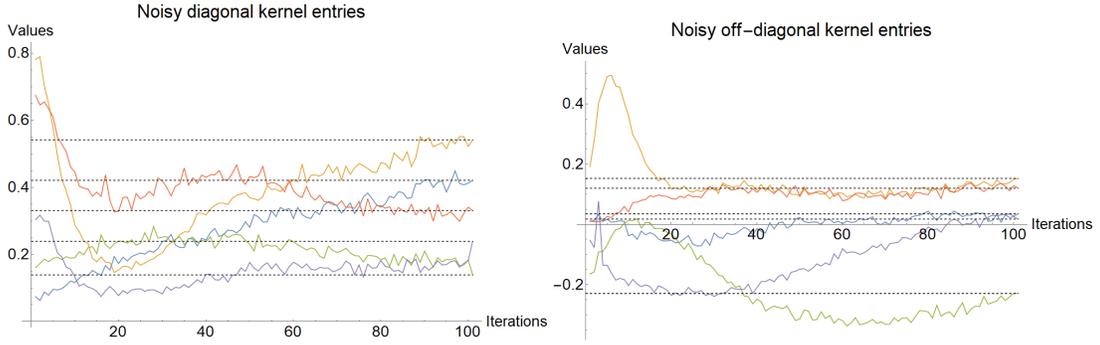

FIG. 2. Noisy diagonal and off-diagonal entries of QNTK during the gradient descent dynamics. We compute the gradient descent evolution of five random diagonal and off-diagonal elements in the QNTK. The solid line is the actual value of the QNTK entries during the simulation of a noisy quantum processor, and the dashed line is the theoretical prediction of the frozen QNTK.

In addition to the plots reported in the main text, here we show more data about the properties of QNTK during the gradient descent. See Fig. 1 and Fig. 2. We could see that, even including a model of device noise, the simulation is successful and the kernel is stable at the late time including the effect of the error mitigation. This will indicate an exponential decay of the residual training error at the late time for noisy quantum circuits.

[1] D. A. Roberts, S. Yaida, and B. Hanin, arXiv preprint arXiv:2106.10165 (2021).
[2] D. A. Roberts and S. Yaida, Deep Learning Theory Summer School at Princeton (2021).
[3] Note that the inverse has to be taken in the over-parametrized regime where the number of variational angles are larger than the training samples to make the formalism simple, which is not the case in our numerical simulation in this paper.
[4] M. Mohri, A. Rostamizadeh, and A. Talwalkar, *Foundations of machine learning* (MIT press, 2018).
[5] The convention factor $1/2$ appearing in $\eta$ is from the worldsheet index defined for holomorphic and anti-holomorphic variables. Similar conventions appear in the two-dimensional conformal field theories, see [9].
[6] L. Zhao, Z. Zhao, P. Rebentrost, and J. Fitzsimons, Quantum Machine Intelligence **3**, 1 (2021).
[7] A. Jacot, F. Gabriel, and C. Hongler, arXiv preprint arXiv:1806.07572 (2018).
[8] J. Sohl-Dickstein, R. Novak, S. S. Schoenholz, and J. Lee, arXiv preprint arXiv:2001.07301 (2020).
[9] J. Polchinski, *String theory: Volume 1, an introduction to the bosonic string* (Cambridge university press, 1998).


}

\end{document}